\documentclass[a4paper,11pt]{article}
\pdfoutput=1 % if your are submitting a pdflatex (i.e. if you have
\usepackage{jheppub} % for details on the use of the package, please
\usepackage{amsmath}
\usepackage{esvect} 
\usepackage{tikz}
\usetikzlibrary{arrows}
\usetikzlibrary{positioning}
\usepackage{amsfonts}
\usepackage{amsthm}

\usepackage{epic} 
\usepackage{mathtools}
\usepackage{url}
\usepackage{hyperref}
\usetikzlibrary{calc,patterns,angles,quotes}
\usepackage{graphicx}
\usepackage{float}
\usepackage{amssymb}
\usepackage{bm}
\usepackage{seqsplit}
\usepackage{mathrsfs}
\usepackage{comment}
\usetikzlibrary{decorations.pathmorphing}
\usepackage{tikz-cd}
\numberwithin{equation}{section}
\usepackage{braket}
\usepackage{mathtools}
\usepackage{leftidx}
\usepackage{commath}
\usepackage{slashed}
\usepackage{cancel}
\allowdisplaybreaks[4]
\usepackage{calc}
\usepackage{accents}

\usetikzlibrary{decorations.pathmorphing}
\usetikzlibrary{arrows.meta}
\usetikzlibrary{decorations.markings}

\definecolor{darkpastelgreen}{rgb}{0.01, 0.75, 0.24 }
\definecolor{hooker\'sgreen}{rgb}{0.0, 0.44, 0.0}
\definecolor{indiagreen}{rgb}{0.07, 0.53, 0.03}
\definecolor{islamicgreen}{rgb}{0.0, 0.56, 0.0}
\definecolor{amber}{rgb}{1.0, 0.75, 0.0}

\usepackage[normalem]{ulem}  % \sout{old text} for strikeout

\title{Gravitational charges and radiation in asymptotically locally de Sitter spacetimes}

\author{Aaron Poole$^{a,b}$,}
\author{Kostas Skenderis$^c$}
\author{and Marika Taylor$^d$}

\affiliation{$^a$Department of Physics and Center for Theoretical Physics, National Taiwan University, Taipei 106, Taiwan.}
\affiliation{$^b$Department of Physics and Research Institute of Basic Science, Kyung Hee University, Seoul 02447, Korea.}

\affiliation{$^c$STAG Research Centre \& Mathematical Sciences, University of Southampton, Highfield, Southampton SO17 1BJ, United Kingdom.} 

\affiliation{$^d$College of Engineering and Physical Sciences, University of Birmingham,  Birmingham B15 2TT, United Kingdom.}

\emailAdd{apoole@khu.ac.kr} 
\emailAdd{K.Skenderis@soton.ac.uk} 
\emailAdd{M.M.Taylor@bham.ac.uk}

\abstract{We present a comprehensive discussion of gravitational charges and radiation in four-dimensional asymptotically locally de Sitter (AldS) spacetimes. Such spacetimes have compact spatial sections and possess spacelike past and future infinities, $\mathscr{I}^\pm$. We show that the variational problem is well-posed if one specifies a conformal class up to three-dimensional diffeomorpshims at $\mathscr{I}^\pm$, provided one adds at $\mathscr{I}^\pm$ suitable local terms. These are analogues of the AdS covariant counterterms of holographic renormalisation. Radiation is possible only when the conformal class at infinity is non-trivial. Bulk diffeomorphisms that asymptote to conformal Killing vectors at $\mathscr{I}^\pm$ lead to gravitational charges that are conserved under radial translations, while transformations that asymptote to a general three-dimensional vector imply that the gravitational charges satisfy flux-balance laws. We also present quantities that are invariant under temporal translations, if gravitational flux is absent, but otherwise satisfy flux-balance laws. We derive these results using {\it first principles} application of Noether's method as well as covariant phase space methods.  
We apply our formalism to several exact solutions, including the Robinson-Trautman-dS class, which we use to demonstrate the existence of conserved charges even in the absence of asymptotic conformal Killing vectors and the existence of monotonic charges, including the Bondi mass.

} 

\begin{document}
\maketitle
\flushbottom

\section{Introduction and summary of results} \label{sec: Introduction}

In a diffeomorphism invariant theory there are no local bulk observables. Instead, if the spacetime possesses asymptotia, the theory has observables associated with structure at infinity. For example, in asymptotically flat gravity, the natural observable is the S-matrix, while in spacetimes with negative cosmological constant the boundary correlators play an analogous role.

Symmetries play an essential role in our understanding of physics. In a non-gravitational setting, symmetries give rise to conserved quantities and these constrain the dynamics of the system. Elementary examples are the energy and momentum carried by excitations whose conservation follows from translational invariance. This principle is quantified by Noether's first theorem \cite{Noether:1918zz}, which states that if the action is invariant under a global symmetry then there is a corresponding conserved current and the conserved charge arises from an appropriate integral of the charge density. 

In diffeomorphism invariant theories, the symmetries are local and we are now dealing with Noether's second theorem. The corresponding charge is now trivial in the bulk, and it may only act non-trivially at asymptotia.
Early papers on this topic in the context of asymptotically flat gravity \cite{Penrose:1962ij} include \cite{Arnowitt:1959ah} for charges at spatial infinity and \cite{Pirani:1956wr, Trautman:2016xic, Trautman:2002zz, Bondi:1962px, Sachs:1962zza} for charges at null infinity which capture the properties of outgoing radiation. The most famous example of such a quantity at null infinity is the monotonically decreasing Bondi mass, developed into a systematic charge using the frameworks of \cite{Ashtekar:1981hw, Ashtekar:1981bq, Wald:1999wa}.

Defining gravitational charges at infinity is challenging as formal expressions for the charges are typically divergent due to the infinite volume of spacetime.  For asymptotically flat gravity, the charges are typically defined relative to a reference spacetime such as the Minkowski metric (see however also \cite{Bombelli:1986sb}, for an example of a definition of charges relative to an off-shell asymptotically locally flat metric). 
With the advent of the AdS/CFT correspondence \cite{Maldacena:1997re,  Gubser:1998bc, Witten:1998qj}, significant attention was paid to the case of negative cosmological constant. Conserved charges for asymptotically locally anti-de Sitter (AlAdS) spacetimes have been worked out in complete generality in \cite{Papadimitriou:2005ii}, building on \cite{Henningson:1998gx, Henningson:1998ey, Balasubramanian:1999re, deHaro:2000vlm, Skenderis:2000in}.

In AlAdS, the computation of charges is more robust than in asymptotically flat space due to holographic renormalisation; this allows for a definition of the charges which is intrinsic to the spacetime without the need for a reference background \cite{Skenderis:2002wp, Papadimitriou:2004ap}. This leads to a first principles derivation of gravitational charges for any AlAdS spacetime \cite{Papadimitriou:2005ii} (and not merely those which have the same conformal boundary as empty AdS spacetime (asymptotically AdS = AAdS)). The charges computed holographically via Noether's theorem \cite{Papadimitriou:2005ii} are equivalent to those computed via covariant phase space techniques \cite{Crnkovic:1986ex, Zuckerman:1989cx, Lee:1990nz, Wald:1993nt, Iyer:1994ys} (see also \cite{Hollands:2005ya}). These techniques supersede the background subtraction approach that was used prior to the development of holographic renormalisation {\it e.g.} \cite{Ashtekar:1984zz, Henneaux:1984xu, Henneaux:1985tv}.

The purpose of this work is to provide a comprehensive analysis of gravitational charges for the case of positive cosmological constant, {\it i.e.} for spacetimes that are asymptotically locally de Sitter (AldS). Such spacetimes have compact spatial slices and as such standard construction of gravitational charges that involve integrals at spatial infinity do not apply. The asymptotia for such spacetimes are past infinity, $\mathscr{I}^-$, and future infinity, $\mathscr{I}^+$. 

A natural ``observable''\footnote{We have put quotation marks in {\it observable} because $\langle f | i \rangle$ cannot be measured by an observer that lives in this spacetime; these were called meta-observables in \cite{Witten:2001kn}. As in early universe cosmology, one may consider the dS universe as a phase to be joined to a subsequent hot Big Bang cosmology. Then the final states become initial conditions for the hot Big Bang cosmology, and as such they are effectively observable, for example via their imprint in the CMB.} in AldS spacetimes is the transition amplitude from past infinity to future infinity, $\langle f | i \rangle$, where the initial and final states, $|i\rangle$ and $| f\rangle$,  are determined by a conformal class of Riemannian three-dimensional metrics up to three-dimensional diffeomophisms, one at past infinity and a second  (in general different) at future infinity. The pairing  $\langle f | i \rangle$ may be computed by evaluating the gravitational path integral with these boundary conditions \cite{Witten:2001kn}. It was argued in \cite{Witten:2001kn} that this construction provides a quantum Hilbert space associated with dS quantum gravity. Here our discussion will be classical and our aim will be to derive the gravitational charges in this setup.

The first step is to specify the boundary conditions that make the variational problem well-posed. 
We show that the variational problem is well-posed for the boundary conditions discussed in the previous paragraph, namely 
we keep fixed a conformal class up to three-dimensional diffeomorphisms\footnote{In this paper we focus on AldS$_4$ but the methodology applies to all dimensions. Odd dimensional AldS$_d$ spacetimes will have additional features due to conformal anomalies \cite{Henningson:1998gx}, see \cite{Papadimitriou:2005ii} for a discussion in the context of AlAdS spacetimes.} on $\mathscr{I}^-$ and $\mathscr{I}^+$, provided we add new boundary terms  at $\mathscr{I}^-$ and $\mathscr{I}^+$ (on top of the usual Gibbons-Hawking-York term), as in the analogous discussion for AlAdS in \cite{Papadimitriou:2005ii}. These are the analogue of the counterterms originally introduced in \cite{Henningson:1998gx, Henningson:1998ey, Balasubramanian:1999re, deHaro:2000vlm, Skenderis:2000in} to remove divergences from the on-shell action, but as discussed in \cite{Papadimitriou:2005ii} they are in fact already required by the variational problem. 

The asymptotic structure of AldS spacetimes near $\mathscr{I}^+$ and $\mathscr{I}^-$  
is given by \cite{Starobinsky:1982mr}
\begin{equation} \label{intro: metric_ADM}
     ds^2 =  - d\tau_{\pm}^2 + \gamma^\pm_{ij} dx^i dx^j\,, \qquad \gamma^\pm_{ij} =  e^{2 \tau_{\pm}} \left[ g^\pm_{(0)ij} + e^{-2\tau} g^\pm_{(2)ij} \mp e^{-3\tau} g^\pm_{(3)ij} + \ldots \right]\,,
\end{equation}
where superscripts $\pm$ indicate whether this refers to $\mathscr{I}^+$ or $\mathscr{I}^-$. The asymptotic form \eqref{intro: metric_ADM} is related  to the near boundary behavior in AlAdS spacetimes \cite{Fefferman:1985zza} by analytic continuation \cite{Skenderis:2002wp}, but we emphasize that all of our results can be obtained by an AldS analysis alone without reference to analytic continuation. The variational problem that we impose is 
\begin{equation} \label{intro: well_posed_bcs}
     \delta g^\pm_{(0)ij} = 2\left(\sigma^\pm g^\pm_{(0)ij} + \nabla^\pm_{(0)(i} \zeta^\pm_{j)}\right)\,,
\end{equation}
where $\nabla^\pm_{(0)}$ is the covariant derivative of $g^\pm_{(0)ij}$, 
$\sigma^\pm(x)$  parametrize  Weyl rescalings at $\mathscr{I}^\pm$ and $\zeta^\pm(x)$ are generic boundary vector fields that parametrize boundary diffeomorphisms at $\mathscr{I}^\pm$. Einstein equations imply that $g^\pm_{(3)ij}$ satisfies,
\begin{equation} \label{intro: conserv}
    \nabla^\pm_{(0)}{}^i g^\pm_{(3)ij} =0\,, \qquad  g^\pm_{(3)ij} g^\pm_{(0)}{}^{ij} =0\, .  
\end{equation}
In the context of the AdS/CFT, $g_{(3)ij} \sim \langle T_{ij} \rangle$, where $T_{ij}$ is  the energy-momentum tensor of the dual CFT \cite{deHaro:2000vlm} and these relations encode its conservation and the fact that in a CFT $T^i_{i} =0$ (modulo conformal anomalies, which are not present in $3d$ (unless the spectrum of the CFT contains operators of suitable dimension to contribute to a conformal anomaly)). While we will not make use of holography in this paper, we will use suggestive notation by introducing
\begin{equation} \label{intro: tij}
T_{ij}^{\pm} = - \frac{3 \ell}{16\pi G} g^{(3)\pm}_{ij}\,.
\end{equation}

Now that we have a well-posed variational problem we can proceed to discuss asymptotic symmetries and conserved charges. As mentioned earlier, AldS spacetimes do not possess a spacelike infinity and the usual discussions of gravitational charges do not apply directly. 
For this reason the notion of charges in AldS is somewhat less established than that of the $\Lambda \leq 0$ cases, but nevertheless much work has still been performed upon which this paper builds. Early work on charges in asymptotically de-Sitter spacetimes can be found in \cite{Abbott:1981ff}, and this field has gathered steam in the wake of our Universe seemingly having $\Lambda >0$ \cite{SupernovaSearchTeam:1998fmf, SupernovaCosmologyProject:1998vns}. 
Recent work to understand various aspects of charges and gravitational wave theory in AldS spacetime includes general properties of radiation relative to flat space \cite{Date:2015kma, Ashtekar:2019khv} (such as the no-incoming radiation condition), the quadrupole formulae \cite{Ashtekar:2015lxa, Date:2016uzr, Dobkowski-Rylko:2022dva, Compere:2023ktn, Dobkowski-Rylko:2024jmh, Compere:2024ekl}, linearised analyses \cite{Bishop:2015kay, Ashtekar:2015lla, Hoque:2018byx, Chrusciel:2020rlz, Kolanowski:2020wfg,Chrusciel:2021ttc, Chrusciel:2023umn, Harsh:2024kcl}, application of Bondi-type coordinates \cite{Chrusciel:2016oux, Poole:2018koa, Compere:2019bua, Compere:2020lrt, Compere:2023ktn,Bonga:2023eml, McNees:2025acf}, attempts to define Bondi-type masses \cite{Szabados:2015wqa, Saw:2016isu, Saw:2017amv, Szabados:2018erf}, classifying radiation via the gravitational Poynting vector \cite{Fernandez-Alvarez:2019kdd, Fernandez-Alvarez:2020hsv, Wylleman:2020ubq, Fernandez-Alvarez:2021yog, Senovilla:2022pym, Fernandez-Alvarez:2023wal, Fernandez-Alvarez:2024bkf, Ciambelli:2024kre, Arenas-Henriquez:2025rpt, Fernandez-Alvarez:2025ivs}, asymptotic structure and symmetries \cite{Anninos:2011jp, Ashtekar:2014zfa, Fernandez-Alvarez:2021uvz, Kaminski:2022tum, Compere:2023ktn,Hoque:2025iar, Mars:2025xyx}, and non-linear analyses using the covariant phase space \cite{Balasubramanian:2001nb, Anninos:2010zf, Kelly:2012zc, PremaBalakrishnan:2019jvz, Kolanowski:2021hwo, Compere:2019bua, Compere:2020lrt, Fiorucci:2020xto, Bonga:2023eml}. The last topic is one of the approaches we will employ in this work and we will follow various aspects of the aforementioned papers. 

In our case our starting point is the action together with the boundary terms that make the variational problem well-posed. The asymptotic symmetries are determined by working out the class of bulk diffeomorphisms that preserve the action, including boundary terms. There are two cases to consider: (1) transformations that preserve the conformal class, and (2) transformations that preserve the conformal class up to three-dimensional diffeomorphisms. 

In the first case the symmetries are associated with bulk diffeomorphisms generated by a bulk vector $\xi$ which approaches a conformal Killing vector $\xi_{(0)}$ of the boundary conformal structure at a specific rate. This is the AldS analogue of the AlAdS discussion in \cite{Papadimitriou:2005ii}.  
This case is the analogue of a global symmetry in a non-gravitational system. We can implement Noether's method to work out from {\it first principles} the corresponding conserved charges. The fact that this is possible is due to having full control over the asymptotic behavior of the gravitational field, which allows us to tame the infinities due to the infinite volume of spacetime. The gravitational charges may also be derived via covariant phase methods \cite{Lee:1990nz, Wald:1993nt, Iyer:1994ys}, and we will see that all methods yield the same charges (as they should). The gravitational charges are now defined using a {\it timelike hypersurface} $C$ that ends on the past and future infinity, and the charge is given by 
\begin{align} \label{intro: H_two_ended}
    H_{\xi} |_C & = Q_{\xi}^+[C] - Q_{\xi}^-[C]\,,
\end{align}
where $Q_{\xi}^{\pm}[C]$ are quantities that are defined at $\mathscr{I}^\pm$, given by
\begin{equation} \label{intro: Q_pm}
    Q_{\xi}^{\pm}[C]  =  \int_{\partial C^{\pm}} d\sigma_i^{\pm} T_{\pm}^{ij} \xi^{\pm}_{j(0)}\,,
\end{equation}
where $d\sigma_i^{\pm}$ is the integration measure for the surface integral at $\mathscr{I}^\pm$, $T^\pm_{ij}$ can be extracted from the asymptotics, \eqref{intro: tij}, and $\xi^{\pm}_{j(0)}$ is the conformal Killing vector that $\xi$ asymptotes to at $\mathscr{I}^\pm$. Properties of the hypersurface $C$ are dictated by the fact that it should intersect the conformal boundary, and thus it must be causal to intersect the spacelike conformal boundary of AldS space. The technical steps of analytic continuation from AlAdS space will force these surfaces to be timelike. (See however \cite{Chrusciel:2020rlz, Chrusciel:2021ttc, Chrusciel:2023umn} for work involving the alternative causal choice of null cones.)

The conservation implies that the charge is invariant under radial shifts of the hypersurface $C$:  
\begin{equation} \label{intro: Hamiltonian_Conservation}
H_{\xi} |_{C_1} = H_{\xi} |_{C_2}\,,
\end{equation}
\textit{i.e.} the charges associated to the timelike slices are independent of the slice.
Classically, past and future infinity are separated by a cosmological horizon, so one may anticipate on causality grounds that conservation laws may be formulated independently at $\mathscr{I}^\pm$. This is indeed the case and one can show using \eqref{intro: conserv}
that $Q_{\xi}^+[C]$ are each conserved separately,
\begin{equation} 
Q^\pm_{\xi} |_{C_1} = Q^\pm_{\xi} |_{C_2}\,.
\end{equation}
Most of the previous literature has focused on conserved charges at future infinity. While this is sufficient in the absence of flux and for classical considerations, the two-ended nature of the timelike surfaces $C$ is an important feature of AldS spacetimes and is likely to play a role in the wider context.   

We next move to the case of the boundary conformal structure not admitting conformal Killing vectors. This case is important for understanding radiation in AldS spacetimes. Our analysis builds on earlier work on AlAdS spacetimes. The AdS Robinson-Trautman solution, which is a prototype of a radiative solution, has been analysed in \cite{Bakas:2014kfa}. An important feature of this solution is that its conformal structure at infinity is non-trivial, meaning that the boundary metric is not conformally flat. More recently, gravitational radiation in AlAdS was examined via an application of the Bondi gauge coordinate system  \cite{Poole:2018koa, Compere:2019bua} and it was found that only spacetimes which are strictly AlAdS and not AAdS can admit non-trivial gravitational radiation. 
In other words, the conformal structure must be non-trivial ({\it i.e.} not conformally flat) for the spacetime to allow radiation, as in the example of the AdS Robinson-Trautman solution. Some recent works, for example \cite{Compere:2020lrt, Fiorucci:2020xto, Grumiller:2023ahv}, have computed charges for AlAdS spacetimes, allowing for ``leaky boundary conditions'' which allow flux through the conformal boundary. We emphasize, however, that leaky boundary conditions are not needed to discuss radiation in Al(A)dS:  Dirichlet boundary conditions, where one fixes a conformal class up to diffeomorphisms, suffice.    

As discussed earlier, the variational problem where one one fixes a conformal class up to three-dimensional diffeomorphisms is well-posed 
(after we add the boundary counterterms). The set of bulk diffeomorphisms that preserve the bulk action, including boundary terms, and are consistent with this Dirichlet problem (this is case (2) above) are the ones that are determined by bulk vectors $\xi$ that asymptote to a three-dimensional vector $\xi_{(0)}^i$ with specific fall-off rate. If $\xi_{(0)}^i$ is a CKV of the boundary metric, then we fall back to case (1), where the boundary conformal structure is preserved exactly. In the case we discuss now the action (including boundary terms) is invariant under a local transformation, so the corresponding charge should vanish. This may be re-expressed as flux-balance law using the same charges \eqref{intro: Q_pm}, but with $\xi_{\pm (0)}^i$ not being a CKV, 
\begin{equation} \label{intro: flux-balance}
   \left. Q^\pm_{\xi} \right |_{C_2} -  \left. Q^\pm_{\xi} \right |_{C_1} = - \int_{B^{\pm}_{12}} \mathbf{F}^\pm_{\xi_{(0)}}\,,
\end{equation}
where, as earlier, the indices $\pm$ refers to $\mathscr{I}^\pm$, the regions $B_{12}^\pm$ are the parts of $\mathscr{I}^\pm$ between $\partial C^\pm_1$ and $\partial C_2^\pm$ (see Figure \ref{fig: dS_both_boundaries}), and the flux is given by
\begin{equation} \label{intro: flux_def}
   \mathbf{F}^\pm_{\xi_{(0)}} = - \sqrt{g^\pm_{(0)}} T_\pm^{ij} \nabla^\pm_i \xi^{\pm (0)}_{j} \, d\mu\,,
\end{equation}
We have derived this formula using Noether's method applied to the action supplemented by the boundary terms appropriate for our boundary conditions. We also derived the same formula using the covariant phase space approach. In this approach, while a Hamiltonian does not exist, a modified Hamiltonian can be constructed following  \cite{Wald:1999wa}, which is not conserved, but rather it satisfies the flux-balance law \eqref{intro: flux-balance}.

While the flux-balance law \eqref{intro: flux-balance} is valid for any $\xi_{(0)}^i$, its physical interpretation is most transparent when $\xi_{(0)}^i$ becomes a CKV in some region of $\mathscr{I}^\pm$. In such regions the charges  \eqref{intro: Q_pm} are conserved and the flux-balance law describes the change in the value of the charge because excitations carrying this conserved quantity arrive or leave $B_{12}^\pm$ through $\partial C_{1, 2}^\pm$. This is the analogue of the flux-balance law in non-gravitational open systems, where particles carrying conserved charges (such as energy, momentum, electric charges, {\it etc.}) arrive or leave a given spatial region. When we discuss examples of gravitational radiation in AldS we will specifically look for such examples. 

It is clear from the flux-balance formula \eqref{intro: Q_pm} that when $\xi_{(0)}^i$ is a CKV the flux vanishes point-wise and the charge is conserved. However, this is only a sufficient condition for conservation. For the charge to be conserved, it is enough that the integral of the flux vanishes, {\it i.e.} that $\mathbf{F}_{\xi_{(0)}}$ is exact over the compact directions of $B_{12}^{\pm}$. We will discuss examples of such conserved charges that are not associated with ACKVs. A trademark of gravitational radiation is the Bondi-mass loss formula, which measures outgoing radiation. With this in mind, one can ask more generally for the existence of {\it monotonic} charges, {\it i.e.} charges that change monotonically under radial shifts. The flux-balance formula \eqref{intro: Q_pm} provides a way to investigate this question systematically. Indeed, a necessary and sufficient condition for such monotonic charges is that the flux has definite sign. We will find several examples of such monotonic charges. 

The discussion so far has focused on properties of charges at future or past infinity and their invariance/flux-balance formula associated with radial translations. However, as discussed earlier, the bulk charges are associated with \textit{two-ended timelike hypersurfaces}, \eqref{intro: H_two_ended}, 
and building on \cite{Poole:2021avh} we also investigate quantities that are invariant under time evolution.  The inclusion of the contribution to the overall Hamiltonian from the ``past end'' of a timelike hypersurface allows us to interpret the effects of gravitational radiation in terms of the temporal change of a quantity, rather than the spacelike separated charges defined at $\mathscr{I}^+$. 
First, if there is flux both at $\mathscr{I}^\pm$,  
\begin{equation} \label{intro: mod_H_fluxes_two_ends}
     \mathcal{H}_{\xi} |_{C_2} - \mathcal{H}_{\xi} |_{C_1} = - \int_{B^{+}_{12}} \mathbf{F}^+_{\xi_{(0)}^+} + \int_{B^{-}_{12}} \mathbf{F}^-_{\xi_{(0)}^-}\,,
\end{equation}
Now, consider first the case of zero net flux, 
$\Delta F^{\pm}_\xi \equiv \int_{B^{+}_{12}} \mathbf{F}^+_{\xi_{(0)}^+}-\int_{B^{-}_{12}} \mathbf{F}^-_{\xi_{(0)}^-} =0$,   
then
\begin{equation}
    \Delta Q^+_{\xi} = \Delta Q^-_{\xi}\, ,
\end{equation}
where $\Delta Q^\pm_\xi \equiv  \left. Q^\pm_{\xi} \right |_{C_2} -  \left. Q^\pm_{\xi} \right |_{C_1}$.
In other words, while the value of the charges $Q^+_\xi$ and $Q^-_\xi$ may be different, the difference of their values between two-ended timelike hypersurfaces is invariant, provided there is a zero net flux, $\Delta F^{\pm}_\xi$. On the other hand, if there is net flux then $\Delta Q^+_{\xi}$  differs 
from $\Delta Q^-_{\xi}$ by the amount of the net flux. Thus we find that in the absence of gravitational radiation, the temporally invariant quantity is the \textit{difference between spacelike separated charges}, and in the presence of radiation, the difference between the charges is equivalent to the flux of radiation in the system.    

One may also formulate a local version of this result by considering the
two-ended timelike hypersurfaces ending at an interior the Cauchy surface, $I_t$, then 
\begin{equation}
    \Delta Q^+_{\xi} = \Delta Q^t_{\xi}\, ,
\end{equation}
provided the net flux between $\mathscr{I}^+$ and $I_t$ is zero.
Charges at $I_t$ may be defined analogously to \eqref{intro: Q_pm} provided that the Cauchy surface admits a Killing vector $\xi^{t}_{j(0)}$
\begin{equation} 
    Q_{\xi}^{t}[C] =  \int_{\partial C^{t}} d\sigma_i^{t} \pi_{t}^{ij} \xi^{t}_{j(0)}\,,
\end{equation}
where $\pi_{ij}^t$ is the canonical momentum. In a quasi-local analysis this charge would be associated with Dirichlet boundary conditions at $I_t$; the induced metric at $I_t$ is kept fixed.

We illustrate our analysis with several examples. These includes spacetimes that possess radial invariant gravitational charges: dS${}_4$, Schwarzschild-dS${}_4$ and Kerr-dS${}_4$, where in particular we compute the charges at both $\mathscr{I}^\pm$. We also discuss in detail an example that exhibits radiation, the
Robinson-Trautman solution \cite{Robinson:1960zzb}. The radiative  Robinson-Trautman de Sitter (RTdS) spacetimes, is the unique family of exact solutions which admits a null geodesic congruence of zero shear and twist and non-vanishing divergence. This is a solution of much interest due to the fact that it describes the idealised emission of gravitation waves arising from a compact source. This solution extends naturally to encompass a cosmological constant of any sign, see {\it e.g.} \cite{Chrusciel:1992tj} for work on $\Lambda = 0$, \cite{Bicak:1995vc} for work on $\Lambda > 0$, and \cite{Bicak:1997ne} for work on $\Lambda < 0$. In recent years, the case of $\Lambda < 0$ has become the topic of much interest due to its role in the AdS/CFT correspondence and non-equilibrium dynamics \cite{deFreitas:2014lia, Bakas:2014kfa, Bakas:2015hdc, Skenderis:2017dnh, Adami:2024mtu}, as well as the connection with holographic fluids \cite{Ciambelli:2017wou, Ciambelli:2018xat, Ciambelli:2018wre, Arenas-Henriquez:2025rpt}. RT metrics appear to be a crucial testing ground for properties of gravitational radiation as they admit a $\Lambda$-insensitive ``Bondi mass loss" \cite{Bondi:1962px} equation, suggesting that they always describe outgoing gravitational waves which carry positive energy away from the system. 
In this work, we focus on the comparatively less-studied $\Lambda > 0$ case, and we illustrate the construction of conserved charges even though 
$\mathscr{I}^+$ do not possess any CKV, as well as the construction of several monotonic charges, including a charge which may be identified as the ``Bondi mass''.

The structure of the paper is as follows: In Section \ref{sec: asymptotics} we will introduce \textit{asymptotically locally de Sitter} (AldS) spacetimes and give definitions for their asymptotic structure using both gauge invariant and coordinate approaches. We will show explicitly that one can move between AdS and dS asymptotics using a (triple) Wick rotation, a tool which will be of great use in constructing the charges of the AldS spacetimes. In Section \ref{sec: covariant} we introduce the \textit{covariant phase space formalism}, along the lines of \cite{Kijowski:1979dj, Crnkovic:1986ex, Lee:1990nz, Wald:1993nt, Iyer:1994ys, Iyer:1995kg, Wald:1999wa} (see also the recent works \cite{Harlow:2019yfa, Chandrasekaran:2021vyu} and references therein, and \cite{Barnich:2001jy} for an alternative but equivalent formalism), including the definitions of the basic quantities used to construct charges and the modifications  necessary to define charges in the presence of gravitational flux \cite{Wald:1999wa}.  We will work in generic $d$ spacetime dimensions here, before specialising to $d=4$ in later sections. A number of useful identities for the formalism are contained in Appendix \ref{sec: CPS_proofs}. Section \ref{sec: Charges} forms the main new content of the paper. Here we derive the asymptotic symmetries and the corresponding gravitational charges. This includes both cases where the charges are conserved as well as cases where they satisfy a flux-balance law.
The derivations are done using both Noether's theorems as well as applying the covariant phase space techniques discussed in Section \ref{sec: covariant} to the case of AldS spacetimes as defined in Section \ref{sec: asymptotics}. The discussion in this section closely parallels the analysis the AlAdS case discussed in \cite{Papadimitriou:2005ii}; we point the reader to that work for a detailed analysis of AlAdS spacetimes and their conserved charges in all spacetime dimension. Here we will focus on the case of pure gravity in $d=4$ dimensions and we will highlight some of the differences present between the AdS and dS cases, including the 
differences due to the change of causal structure and the fact that AldS spacetimes have a conformal boundary consisting of two components.

In Section \ref{sec: Examples} we give several applications of our formalism to well-known exact solutions of Einstein's equations where no radiation is present. Much of the focus in this section is on the contribution arising from the past end of the timelike hypersurfaces used to construct the Hamiltonians. In Section \ref{sec: RTdS} we give a detailed discussion of an AldS spacetime that does contain non-trivial radiation, namely the \textit{Robinson-Trautman de Sitter} (RTdS) class. We apply our formalism to the RTdS class at $\mathscr{I}^+$ and compute several interesting examples of gravitational charges, including quantities which are radially invariant as well as others which decrease monotonically with outgoing radiation. We conclude and discuss future directions in Section \ref{sec: conclusions}. We also include several appendices, with Appendices \ref{sec: CPS_proofs}-\ref{app: pullback_omega} containing technical details, and Appendix \ref{app: Quadratic_flux} a comparison of our results concerning the RTdS class with the existing literature.   

\textbf{Comment on notation}: We use the notation $Q^\pm_{\xi}$, $H_{\xi}$ and $\mathcal{H}_{\xi}$ to denote the charges:\footnote{We also used $\mathcal{Q}_{\xi}$ in Section \ref{sec: Noet2} to denote the charge derived using Noether's second theorem, which turns out to be equal to $Q_{\xi}$.} $Q^\pm_{\xi}$ denotes the charge at $\mathscr{I}^\pm$, $H_{\xi}$ is the Wald Hamiltonian that depends on the two-ended timelike hypersurface $C$, and $\mathcal{H}_{\xi}$ is the modified Wald Hamiltonian. As we find that all  give the same expression at $\mathscr{I}^+$, in later sections we mainly use the Hamiltonian notation.

\section{(A)dS asymptotics and analytic continuation} \label{sec: asymptotics}

\subsection{Asymptotically locally de Sitter spacetimes} \label{sec: AldS}

We begin by establishing our definition of \textit{asymyptotically locally de Sitter} (AldS) spacetimes. We follow in the footsteps of \cite{Skenderis:2002wp} in defining an AldS spacetime $(M, g)$ as a \textit{conformally compact Einstein metric}, terms which we will now explain the meanings of more precisely. 

 \textit{Conformally compact:} In defining this property we follow \cite{penrose_rindler_1986, Skenderis:2002wp}. Consider a manifold with boundary $\bar{M} = M \cup \partial M$ where $\partial M$ is the boundary. The metric $g$ is conformally compact if it has a second order pole at $\partial M$ and there exists a defining function $\Omega$ which satisfies 
 \begin{equation} \label{eq: defining_function}
     \left. \Omega\right|_{\partial M} = 0\,, \qquad \left. d \Omega\right|_{\partial M} \neq 0\,, \qquad \left. \Omega\right|_{M} > 0\,,
 \end{equation}
and the metric $\bar{g}$ defined by 
\begin{equation}
    \bar{g}_{\mu \nu} = \Omega^2 g_{\mu \nu}\,,
\end{equation}
extends smoothly to $\partial M$. Furthermore, the extension of $\bar{g}$ to $\partial M$ 
\begin{equation} \label{eq: g_bar}
     \left. \bar{g}_{\mu \nu} \right|_{\partial M} = g^{(0)}_{\mu \nu}\,,
\end{equation}
must be non-degenerate. We note that at $\partial M$ one cannot induce a unique metric $g_{(0)}$ but rather a conformal class of metrics $[ g_{(0)} ]$. This is because of the inherent ambiguity in choosing the defining function $\Omega$: for any function $w$ with no poles on $\partial M$, $e^w \Omega$ would also provide a perfectly good choice of defining function. 

\textit{Einstein}: The metric $g$ satisfies the vacuum\footnote{The extension to include matter fields is straightforward but it will not be discussed here. From the effective field theory perspective the cosmological constant is the lowest dimension operator and the dynamics will be controlled by it at sufficiently late times, so the discussion here will also be the leading order behaviour in the presence of matter fields.
} Einstein equations in the presence of a positive cosmological constant 
\begin{equation} \label{eq: Einstein}
    R_{\mu \nu} = \Lambda g_{\mu \nu}\,, \qquad \Lambda > 0\,. 
\end{equation}
One can show (see {\it e.g.} \cite{Ashtekar:2014zfa}) that the imposition of (\ref{eq: Einstein}) forces the boundary $\partial M$ to be a \textit{spacelike} hypersurface. In general, AldS spacetimes have both future and past boundaries, $\mathscr{I}^+$ and $\mathscr{I}^-$  respectively, with the total boundary being 
\begin{equation} \label{eq: disjoint_boundary}
    \partial M = \mathscr{I}^+ \sqcup \mathscr{I}^-\,. 
\end{equation}
As a result of having a disjoint boundary consisting of two components, we define spacetimes to be asymptotically locally de Sitter at $\mathscr{I}^+$ ($\mathscr{I}^-$) if the above properties hold with $\mathscr{I}^+$ ($\mathscr{I}^-$) replacing $\partial M$ in equations (\ref{eq: defining_function})-(\ref{eq: g_bar}). For simplicity of discussion, we will often consider a single boundary (usually $\mathscr{I}^+$) but we will also see how the asymptotics of both boundaries play an important role in the charge prescription of Section \ref{sec: Charges}.
 
These two properties define AldS spacetimes. We now define a spacetime to be \textit{asymptotically de Sitter} if, in addition to being AldS, the following two properties hold:

\begin{enumerate}
    \item $g^{(0)}_{\mu \nu}$ is \textit{conformally flat} {\it i.e.} $[ g^{(0)} ] = [ \delta ]$, where $\delta_{i j}$ is a flat 3-metric 

    \item The conformal boundary $\partial M$ is topologically equivalent to any of $\{ S^3, \mathbb{R}^3, \mathbb{R} \times S^2\}$
\end{enumerate}
We note that the second property ensures that spacetimes such as global de Sitter ($\mathscr{I}^+ \cong S^3$), de Sitter in a Poincar\'e patch ($\mathscr{I}^+ \cong \mathbb{R}^3$) and Schwarzchild de Sitter ($\mathscr{I}^+ \cong \mathbb{R} \times S^2$) all fall under the same umbrella of being asymptotically de Sitter spacetimes. We note that this definition is in contrast to that of \cite{Ashtekar:2014zfa}, where each of these spacetimes are treated with separate asymptotic definitions. For our purposes there will be no need to distinguish between such topologies and thus we will keep them all on the same footing. 

\subsection{Triple Wick rotation between AdS and dS asymptotics} \label{sec: Triple_Wick}

In definining AldS spacetimes, we have employed very similar notions to those used to define \textit{asymptotically locally anti-de Sitter} (AlAdS) spacetimes \cite{Fefferman:1985zza, penrose_rindler_1986,  Graham:1999jg, Skenderis:2002wp} (see also \cite{Poole:2018koa, Compere:2019bua, Compere:2020lrt} for more recent work employing this definition). In fact, the difference between the two asymptotic classes is entirely contained in the sign of the cosmological constant, as AlAdS spacetimes are defined entirely analogously with $\Lambda < 0$ in equation (\ref{eq: Einstein}). It is thus unsurprising that the form of the line elements are intimately related near the conformal boundaries, a relation which we shall now explain explicitly. 

In order to set our conventions for the respective actions, we begin by giving the bulk action for AlAdS gravity in four spacetime dimensions
\begin{equation} \label{eq: AdS_action}
    S_{\text{bulk}} =  \frac{1}{16 \pi G} \int d^4x \, \sqrt{-g} \left( R + \frac{6}{\ell_{\text{AdS}}^2} \right) \,.
\end{equation}
 We begin with the well-known result \cite{Fefferman:1985zza} that in a neighbourhood of the conformal boundary $\mathscr{I}$ of an AlAdS spacetime, one can always write the metric in the Fefferman-Graham (FG) form
    \begin{equation} \label{eq: FG gauge}
ds^2 = \frac{\ell^2_{\text{AdS}}}{\rho^2} \left\{ d \rho^2 + \left[g^{(0)}_{ij}(x^c) + \rho^2 g^{(2)}_{ij}(x^k) + \rho^3 g^{(3)}_{ij}(x^k) + \ldots\right] dx^i dx^i \right\}\,,
\end{equation}
where $\ell^2_{\text{AdS}} = -3/\Lambda$ is the (squared) AdS radius of the spacetime. The coordinate $\rho \in (0, \infty)$ is a radial coordinate and the conformal boundary $\mathscr{I}$ is located at $\rho = 0$. The coordinates $x^i$ are taken to run over the boundary directions. In writing the expression above for the metric, we have used dimensionless coordinates s.t. the only dimensionful object is the AdS radius.

This form of the line element is of particular use in studying the AdS/CFT correspondence \cite{Maldacena:1997re, Gubser:1998bc, Witten:1998qj} as it provides an explicit map between gravitational and CFT data \cite{Henningson:1998gx, Henningson:1998ey, deHaro:2000vlm, Skenderis:2002wp}. The term $g_{(0)}$ acts as a representative of the conformal class induced at $\mathscr{I}$ and also the background metric for the dual CFT. $g_{(2)}$ is determined locally by $g_{(0)}$, with the explicit relation being 
\begin{equation} \label{eq: g_2_AdS}
    g^{(2)}_{ij} = -R^{(0)}_{ij} + \frac{1}{4} R_{(0)} g^{(0)}_{ij}\,.
\end{equation}
Finally, we note that $g^{(3)}_{ij}$ is a (partially) undetermined term which is traceless and conserved with respect to $g^{(0)}_{ij}$
\begin{equation}
    g_{(0)}^{ij} g^{(3)}_{ij} = 0 \,, \qquad \nabla^{(0)}_{i} g_{(3)}^{ij} = 0\,,
\end{equation}
and acts as the energy-momentum tensor for the dual CFT via 
\begin{equation} \label{eq: EM_tensor_AdS}
    \langle T_{ij} \rangle = \frac{3 \ell_{\text{AdS}}}{16 \pi G} g^{(3)}_{ij}\,.
\end{equation}

As first discussed in \cite{Starobinsky:1982mr}, a very similar expansion to (\ref{eq: FG gauge}) can be employed for one of the components of the conformal boundary of an AldS spacetime, which we can take w.l.o.g. to be $\mathscr{I}^+$. The metric expansion reads
\begin{equation} \label{eq: Starobinsky_gauge}
    ds^2 = \frac{\ell^2_{\text{dS}}}{\rho_+^2} \left\{ -d \rho_+^2 + \left[g^+_{(0)ij}(\tilde{x}^k) + \rho_+^2 g^+_{(2)ij}(\tilde{x}^k) + \rho_+^3 g^+_{(3)ij}(\tilde{x}^k) + \ldots\right] d\tilde{x}^i d\tilde{x}^j \right\}\,,
\end{equation}
where $\ell_{\text{dS}}^2 = 3/\Lambda$, $\rho_+ \in (-\infty,0)$ is a time coordinate and $\rho_+ =0$ is the location of $\mathscr{I}^+$. We have again used the conventions of $\tilde{x}^i$ denoting boundary directions and all coordinates being dimensionless. Clearly, the expansions (\ref{eq: FG gauge}) and (\ref{eq: Starobinsky_gauge}) appear to be very similar but it remains to relate the spacetimes more precisely. In order to do this, we follow Appendix A of \cite{Skenderis:2002wp}, which presents the explicit analytic continuation between Euclidean AdS and Lorentzian dS. Here we will generalise this to a Lorentzian-Lorentzian mapping. We also use this as an opportunity to describe the analytic continuation using our coordinate conventions.

Before studying the metrics explicitly, it will help to gain intuition about how to take the rotation. We write the vacuum Einstein equations as 
\begin{equation}
R_{\mu \nu} = \Lambda g_{\mu \nu} \,,
\end{equation}
\noindent where $\Lambda < 0$ for AlAdS and $\Lambda > 0$ for AldS. We recall the characteristic length scales for each  
\begin{equation}
\ell_{\text{AdS}}^2=-\frac{3}{\Lambda}\,, \qquad \ell_{\text{dS}}^2=\frac{3}{\Lambda}\,,
\end{equation}
\noindent and thus the Einstein equations for an AlAdS and AldS spacetime respectively read
\begin{align}
\begin{split}
R_{\mu \nu}&= -\frac{3}{\ell_{\text{AdS}}^2} g_{\mu \nu}\,,  \\
R_{\mu \nu}&= \frac{3}{\ell_{dS}^2}  g_{\mu \nu}\,,
\end{split}
\end{align}
\noindent from which we see that we can map between the two equations by taking $\ell_{\text{AdS}}^2 \leftrightarrow -\ell_{dS}^2$. Thus this is one of the Wick rotations we perform when analytically continuing from AlAdS to AldS. \par 

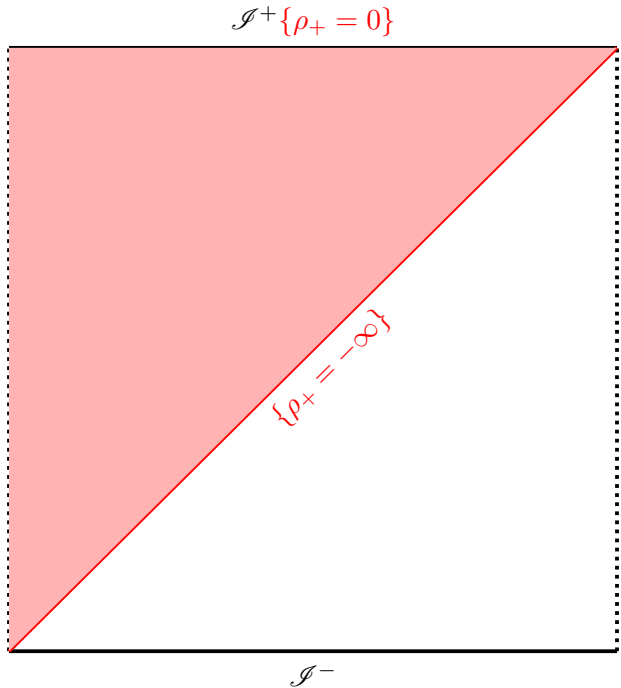
\begin{figure}[ht] 
  \begin{center}
\begin{tikzpicture}
\node (I)    at (0,0)   {};

\draw[line width=0.5mm] (-4,4) -- 
 node[midway, above]   {$\mathscr{I}^+ \color{red}{\{\rho_+=0\}}$} (4,4) ; %The conformal Boundary% 
        \draw[line width=0.5mm] (-4,-4) -- node[midway, below]   {$\mathscr{I}^-$} (4,-4);

      \draw[dotted, line width=0.5mm] (-4,-4) -- (-4,4);
      \draw[dotted, line width=0.5mm] (4,-4) -- (4,4);
    %Lines either side of the diagram%
      
      \draw[red, line width=0.5mm] (-4,-4) -- node[midway,  below, sloped] {$\{\rho_+=-\infty\}$} (4,4);
      \fill[red!30] (-4,4) -- (4,4) -- (-4,-4) -- cycle; 
  
\end{tikzpicture}
\end{center}
\caption{Penrose diagram for global de Sitter spacetime showing the region covered by the ``big bang'' metric. The dotted line on the left hand side is a coordinate singularity where $T=x_1=x_2=0$.}
\label{fig: Penrose_diagram_big_bang}
\end{figure}

To illustrate the rotations of the other variables, it will help to consider the specific example of Lorentzian $AdS_4$. We write the metric in dimensionless Poincar\'e coordinates $(\rho, t, x_1, x_2)$ as
\begin{equation}
ds^2=\frac{\ell_{\text{AdS}}^2}{\rho^2}(d\rho^2-dt^2+dx_1^2 +dx_2^2)\,,
\end{equation}
\noindent where we note by comparison with (\ref{eq: FG gauge}) that this metric has $g^{(0)}_{ij}=\eta_{ij}$ with all other $g_{(n)}$ vanishing. We perform the rotation $\ell_{\text{AdS}}^2 \rightarrow - \ell_{\text{dS}}^2$ as well as the following rotations in the other coordinates 
\begin{equation}
\rho^2 \rightarrow -\rho_+^2\,, \qquad t^2 \rightarrow -T^{\,2}\,,
\end{equation} 
\noindent which bring the line element into the form 
\begin{equation} \label{eq: Poincare dS} 
ds^2 \rightarrow \frac{\ell_{\text{dS}}^2}{\rho_+^2}(-d\rho_+^2+dT^{\,2}+dx_1^2 +dx_2^2)\,. 
\end{equation}
\noindent This metric is a line element for de Sitter spacetime which we will follow the conventions of \cite{Skenderis:2002wp} in naming this the ``big bang'' metric.  $\rho_+$ is now a time coordinate (the conformal time) and the surfaces of $\rho_+=\text{constant}$ are spacelike 3-planes. These coordinates cover half of the global geometry of $dS_4$, see for example \cite{Anninos:2012qw}, and depending upon the ranges of the coordinates one can choose which of the two spacelike boundaries of de Sitter space is covered: choosing $\rho_+ \in (-\infty, 0)$ covers $\mathscr{I}^+$ and $\rho_+ \in (0, \infty)$ $\mathscr{I}^-$ where in both cases the boundary is located at $\rho_+=0$. The case of covering $\mathscr{I}^+$ is displayed in the Penrose diagram of Figure \ref{fig: Penrose_diagram_big_bang}.
 \par 

This example shows that we can transform from $AdS_4$ to $dS_4$ by performing the triple Wick rotation  
\begin{equation} \label{eq: Triple Wick}
(\ell_{\text{AdS}}^2, \rho^2, t^2) \rightarrow -(\ell_{\text{dS}}^2, \rho_+^2, T^{\,2})\,,
\end{equation}
which will turn out to be the general map from (\ref{eq: FG gauge}) to (\ref{eq: Starobinsky_gauge}). We will explicitly implement this map in a general way via 
\begin{equation} \label{eq: Triple_Wick_explicit}
    \ell_{\text{AdS}} \rightarrow i \eta_{\ell} \ell_{\text{dS}}\,, \qquad \rho \rightarrow i \eta_{\rho} \rho_+\,, \qquad t \rightarrow i \eta_t T\,, \qquad \eta_{\ell}^2 = \eta_{\rho}^2 = \eta_t^2 = 1\,,
\end{equation}
as we note that taking any combination of the rotations in the clockwise/anticlockwise directions is valid. Working in this way is the most general form and we will see that a number of features of this choice are helpful in providing comparison with the literature, particularly regarding the sign convention in the energy-momentum tensor. 

To explicitly demonstrate the mapping from AdS to dS asymptotics, we perform the triple Wick rotation (\ref{eq: Triple_Wick_explicit}) on the generic line element of an AlAdS metric given in (\ref{eq: FG gauge}). This operation sends 
\begin{align}
    ds^2 \rightarrow \frac{\ell^2_{\text{dS}}}{\rho_+^2} \left\{ -d \rho_+^2 + \left[\bar{g}^{(0)}_{ij}(\tilde{x}^k) - \rho_+^2 \bar{g}^{(2)}_{ij}(\tilde{x}^k) - i \eta_{\rho} \rho_+^3 \bar{g}^{(3)}_{ij}(\tilde{x}^k) + \ldots\right] d\tilde{x}^i d\tilde{x}^j \right\}\,,
\end{align}
where we have introduced $\bar{g}^{(n)}_{ab}$ to denote the effect of (\ref{eq: Triple_Wick_explicit}) on the boundary directions of the expansion. Comparing with (\ref{eq: Starobinsky_gauge}), this allows us to read off
\begin{equation} \label{eq: wick_rotated_gs}
    \bar{g}^{(0)}_{ij} = g^+_{(0)ij}\,, \qquad \bar{g}^{(2)}_{ij} = -g^+_{(2)ij}\,, \qquad \bar{g}^{(3)}_{ij} = i \eta_{\rho} g^+_{(3)ij}\,, 
\end{equation}
expressions which give us all of the necessary information about the asymptotic expansion of AldS spacetimes. First, we note that the representative of the conformal class $g^+_{(0)ij}$ can be obtained directly from the AlAdS representative via the transformation (\ref{eq: Triple_Wick_explicit}). The second non-trivial component, $g^+_{(2)ij}$, is obtained from the AlAdS case by performing (\ref{eq: Triple_Wick_explicit}) and multiplying by an overall factor of $-1$, resulting in the following local expression in terms of $g^+_{(0)ij}$
\begin{equation} \label{eq: g_2_dS}
    g^+_{(2)ij} = R^+_{(0)ij} - \frac{1}{4} R^+_{(0)} g^+_{(0)ij}\,,
\end{equation}
an expression which can also be derived by explicitly transforming (\ref{eq: g_2_AdS}) and using (\ref{eq: wick_rotated_gs}). The final term of interest is $g^+_{(3)ij}$, which at first sight appears to be imaginary due to the additional factor of $i$ induced in performing (\ref{eq: Triple_Wick_explicit}). This turns out not to be the case, as dimensional analysis (see {\it e.g.} \cite{Poole:2018koa}) implies that the components of $g^{(3)}_{ij}$ transform with odd parity under the Wick rotation (\ref{eq: Triple_Wick_explicit}), producing an overall odd power of $i$ in order to ensure reality. 

We also use the rotation (\ref{eq: Triple_Wick_explicit}) in order to study the analytic continuation of the energy-momentum tensor $\langle T_{ij} \rangle$. Starting from (\ref{eq: EM_tensor_AdS}) we obtain\footnote{We note that an alternative approach to analytically continuing in $\ell^2$ is to analytically continue Newton's constant $G_{\text{AdS}} \rightarrow - G_{\text{dS}}$. For a recent discussion of this alternative approach, see \cite{Bzowski:2023nef}.}
\begin{align} \label{eq: EM_rotation}
    \langle T_{ij} \rangle \rightarrow  \frac{3 i \eta_{\ell} \ell_{\text{dS}}}{16 \pi G} \bar{g}_{(3)ij} = - \eta_{\ell} \eta_{\rho} \frac{3 \ell_{\text{dS}}}{16 \pi G} g^+_{(3)ij}\,,
\end{align}
where we will make expicit the relationship between this analytic continuation and the energy-momentum tensor for an AldS spacetime in Section \ref{sec: Charges}. For now we will work without choosing an explicit rotation and leave the factors of $\eta$ as a way to bookkeep the signs.

With the basic asymptotic definitions and notation established, we are ready to address the description of gravitational radiation in AldS spacetimes, a field where progress has been made \cite{Poole:2018koa, Compere:2019bua, Bonga:2023eml} by direct comparison between the Starobinsky \eqref{eq: Starobinsky_gauge} and Bondi-Sachs coordinates often used to study asymptotically flat spacetime near null infinity \cite{Bondi:1962px, Sachs:1962wk}. In this work we will not utilise Bondi type coordinates and will instead work directly in Starobinsky coordinates in order to compute asymptotic charges and fluxes \cite{Wald:1999wa}. This strategy follows previous applications to Al(A)dS spacetimes contained in works such as \cite{Balasubramanian:1999re, deHaro:2000vlm, Skenderis:2000in, Papadimitriou:2005ii, Compere:2020lrt, Kolanowski:2021hwo}. We will now introduce the main technology of the covariant phase space, a formalism which will allow us to construct charges as suitable integrals over cross-sections of the conformal boundary.

\section{Covariant phase space formalism} \label{sec: covariant}

\subsection{Basics - Definitions and charges}

We consider diffeomorphism invariant theories admitting a Lagrangian description on a $d$-dimensional manifold $M$. We will consider theories of the form $\mathbf{L}(\psi)$ where $\mathbf{L}$ is the Lagrangian $d$-form and $\psi$ denotes the dynamical fields of the theory (for the case of general relativity, $\psi$ is the spacetime metric $g_{\mu \nu}$). We will follow the standard procedure established in \cite{Lee:1990nz, Wald:1993nt, Iyer:1994ys} of assuming that $\mathbf{L}(\psi)$ is \textit{diffeomorphism covariant}, explicitly 
\begin{equation}
    \mathbf{L}(f^*(\psi)) = f^* \mathbf{L} (\psi)\,,
\end{equation}
where $f: M \rightarrow M$ is any diffeomorphism and $f^*$ denotes the action of $f$ on arbitrary rank tensor fields \cite{Wald:1984rg}. We also follow \cite{Lee:1990nz, Wald:1993nt, Iyer:1994ys} in assuming that $\mathbf{L}$ is locally constructed from $\psi$, finite number of derivatives of $\psi$ and any additional background structure, which is invariant under variations of $\psi$. 

With this general framework in mind, we now consider variations of $\mathbf{L}$ with respect to $\psi$. The first variation always takes the form \cite{Lee:1990nz, Wald:1993nt}
\begin{equation} \label{eq: Lagrangian_variation}
\delta \mathbf{L}(\psi) = \mathbf{E}(\psi) \delta \psi + d \bm{\Theta}(\psi, \delta \psi)\,,
\end{equation}
where $\mathbf{E}$ is referred to as the equation of motion $d$-form and the equations of motion are given by $\mathbf{E} = 0$. $\bm{\Theta}$ is a spacetime $(d-1)$-form (defined up to addition of an exact form) called the \textit{symplectic potential}\footnote{Strictly speaking, this object is a \textit{pre}-symplectic potential as there is no guarantee that it will give rise to an invertible symplectic form on phase space. We will ignore these subtleties as they will play no role in our analysis.} and we also note that this object is a $1$-form on the \textit{phase space}, {\it i.e.} the space of variations of the dynamical fields. On-shell, equation (\ref{eq: Lagrangian_variation}) reduces to 
\begin{equation} \label{eq: Lagrangian_variation_on_shell}
\delta \mathbf{L}(\psi) \approx d \bm{\Theta}(\psi, \delta \psi)\,,
\end{equation}
where we have introduced $\approx$ to denote equivalence on-shell.\footnote{We will also use this symbol to denote equivalence when both the equations of motion $\mathbf{E} = 0$ are satisfied \textit{and} $\delta \psi$ satisfies the linearised equations of motion.}

Using the symplectic potential we can construct  the \textit{symplectic current} $\boldsymbol{\omega}$, given as the antisymmetrised variation of $\bm{\Theta}$ 
\begin{equation}\label{eq: omega_def}
\boldsymbol{\omega}(\psi, \delta_1 \psi, \delta_2 \psi) = \delta_1 \bm{\Theta}(\psi, \delta_2 \psi)-\delta_2 \bm{\Theta}(\psi, \delta_1 \psi)\,,
\end{equation}
where $\delta_{1,2}$ are arbitrary (but distinct) variations. We note that $\boldsymbol{\omega}$ is a spacetime $(d-1)$-form and an exact phase space 2-form.  
A useful property of $\boldsymbol{\omega}$ is that it is closed when $\psi$ satisfies the equations of motion and $\delta \psi$ the linearised equations of motion, {\it i.e.} 
\begin{equation} \label{eq: omega_closed}
 d \boldsymbol{\omega} \approx 0\, ,
\end{equation}
a proof of which is given in Appendix \ref{sec: symplectic_current_closed}. 

With these objects established, we can now define the \textit{symplectic form} $\Omega_C$ via  
\begin{equation} \label{eq: symplectic_form}
\Omega_C(\psi, \delta_1 \psi, \delta_2 \psi) = \int_{C}  \boldsymbol{\omega}(\psi, \delta_1 \psi, \delta_2 \psi)\,,
\end{equation}
where $C$ is a spacetime hypersurface, often taken to be a Cauchy surface \cite{Wald:1993nt, Iyer:1994ys} when such a surface exists. When the spacetime possesses no Cauchy surfaces, such as the case of AlAdS spacetimes \cite{Hollands:2005wt, Papadimitriou:2005ii}, $C$ is typically a spacelike hypersurface suitably chosen so as to represent ``an instant of time''. We will revisit the causal nature of such a hypersurface in the context of AldS spacetimes in the next section. We note that $\Omega_C$ is a spacetime scalar (0-form) and a phase space 2-form. 

We will often be interested in the case when one of the variations is generated by a diffeomorphism of the spacetime manifold. We consider the case of a diffeomorphism generated by a smooth vector field $\xi$ which is taken to be a fixed quantity on the phase space $\delta \xi = 0$\footnote{For a relaxation of this condition in various contexts, see {\it e.g.} \cite{Barnich:2011mi, Compere:2020lrt, Capone:2023roc}.} and $\delta_{\xi} = \mathcal{L}_{\xi}$, where $\mathcal{L}_{\xi}$ is the Lie derivative in the direction of $\xi$. Given such a diffeomorphism, one can define the \textit{Noether current} $(d-1)$-form $\mathbf{J}$ \cite{Iyer:1994ys} via 
\begin{equation} \label{eq: Noether_current}
\mathbf{J} [ \xi ]=\bm{\Theta}(\psi, \mathcal{L}_{\xi}\psi)- i_{\xi} \mathbf{L}\,,
\end{equation}
where $i_{\xi}\mathbf{L}$ denotes the contraction of $\xi$ into the first index of $\mathbf{L}$. When on-shell, the Noether current is closed 
\begin{align}
\begin{split} \label{eq: Current_closed}
d \mathbf{J} [ \xi ]  = d\bm{\Theta}(\psi, \mathcal{L}_{\xi}\psi) - d ( i_{\xi} \mathbf{L} )  = \mathcal{L}_{\xi} \mathbf{L} - \mathbf{E}  \mathcal{L}_{\xi} \psi- (\mathcal{L}_{\xi} \mathbf{L} - i_{\xi} d \mathbf{L}) =  - \mathbf{E}  \mathcal{L}_{\xi} \psi \approx 0\,,
\end{split}
\end{align}
where we used $d \mathbf{L} = 0$, which is of course true of the exterior derivative acting on any $d$-form. Since $\mathbf{J}$ is closed on-shell, it is locally exact and thus we can write
\begin{equation} \label{eq: Q_def}
\mathbf{J}[\xi] \approx d \mathbf{Q}[\xi]\,,
\end{equation}
where $\mathbf{Q}[\xi]$ is the \textit{Noether charge} ($d-2$)-form. Notice that in a similar vein to the symplectic potential form, this equation only defines $\mathbf{Q}$ up to addition of an exact form. We can relate the variation of the Noether current form, $\delta \mathbf{J}$, to the symplectic current form 
\begin{align} \label{eq: J_omega_rel}
\begin{split}
\delta \mathbf{J} [\xi] \approx \boldsymbol{\omega} (\psi, \delta \psi, \mathcal{L}_{\xi} \psi) + d i_{\xi} \bm{\Theta}(\psi, \delta \psi)\,,
\end{split}
\end{align}
and thus we can utilise equation (\ref{eq: Q_def}) in order to write 
\begin{align} \label{eq: on_shell_omega}
\begin{split}
\boldsymbol{\omega} (\psi, \delta \psi, \mathcal{L}_{\xi} \psi) \approx d [\delta \mathbf{Q}[\xi] - i_{\xi} \bm{\Theta}(\psi, \delta \psi)]\,,
\end{split}
\end{align}
the derivation of which is given in Appendix \ref{sec: Noether_current_variation}. Notice that $\boldsymbol{\omega}$ is exact for these field configurations, thus providing us with an alternative way of proving (\ref{eq: omega_closed}).

With all of this technology established, we are finally at the stage where we can introduce the concept of Wald Hamiltonians. We will first give the basic definition of the Hamiltonian and then give two (equivalent) necessary and sufficient conditions for existence of such a Hamiltonian. Again, we consider a diffeomorphism of the spacetime $M$ generated by a fixed vector field $\xi$ and we define the \textit{Hamiltonian} conjugate to the vector field $\xi$, $H_{\xi}$, as the function which satisfies the following equation
\begin{equation} \label{eq: Hamiltonian_def}
\delta H_{\xi} = \Omega_{C} (\psi, \delta \psi, \mathcal{L}_{\xi} \psi)= \int_{C} \boldsymbol{\omega} (\psi, \delta \psi, \mathcal{L}_{\xi} \psi)\,,
\end{equation}
when we take $C$ to be a hypersurface, often referred to as a `slice' following the terminology of \cite{Wald:1999wa}. If $C$ is a spacelike hypersurface, we can view $H_{\xi}$ as giving us a natural definition of a conserved quantity associated with the diffeomorphism generated by $\xi$ at the ``instant of time'' $C$. We will follow \cite{Wald:1999wa} in assuming that the integral on the right hand side of (\ref{eq: Hamiltonian_def}) converges for all $\psi$ which solve the equations of motion and $\delta \psi$ which solve the linearised equations of motion and we will now restrict our consideration to this case. Henceforth, we will drop the notation $\approx$ for brevity.

In order to derive the existence conditions for the Hamiltonian we start by using (\ref{eq: on_shell_omega}) to write the variation of the Hamiltonian as
\begin{equation} \label{eq: on_shell_Hamiltonian}
\delta H_{\xi} = \int_{C} d [\delta \mathbf{Q}[\xi] - i_{\xi} \bm{\Theta}(\psi, \delta \psi)]= \int_{\partial C} \delta \mathbf{Q}[\xi] - i_{\xi} \bm{\Theta}(\psi, \delta \psi)\,,
\end{equation}
where we have applied Stokes' theorem and the interpretation of the integral over $\partial C$ is that one takes the integral over the co-dimension 2 manifold given by a cut of $C$ in the limit to asymptotic infinity \cite{Wald:1999wa}. We will remain agnostic about the character ({\it i.e.} timelike, null or spacelike) of both the hypersurface $C$ and the conformal boundary of the spacetime (which we will generically denote as $\mathscr{I}$) and we will write $\partial C= C \cap \mathscr{I}$. 

The Hamiltonian, $H_{\xi}$, exists if (\ref{eq: on_shell_Hamiltonian}) can be integrated to give $H_{\xi}$, {\it i.e.} if we can write the right hand side as a total variation. The first term on the right hand side of (\ref{eq: on_shell_Hamiltonian}) is already a variation so it is only the second term which we need to worry about. By inspecting this term, we can immediately write a necessary and sufficient condition for existence of $H_{\xi}$, namely that there exists a $(d-1)$-form $\mathbf{B}$ such that
\begin{equation} \label{eq: B}
\int_{C \cap \mathscr{I}} i_{\xi} \bm{\Theta}(\psi, \delta \psi) = \delta \int_{C \cap \mathscr{I}} i_{\xi} \mathbf{B}(\psi)\,,
\end{equation} 
and thus  
\begin{equation} \label{eq: Hamiltonian_integrable}
H_{\xi}= \int_{C \cap \mathscr{I}} \mathbf{Q}[\xi] - i_{\xi} \mathbf{B}(\psi)\,,
\end{equation}
up to addition of terms contained in the kernel of the variation operator $\delta$ \cite{Papadimitriou:2005ii}.

An alternative condition for existence of $H_{\xi}$ can be derived by considering $[\delta_1, \delta_2] H_{\xi}$, which of course must vanish by the commutativity of the variational derivatives. 
\begin{align} \label{eq: H_commutator}
\begin{split} 
[\delta_1, \delta_2] H_{\xi} & = \delta_1  \int_{C \cap \mathscr{I}}[ \delta_2 \mathbf{Q}[\xi] - i_{\xi} \bm{\Theta}(\psi, \delta_2 \psi)]- \delta_2   \int_{C \cap \mathscr{I}}[ \delta_1 \mathbf{Q}[\xi] - i_{\xi} \bm{\Theta}(\psi, \delta_1 \psi)] \\
& = \int_{C \cap \mathscr{I}} i_{\xi} \delta_2 \bm{\Theta}(\psi, \delta_1 \psi) - \int_{C \cap \mathscr{I}} i_{\xi} \delta_1 \bm{\Theta}(\psi, \delta_2 \psi) \\
& = \int_{C \cap \mathscr{I}} i_{\xi} \boldsymbol{\omega} (\psi, \delta_2 \psi, \delta_1 \psi)\,,
\end{split}
\end{align}
and thus the integrability condition becomes 
\begin{equation} \label{eq: H_existence}
\int_{C \cap \mathscr{I}} i_{\xi} \boldsymbol{\omega} (\psi, \delta_2 \psi, \delta_1 \psi) = 0\,.
\end{equation}
It is clear that this condition is necessary for existence but it is also sufficient (proved in \cite{Wald:1999wa}) and is thus equivalent to (\ref{eq: B}). In Section \ref{sec: Charges} we will use this formalism to compute the Wald Hamiltonians (when they exist!) for AldS spacetimes but first we will provide an analysis of the integrability conditions and a brief discussion of the cases when the Hamiltonians both do and do not exist.  

\subsection{Fluxes - Modification procedure} \label{sec: Modified_H}

A particularly interesting topic of study in the covariant phase space formalism is that of non-existence of the Wald Hamiltonians, a topic which was instigated in \cite{Wald:1999wa} (and therein applied to asymptotically flat spacetimes at null infinity) and has since been one of great interest in various contexts, including in Al(A)dS spacetimes \cite{Anninos:2010zf, Anninos:2011jp, Compere:2019bua, Compere:2020lrt, Fiorucci:2020xto, Kolanowski:2021hwo}. Here we will give a brief review of the techniques required to define a ``modified Hamiltonian" when the integrability condition (\ref{eq: H_existence}) fails and discuss the physics behind this failure of existence. 

\begin{figure}[H]
    \centering
    \begin{minipage}{0.5\textwidth}
    \centering
    \includegraphics[height=6cm]{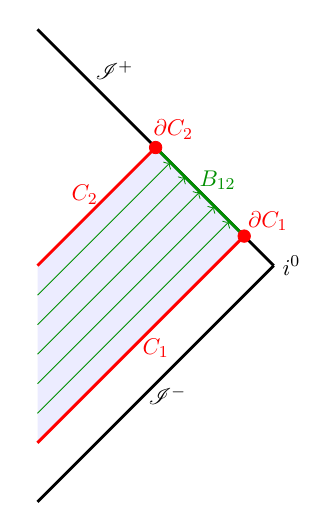}
    \end{minipage}\hfill 
    \begin{minipage}{0.5\textwidth}
    \centering
    \includegraphics[height=6cm]{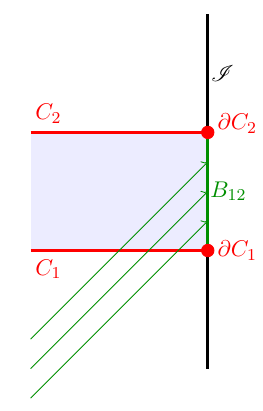}
    \end{minipage}
    \caption{Penrose diagrams corresponding to two examples of the setup described in (\ref{eq: H_difference}), for asymptotically flat and AlAdS spacetimes respectively. In both figures the green arrows represent outgoing gravitational radiation, leading to the non-existence of the Hamiltonians $H_{\xi}$.}
    \label{fig: Flux_diagrams}
\end{figure}

We first consider the difference between (the variations of) two Hamiltonians: consider two hypersurfaces $C_{1,2}$ which intersect  the conformal boundary at different places, {\it i.e.} $\partial C_{1} \neq \partial C_2$. Using equation (\ref{eq: on_shell_Hamiltonian}), we can use Stokes' theorem in reverse to write 
\begin{equation} \label{eq: H_difference}
    \left. \delta H_{\xi} \right|_{\partial C_2} -  \left. \delta H_{\xi} \right|_{\partial C_1} = - \int_{B_{12}} \boldsymbol{\omega} (\psi, \delta \psi, \mathcal{L}_{\xi} \psi),
\end{equation}
where $B_{12}$ is a portion of the conformal boundary with $\partial B_{12} = \partial C_1 \sqcup  \partial C_2$ and the minus sign on the right hand side arises due to the opposite orientation induced on the cuts $\partial C_{1,2}$ when embedded in the conformal boundary as opposed to the hypersurface $C$. Two examples of this setup are shown in Figure \ref{fig: Flux_diagrams} below

There are a number of interesting cases of the above formula, a couple of which we will recap here. The first possibility is simply $\left. \boldsymbol{\omega} (\psi, \delta \psi, \mathcal{L}_{\xi} \psi) \right|_{B_{12}} = 0$, in which case the right hand side of (\ref{eq: H_difference}) vanishes and the integrability condition (\ref{eq: H_existence}) is also satisfied. Physically speaking, this means that the Hamiltonians $H_{\xi}$ exist and are independent of the slice $C$. 
In other words, they are conserved under evolution that takes $C_1$ to $C_2$. When $C_i$ are spacelike Cauchy surfaces, this is the GR version of conservation under time evolution. This is the case that one encounters at spatial infinity in asymptotically flat spacetimes \cite{Wald:1993nt} (as all spacelike Cauchy surfaces end at the two-sphere at spatial infinity so all $\partial C_i$ are the same) and AlAdS spacetimes \cite{Papadimitriou:2005ii} when the spacetime exhibits suitable asymptotic symmetries. We will discuss the extension of this case to AldS spacetimes in Section \ref{sec: Charges}. 

There is a second case of existence when $\xi$ is everywhere tangent to $\partial C_i$, where \eqref{eq: H_existence} is satisfied as the integrand can no longer be a non-trivial $(d-2)$-form on $C \cap \mathscr{I}$. This case was discussed in \cite{Wald:1999wa} and describes the case of Hamiltonians which exist but are not conserved (the right hand side of \eqref{eq: H_difference} no longer vanishes). An example of this type of Hamiltonian is the angular momentum of a system, of which we will see several examples of in Section \ref{sec: RTdS}.

The final case is that of non-existence, {\it i.e.} $\left. \boldsymbol{\omega} (\psi, \delta \psi, \mathcal{L}_{\xi} \psi) \right|_{B_{12}} \neq 0$ and $\xi$ is not tangential to $\partial C_i$. In this case, the integrability condition (\ref{eq: H_existence}) fails and $H_{\xi}$ fails to exist. In order to understand the physics of such a situation the authors of \cite{Wald:1999wa} defined a \textit{modified Hamiltonian} $\mathcal{H}_{\xi}$ via 
\begin{equation} \label{eq: mod_H}
 \delta \mathcal{H}_{\xi} = \int_{\partial C} \delta \mathbf{Q}[\xi] - i_{\xi} [ \bm{\Theta}(\psi, \delta \psi) - \bm{\theta}(\psi, \delta \psi) ]\,,
\end{equation}
where $\bm{\theta}$ is a symplectic potential of $\bar{\bm{\omega}}$, 
\begin{equation} \label{eq: modification_criterion}
    \bar{\bm{\omega}} (\psi, \delta_1 \psi, \delta_2 \psi) = \delta_1  \bm{\theta}(\psi, \delta_2 \psi) - \delta_2  \bm{\theta}(\psi, \delta_1 \psi)\,.
\end{equation}
with $\bar{\bm{\omega}}$ the pullback of $\bm{\omega}$ 
to $B_{12}$.
Using equations (\ref{eq: H_commutator}) and (\ref{eq: mod_H}), one can now check that $[\delta_1, \delta_2] \mathcal{H}_{\xi} \equiv 0$ and thus the existence property is satisfied. However, one now finds that the modified Hamiltonians are no longer independent of the slice: taking the difference one finds 
\begin{equation} \label{eq: flux}
    \left. \delta \mathcal{H}_{\xi} \right|_{\partial C_2} -  \left. \delta \mathcal{H}_{\xi} \right|_{\partial C_1} = - \int_{B_{12}} \bar{\boldsymbol{\omega}} (\psi, \delta \psi, \mathcal{L}_{\xi} \psi) + d i_{\xi} \bm{\theta}(\psi, \delta \psi) = - \int_{B_{12}} \delta \bm{\theta} (\psi, \mathcal{L}_{\xi} \psi)\,,
\end{equation}
a formula which allows for identification of the \textit{flux form} $\mathbf{F}_{\xi}$ as
\begin{equation} \label{eq: flux_form}
    \mathbf{F}_{\xi} = \bm{\theta} (\psi, \mathcal{L}_{\xi} \psi)\,,
\end{equation}
a quantity which captures the symplectic flux passing through the boundary region $B_{12}$ of $\mathcal{H}_{\xi}$. This term is well known to be non-zero in many cases of interest, in particular that of an asymptotically flat spacetime at null infinity where $\xi$ is an element of the BMS group \cite{Bondi:1962px, Wald:1999wa, Barnich:2011mi}. We will see in Section \ref{sec: Charges} that such a term also appears in certain AldS spacetimes admitting non-trivial gravitational radiation \cite{Anninos:2010zf, Compere:2020lrt, Kolanowski:2021hwo, Poole:2021avh}.

We conclude this section by recalling the observation of \cite{Wald:1999wa} that this prescription to define a modified Hamiltonian is not unique. In particular, $\bm{\theta}$ is only defined up to the ambiguity
\begin{equation}
    \bm{\theta}(\psi, \delta \psi) \rightarrow  \bm{\theta}(\psi, \delta \psi) + \delta \mathbf{W} (\psi)
\end{equation}
where $\mathbf{W}$ is a $(d-1)$-form on $B_{12}$. In order to uniquely define $\bm{\theta}$ one needs to specify additional conditions which will select $\mathbf{W}$. Such conditions were discussed in the context of asymptotically flat spacetimes in \cite{Wald:1999wa} and in AlAdS are related to the addition of scheme dependent terms in the dual QFT. Such a discussion carries over to the AldS case and has been recently analysed in \cite{Kolanowski:2021hwo}. We will discuss this ambiguity and its resolution more explicitly in the next section. 
 
\section{Gravitational charges in asymptotically locally dS spacetimes} \label{sec: Charges}

\subsection{Theory and holographic renormalisation} \label{sec: Theory and holographic renormalisation}

First we set up the variational problem for AldS spacetimes and show that one can renormalise the action via the addition of covariant and local counterterms \cite{Henningson:1998gx, Henningson:1998ey, Balasubramanian:1999re, deHaro:2000vlm, Skenderis:2002wp, Skenderis:2000in, Papadimitriou:2004ap, Papadimitriou:2004rz, Papadimitriou:2005ii, Papadimitriou:2016yit}. In doing this, we will use the \textit{Hamiltonian} approach to holographic renormalisation introduced for AlAdS spacetimes in \cite{Papadimitriou:2004ap} (see also the exposition in \cite{Papadimitriou:2005ii, Kanitscheider:2008kd, Papadimitriou:2016yit}). In this section we will extend this method to AldS asymptotics, making comparison with AlAdS via the analytic continuation (\ref{eq: Triple_Wick_explicit}). This will serve as an alternative (but physically equivalent) approach to the well-known \textit{Lagrangian} method of holographic renormalisation of AldS spacetimes \cite{deHaro:2000vlm, Skenderis:2002wp}, a technique which has also been recently utilised to derive charges  of AldS spacetimes \cite{Compere:2020lrt, Kolanowski:2021hwo}. 

We consider pure gravity in $d=4$ in the presence of a positive cosmological constant $\Lambda = 3/\ell^2 >0$ and thus the only dynamical field of the theory is the metric tensor $\psi = g_{\mu \nu}$. The bulk action reads 
\begin{equation} \label{eq: bulk_action}
S_{\text{bulk}} = \frac{1}{16 \pi G} \int d^4x \, \sqrt{-g} \left(R - \frac{6}{\ell^2} \right)\,,
\end{equation}
from which we can extract the bulk Lagrangian as 
\begin{equation} \label{eq: Lagrangian}
    \mathbf{L}_{\text{bulk}} = \frac{1}{16 \pi G} \bm{\epsilon} (R - 2 \Lambda) \,,
\end{equation}
where the volume form $\bm{\epsilon}$ is defined via $\epsilon_{0123} = \sqrt{-g}$ such that
\begin{equation} \label{eq: volume_form}
    \bm{\epsilon} = \frac{1}{4!} \epsilon_{\mu \nu \rho \sigma} dx^{\mu} \wedge dx^{\nu} \wedge dx^{\rho} \wedge dx^{\sigma} = \sqrt{-g}\, dx^0 \wedge dx^{1} \wedge dx^{2} \wedge dx^{3}\,.
\end{equation}
As explained in equation (\ref{eq: Lagrangian_variation}) we find
\begin{equation}
    \delta \mathbf{L}_{\text{bulk}} = \textbf{E}^{\mu \nu} \delta g_{\mu \nu} + d \bm{\Theta}\,,
\end{equation}
where 
\begin{align}
    \textbf{E}^{\mu \nu} & = \frac{1}{16 \pi G} \bm{\epsilon} \left(- R^{\mu \nu} +\frac{1}{2} g^{\mu \nu} R - \frac{3}{\ell^2} g^{\mu \nu} \right)\,,  \\
    \bm{\Theta} & = *\, \bm{v} = i_{v} \bm{\epsilon}\,,  \qquad v^{\mu} = \frac{1}{16 \pi G} \left( g^{\mu \nu} \nabla^{\rho} \delta g_{\nu \rho} - g^{\lambda \rho} \nabla^{\mu} \delta g_{\lambda \rho} \right) \label{eq: sym_pot}\,.
\end{align}
The equations of motion are given by Einstein's equations (\ref{eq: Einstein})
\begin{equation} \label{eq: eom}
    \mathbf{E}^{\mu \nu} = 0 \iff R_{\mu \nu} = \frac{3}{\ell^2} g_{\mu \nu}\,,
\end{equation}
and we will be interested in studying solutions both at the exact and linearised level $\mathbf{E}^{\mu \nu} = \delta \mathbf{E}^{\mu \nu} = 0$. 

We will begin our analysis of this theory by looking for solutions in a neighbourhood of the conformal boundary $\mathscr{I}^+$. As discussed in Section \ref{sec: asymptotics}, one can always bring the metric into the form (\ref{eq: Starobinsky_gauge}), a form which we will move away from via the coordinate transformation
\begin{equation}
    \rho_+ = - \exp\left(-\frac{\tau}{\ell}\right)\,
\end{equation}
with the signs chosen such that $\tau = -\infty$ is $\rho_+ = - \infty$ and $\tau=\infty$ is $\rho_+ =0$. This transformation brings the line element (\ref{eq: Starobinsky_gauge}) into the form
\begin{equation} \label{eq: synchonous_gauge}
    ds^2 = -d\tau^2 + \gamma_{ij} dx^i dx^j\,, \qquad  \gamma_{ij} = \ell^2 e^{2 \tau/\ell} \left[ g^+_{(0)ij} + e^{-2\tau/\ell} g^+_{(2)ij} - e^{-3\tau/\ell} g^+_{(3)ij} + \ldots \right]\,,
\end{equation}
where we note that $\tau$ is a dimensionful time coordinate. This metric is now in a particularly convenient gauge in order to perform a Hamiltonian analysis: $\gamma_{ij}$ is the induced metric on  hypersurfaces $\Sigma$ of constant $\tau$ time. The analysis we perform will be similar to \cite{Papadimitriou:2005ii} which was for a \textit{radial} evolution in AlAdS spacetime. We note that under the analytic continuation (\ref{eq: Triple_Wick_explicit}) we have 
\begin{equation} \label{eq: analytic_cont_time_radius}
    \tau  = - i \eta_{\ell} r + \ell  \log ( - i \eta_{\rho} )\,,
\end{equation}
where $r$ is the radial coordinate used in \cite{Papadimitriou:2005ii} and is related to the AdS FG radial coordinate via $\rho = \exp(-r/\ell_{\text{AdS}})$. In order to avoid working explicitly with several factors of $\ell$, we introduce the dimensionless time coordinate
\begin{equation}
    \bar{\tau} = \frac{\tau}{\ell}\,,
\end{equation}
in which the metric takes the form 
\begin{equation}
    ds^2 = \ell^2 \left[ - d\bar{\tau}^2 + \bar{\gamma}_{ij} dx^i dx^j \right]\,, \quad \bar{\gamma}_{ij} =  e^{2 \bar{\tau}} \left[ g^+_{(0)ij} + e^{-2\bar{\tau}} g^+_{(2)ij} - e^{-3\bar{\tau}} g^+_{(3)ij} + \ldots \right]\,,
\end{equation}
we now set $\ell =1$ and note that this quantity can always be reinstated via dimensional analysis and then continued to AlAdS case via use of (\ref{eq: analytic_cont_time_radius}). In summary, we will perform a Hamiltonian analysis on the following spacetime
\begin{equation} \label{eq: metric_ADM}
     ds^2 =  - d\tau^2 + \gamma_{ij} dx^i dx^j\,, \qquad \gamma_{ij} =  e^{2 \tau} \left[ g^+_{(0)ij} + e^{-2\tau} g^+_{(2)ij} - e^{-3\tau} g^+_{(3)ij} + \ldots \right]\,,
\end{equation}
where we have dropped the bar notation for convenience. We note that the key difference between the analysis in AldS and AlAdS is in the character of the evolution, {\it i.e.} time \textit{vs} radial evolution. In the AldS case we evolve with respect to time and thus we can consider the equations of motion as equations relating data on a spacelike constant $\tau$ hypersurface, $\Sigma_{\tau}$, to data on the next constant $\tau$ hypersurface, $\Sigma_{\tau+\epsilon}$. This is the more familiar Hamiltonian evolution from the perspective of classical relativity \cite{Wald:1984rg} (see also \cite{Salopek:1990jq, McFadden:2010na, McFadden:2011kk} for applications of the formalism to perturbations of dS), but differs from the standard technique in AlAdS spacetimes \cite{Papadimitriou:2004ap, Papadimitriou:2005ii} which evolves data from one timelike hypersurface of constant radius, $\hat{\Sigma}_{r}$, to the next, $\hat{\Sigma}_{r+\varepsilon}$ \cite{Kraus:1999di, deBoer:1999tgo, Martelli:2002sp}. 

Focusing now strictly on the AldS case, we have written the metric in the usual ``ADM'' form \cite{Arnowitt:1962hi} with the lapse function $N$ and shift vector $N^i$ being
\begin{equation}
    N = 1\,, \qquad N^{i} = 0\,,
\end{equation}
and the future pointing unit normal vector $n^{\mu}$ to the hypersurface $\Sigma$ 
\begin{equation}
    n^{\mu} \partial_{\mu} = \partial_{\tau}\,, \qquad n_{\mu} dx^{\mu} = -d{\tau}\,,
\end{equation}
and thus 
\begin{equation}
    \gamma_{\mu \nu} = g_{\mu \nu} + n_{\mu} n_{\nu}\,.
\end{equation}
The extrinsic curvature of the hypersurface $\Sigma$ is given by 
\begin{equation} \label{eq: extrinsic_curvature}
   K_{\mu \nu} = \frac{1}{2} \mathcal{L}_{n} g_{\mu \nu} = \frac{1}{2} \partial_{\tau} \gamma_{\mu \nu} = \frac{1}{2} \dot{\gamma}_{\mu \nu}\,,
\end{equation}
and the momentum conjugate to the induced metric is 
\begin{equation} \label{eq: momenta}
    \pi^{ij} = \frac{\partial L_{\text{bulk}}}{\partial \dot{\gamma}_{ij}} = \frac{1}{16\pi G}\sqrt{\gamma} (K^{ij} - \gamma^{ij} K)\,,
\end{equation}
which can be seen explicitly by rewriting the Lagrangian (\ref{eq: Lagrangian}) in terms of ADM variables see {\it e.g.} \cite{Wald:1984rg}. We note that our orientation convention \eqref{eq: volume_form} is opposite to that of \cite{Papadimitriou:2004ap, Papadimitriou:2005ii}, resulting in an additional sign in the action (\ref{eq: bulk_action}) and thus our momenta (\ref{eq: momenta}) also carry the opposite sign. Using the decomposition of the metric (\ref{eq: metric_ADM}), the Gauss-Codazzi equations (see {\it e.g.} (10.2.23) and (10.2.24) of \cite{Wald:1984rg}) may be applied in order to write the equations of motion (\ref{eq: eom}) as
\begin{align}
 \label{eq: Einstein_ADM}
K_{ij} K^{ij} - K^2 & =  R[\gamma] - 6 \,,  \\
D_{i}K^{i}_j - D_j K & = 0\,, \label{eq: Constraint_2} \\
-\dot{K}^i_j-KK^i_j & = R^i_j[\gamma] - 3 \delta^i_j \,,
\end{align}
where $R_{ij}[\gamma]$ and $D_i$ are respectively the Ricci tensor and covariant derivative of the induced metric $\gamma_{ij}$. We also note that the Christoffel symbols for the metric (\ref{eq: metric_ADM}) are 
\begin{equation} \label{eq: Christoffel_symbols}
\Gamma^{\tau}_{i j} = K_{i j}\,, \qquad \Gamma^i_{\tau j} = K^i_j\,, \qquad \Gamma^{i}_{j k}[g] = \Gamma^i_{j k}[\gamma]\,,
\end{equation}
with all other symbols vanishing.

We will use this form of the Einstein equations in order to perform the holographic renormalisation proecure and compute the renormalised on-shell action. This procedure will follow closely the techniques of \cite{Papadimitriou:2004ap, Papadimitriou:2005ii} and will serve as an extension of the AlAdS result to AldS. We begin by regulating the on-shell action by introducing a large-$\tau$ cutoff surface $\Sigma_{\tau_0}:=\{\tau=\tau_0\}$, expressing the on-shell action as 
\begin{equation}
    S_{\text{bulk}} = \frac{3}{8 \pi G} \int d^4 x \,  \sqrt{-g}  = \frac{3}{8 \pi G} \int_{-\infty}^{\tau_0} d\tau d^3x \, \sqrt{-g}  = \frac{1}{8 \pi G} \int_{\Sigma_{\tau_0}} d^3x \, \sqrt{\gamma} \lambda\,,
\end{equation}
where $\lambda$ is a $\Sigma_{\tau}$-covariant function. Taking the $\tau$-derivative of both sides of this equation we find that $\lambda$ satisfies the differential equation
\begin{equation}
    \dot{\lambda} + \lambda K - 3  = 0 \,,
\end{equation}
and the regulated action including the Gibbons-Hawking-York (GHY) term takes the form 
\begin{equation} \label{eq: S_reg}
S_{\text{reg}} = S_{\text{bulk}} + S_{\text{GHY}} = - \frac{1}{8\pi G} \int_{\Sigma_{\tau_0}} d^3x \sqrt{\gamma} (K - \lambda)\,.
\end{equation}

We now want to analyse the asymptotic structure of this term in order to identify the  divergent part of the regulated action $S_{\text{reg}}$ in a fully covariant manner. In order to do this, we expand the extrinsic curvature and $\lambda$ in eigenfunctions of the \textit{dilatation} operator 
\begin{equation} \label{eq: dilatation}
    \delta_D = \int d^3 x \, 2 \gamma_{ij} \frac{\delta}{\delta \gamma_{ij}}\,,
\end{equation}
where the expansions of the fields take the form 
\begin{align}
\begin{split} \label{eq: dilatation_expansions}
    \pi^i_j & = \sqrt{\gamma} \left( \pi_{(0)j}^{\phantom{(0)}i} +  \pi_{(2)j}^{\phantom{(0)}i} + \pi_{(3)j}^{\phantom{(0)}i} + \ldots \right)\,, \\
    K^i_j[\gamma] & = K_{(0)j}^{\phantom{(0)}i} + K_{(2)j}^{\phantom{(0)}i} + K_{(3)j}^{\phantom{(0)}i} + \ldots\,, \\
    \lambda[\gamma] & = \lambda_{(0)} + \lambda_{(2)} + \lambda_{(3)} + \ldots\,,
    \end{split}
\end{align}
and the subscript indicates the eigenvalues\footnote{In odd bulk dimensions (odd $d$) there are additional logarithmic terms in the variation of the coefficients of weight $d-1$ \cite{Papadimitriou:2004ap}: these are related to bulk IR divergences, which in the AdS/CFT correspondence are linked with the holographic conformal anomaly \cite{Henningson:1998gx}. As we focus on $d=4$ in this paper, this subtlety does not play a role here.}
\begin{equation}
   \delta_D  K_{(n)j}^{\phantom{(0)}i} = - n  K_{(n)j}^{\phantom{(0)}i}\,, \qquad  \delta_D \lambda_{(n)} = -n \lambda_{(n)}\,.
\end{equation}

In order to solve the Einstein equations (\ref{eq: Einstein_ADM}) we insert the series expansions (\ref{eq: dilatation_expansions}) into the equations and solve order by order by comparing terms of identical dilatation weight. Note that the Einstein equations do not contain $\lambda$ and thus we will first derive an expression for the coefficients $\lambda_{(n)}$ in terms of the extrinsic curvature $K_{(n)}$. To do this, we start by noting the relationship \cite{Papadimitriou:2004ap, Papadimitriou:2005ii} 
\begin{equation} \label{eq: momenta_on_shell}
    \pi^{ij} = \frac{\delta S_{\text{reg}}}{\delta \gamma_{ij}}\,,
\end{equation}
which is a generic statement for all field theories and allows for an alternative expression for the momenta. We  apply (\ref{eq: momenta_on_shell}) in order to write 
\begin{equation} \label{eq: momenta_variation_no_integral}
  \int_{\Sigma_{\tau_0}}  \pi^{ij} \delta \gamma_{ij} = - \frac{1}{8 \pi G} \int_{\Sigma_{\tau_0}} \delta \left[ \sqrt{\gamma} (K - \lambda) \right]\,,
\end{equation}
 and replace the generic variation $\delta$ with the dilatation operator $\delta_D$ (\ref{eq: dilatation}), together with the field expansions for $K$ and $\lambda$ (\ref{eq: dilatation_expansions}). Comparing terms of equivalent dilatation weight results in\footnote{Note that total derivative terms do not contribute to integrals over $\Sigma_{\tau_0}$ since $\Sigma_{\tau_0}$ is a compact hypersurface.}
\begin{equation} \label{eq: lambda_n}
    \lambda_{(n)} = \frac{1-n}{3-n} K_{(n)} + D_i Y^i\,, \qquad n < 3\,,
\end{equation}
 which means that we can rewrite the regulated on-shell action as 
\begin{align}
\begin{split}
    S_{\text{reg}} & = - \frac{1}{8\pi G}  \int_{\Sigma_{\tau_0}} d^3x  \sqrt{\gamma} \sum_{n=0}(K_{(n)} - \lambda_{(n)}) \\
    & =  - \frac{1}{8\pi G} \int_{\Sigma_{\tau_0}} d^3x  \sqrt{\gamma} \left( \frac{2}{3}K_{(0)} + 2 K_{(2)} + K_{(3)} - \lambda_{(3)} \right), 
    \end{split}
\end{align}
and thus we identify the counterterm action required to renormalise the theory as (minus) the divergent part of the above expression as one takes $\tau_0 \rightarrow \infty$
\begin{align} \label{eq: S_ct}
    S_{\text{ct}} = \frac{1}{8 \pi G}  \int_{\Sigma_{\tau_0}} d^3x  \sqrt{\gamma} \left( \frac{2}{3}K_{(0)} + 2 K_{(2)}\right) 
    =  \frac{1}{8 \pi G}  \int_{\Sigma_{\tau_0}} d^3x  \sqrt{\gamma} \left( 2 - \frac{R[\gamma]}{2}\right)\,,
\end{align}
where the explicit expressions for the eigenfunctions $K_{(0)}, K_{(2)}$ are derived in Appendix \ref{sec: counterterms}.
Using these results, we define the renormalised action as 
\begin{equation} \label{eq: S_ren}
    S_{\text{ren}} = \lim_{\tau_0 \rightarrow \infty} (S_{\text{reg}} + S_{\text{ct}})\,.
\end{equation}

We conclude this section by giving an expression for the energy-momentum tensor in terms of the metric data in (\ref{eq: metric_ADM}). In order to do this, we utilise the important result \cite{Papadimitriou:2004ap, Papadimitriou:2005ii} that to leading order the dilatation operator can be identified with the time derivative
\begin{equation} \label{eq: time_der_dilatation}
    \partial_{\tau} = \int_{\Sigma_{\tau_0}}d^3x \, \partial_{\tau} \gamma_{ij} \frac{\delta}{\delta \gamma_{ij}} =  \int_{\Sigma_{\tau_0}}d^3x \, 2K_{ij} \frac{\delta}{\delta \gamma_{ij}} = \delta_D + \mathcal{O}(e^{-2\tau})\,,
\end{equation}
and since we work in sufficiently low dimension it will be sufficient to ignore the $\mathcal{O}(e^{-2\tau})$ tail on the right hand side above and make the strict identification $\partial_{\tau} = \delta_D$. We write the extrinsic curvature (\ref{eq: extrinsic_curvature}) as 
\begin{equation}
    K^i_{j} = \delta^i_j - \gamma^{ik} g^+_{(2)kj} +\frac{3}{2} e^{-\tau}  \gamma^{ik}g^+_{(3)k j} + \ldots \,,
\end{equation}
which allows us to immediately read off 
\begin{equation} \label{eq: K_eigenfunctions}
  K_{(0)j}^{\phantom{(0)}i} = \delta^i_j \,, \qquad K_{(2)j}^{\phantom{(0)}i} = - e^{-2\tau} g_{(2) j}^{+ \phantom{a} i}\,, \qquad K_{(3)j}^{\phantom{(0)}i} =  \frac{3}{2} e^{-3 \tau} g_{(3) j}^{+ \phantom{a} i}\,,
\end{equation}
where we note that we have raised/lowered the indices of the $g^+_{(2),(3)}$
terms using $g^+_{(0)}$. By inserting these terms into the Einstein equations (\ref{eq: Einstein_ADM}) and comparing terms of the same dilatation weight we are able to derive the important relations
\begin{align}
    K_{(3)} & = 0  \qquad  \iff \qquad  g_{(0)}^{+ \; ij} g^+_{(3)ij} = 0\,, \label{eq: em_traceless}\\
    D_i K_{(3)j}^{\phantom{(0)}i} & = 0 \qquad \iff \qquad \nabla^+_{(0) i} g_{(3) j}^{+ \phantom{a}i} = 0\,, \label{eq: em_conserved}
\end{align}
which, as we shall shortly see, can be interpreted as the tracelessness and conservation of the dual energy-momentum tensor. To explicitly derive this tensor, we recall that the energy-momentum tensor is defined via 
\begin{equation}
\langle T_{ij} \rangle^{\mathscr{I}^+} = \frac{2}{\sqrt{g^+_{(0)}}} \frac{\delta S_{\text{ren}}}{\delta g_{(0)}^{+ \; ij}}\,,
\end{equation}
and utilising this together with equations (\ref{eq: dilatation_expansions}), (\ref{eq: momenta_on_shell}) we find the energy-momentum tensor is given by 
\begin{equation} \label{eq: EM_future}
    \langle T_{ij} \rangle^{\mathscr{I}^+} = -2 \pi_{(3)ij} = - \frac{3 \ell}{16 \pi G} g^+_{(3)ij}\,,
\end{equation}
where we have reinstated the dimensionful factor of the dS radius $\ell$ via dimensional analysis. We note that this sets our sign convention for the energy-momentum tensor at future infinity $\mathscr{I}^+$, where different signs have appeared in previous literature: in \cite{Skenderis:2002wp, Compere:2020lrt} the convention above is chosen whereas in \cite{Compere:2019bua, Kolanowski:2021hwo} the choice of an additional minus sign is utilised. We note that the energy-momentum tensor at $\mathscr{I}^-$ has the form 
\begin{equation} \label{eq: EM_past}
    \langle T_{ij} \rangle^{\mathscr{I}^-} = - \frac{2}{\sqrt{g^-_{(0)}}} \frac{\delta S_{\text{ren}}}{\delta g_{(0)}^{- \; ij}} = - \frac{3 \ell}{16 \pi G} g^-_{(3)ij}\,,
\end{equation}
where $g_{(n)}^-$ represent the terms in the Starobinsky expansion at $\mathscr{I}^-$. The signs in the equation above are explained in Appendix \ref{sec: scri_-}. In Section \ref{sec: past_contribution} we will include the presence of both boundaries and \eqref{eq: EM_past} will be a necessary consideration. 

\subsubsection{Comparison with AlAdS energy-momentum tensor}
We conclude this section with a brief comment on the relation between these results and the analytic continuation of the \textit{single} energy-momentum tensor induced at the conformal boundary of an AlAdS spacetime. We begin by emphasising that the analysis of this section is entirely self-contained in considering AldS spacetimes and thus the AldS energy-momentum tensors (\ref{eq: EM_future}), (\ref{eq: EM_past}) are not defined as analytic continuation of the analogous AlAdS results  (\ref{eq: EM_tensor_AdS}). However, comparison with equation (\ref{eq: EM_rotation}) shows 
\begin{equation}
    \langle T_{ij} \rangle \rightarrow  \eta_{\ell} \eta_{\rho}  \langle T_{ij} \rangle^{\mathscr{I}^+} \,,
\end{equation}
and thus the AlAdS energy-momentum tensor analytically continues to the AldS energy-momentum tensor, up to a sign factor which depends on the directions taken in the Wick rotation of the $\ell$ and $\rho$ variables. We also note that one could also consider the past energy-momentum tensor of an AldS spacetime \eqref{eq: EM_past} being obtained from analytic continuation of an AlAdS spacetime
\begin{equation}
     \langle T_{ij} \rangle \rightarrow  \eta_{\ell} \eta_{\rho}  \langle T_{ij} \rangle^{\mathscr{I}^-}\,,
\end{equation}
but we should be careful to note that such a relation does not force any relationship between $\langle T_{ij} \rangle^{\mathscr{I}^+}$ and $\langle T_{ij} \rangle^{\mathscr{I}^-}$. Locally, both boundaries of an AldS spacetime can be obtained from analytic continuation of an AlAdS spacetime, but the global structure of AldS spacetimes is markedly different. In general, the expansions at $\mathscr{I}^{\pm}$ will be independent. However, since $\mathscr{I}^+$ lies to the future of $\mathscr{I}^-$, the metric expansions near the hypersurfaces may be related via causality arguments (such as in empty de Sitter space) or by evolution of sufficiently broadly specified initial data. We note that such relationships depend purely on the dynamics of AldS spacetimes, and not in the details of analytic continuation from AlAdS.

\subsection{Variational problem} \label{sec: variational}
 Now that we have established our conventions in studying the theory and holographic renormalisation of $d=4$ AldS gravity, we are almost ready to construct explicit expressions for the charges using the techniques described in Section \ref{sec: covariant}. In constructing these charges, we will begin by considering a single boundary of an AldS spacetime, which we will take to be future spacelike infinity $\mathscr{I}^+$. The reason for this is that, when treated independently, the analysis applied to $\mathscr{I}^-$ is structurally identical to $\mathscr{I}^+$. In Section \ref{sec: past_contribution} we will treat the global picture which includes the contribution of $\mathscr{I}^-$, and give formulae for the charges including the contribution of the past end.
 
 First, we determine the space of variations which give a \textit{well posed variational problem} {\it i.e.} determine the variations which satisfy 
 \begin{equation}
     \delta S_{\text{ren}} = 0 \iff \mathbf{E} = 0\,, 
\end{equation}
a result which will be crucial in selecting the boundary conditions at the conformal boundary which give rise to the existence of Wald Hamiltonians (\ref{eq: Hamiltonian_def}) \cite{Iyer:1995kg, Papadimitriou:2005ii}. 

In order to determine the space of well-posed variations, we follow \cite{Papadimitriou:2005ii} by studying the general expression for the symplectic potential (\ref{eq: sym_pot}) for variations which preserve the gauge (\ref{eq: metric_ADM}). Elementary manipulations show that the pullback of $\bm{\Theta}$ to $\Sigma_{\tau}$ is given by 
\begin{equation} \label{eq: theta_pullback_1}
     \bm{\Theta}_{ijk} =  (i_v \bm{\epsilon})_{ijk}  = \frac{1}{16 \pi G} \sqrt{\gamma}  \left( g^{\tau \nu} \nabla^{\rho} \delta g_{\nu \rho} - g^{\lambda \rho} \nabla^{\tau} \delta g_{\lambda \rho} \right) \varepsilon_{\tau i j k }\,,
\end{equation}
where $\varepsilon$ is the Levi-Civita symbol and using our orientation (\ref{eq: volume_form}) for the volume form we write
\begin{equation}
    \frac{1}{3!} \varepsilon_{\tau i j k} dx^i \wedge dx^j \wedge dx^k = \frac{1}{3!} \varepsilon_{i j k} dx^i \wedge dx^j \wedge dx^k = d\mu\,.
\end{equation} 
We now apply the explicit form of the metric (\ref{eq: metric_ADM}) and the Christoffel symbols (\ref{eq: Christoffel_symbols}), together with the definitions of extrinsic curvature (\ref{eq: extrinsic_curvature}) and canonical momenta (\ref{eq: momenta}) in order to write \eqref{eq: theta_pullback_1} as 
\begin{equation} \label{eq: theta_pullback_spacelike}
    \bm{\Theta} =\left( \frac{1}{8\pi G} \delta (K \sqrt{\gamma}) + \pi^{ij} \delta \gamma_{ij} \right) d\mu\,,
\end{equation}
and then using (\ref{eq: dilatation_expansions}) we have 
\begin{equation} 
     \bm{\Theta} = \left( \frac{1}{8\pi G} \delta (K \sqrt{\gamma}) + \sqrt{\gamma} \left[ \pi_{(0)}^{ij} + \pi_{(2)}^{ij} + \pi_{(3)}^{ij} +\ldots \right]  \delta \gamma_{ij} \right) d\mu\,,
\end{equation}
finally applying (\ref{eq: momenta_variation_no_integral}) and grouping the terms according to their dilatation weight we obtain that as $\tau_0 \rightarrow \infty$,
\begin{align}
\begin{split}
    \int_{\Sigma_{\tau_0}} \bm{\Theta} & = \int_{\Sigma_{\tau_0}} \left( \frac{1}{8\pi G} \delta \left[ \sqrt{\gamma} \left\{  K - (K_{(0)} - \lambda_{(0)}) - (K_{(2)} - \lambda_{(2)}) \right\} \right] + \sqrt{\gamma} \pi_{(3)}^{ij} \delta \gamma_{ij} \right)   \\
    & =  \int_{\Sigma_{\tau_0}}  \left( \frac{1}{8\pi G} \delta \left[ \sqrt{\gamma} \left\{  K - \frac{2}{3} K_{(0)} - 2 K_{(2)}  \right\} \right] + \sqrt{\gamma} \pi_{(3)}^{ij} \delta \gamma_{ij}  \right)\,, 
    \end{split}
\end{align}
which will be the most useful form for analysing the variational problem. We use (\ref{eq: S_ren}) to write the variation of the renormalised on-shell action as
\begin{align} 
\begin{split} \label{eq: well_posed_variational}
    \delta S_{\text{ren}} & = \lim_{\tau_0 \rightarrow \infty} \delta \left( S_{\text{bulk}} + S_{\text{GHY}} + S_{\text{ct}} \right) \\
    & = \lim_{\tau_0 \rightarrow \infty} \left( \int_{\Sigma_{\tau_0}}  \bm{\Theta} -  \frac{1}{8\pi G} \int_{\Sigma_{\tau_0}} \delta \left[ \sqrt{\gamma} \left\{  K - \frac{2}{3} K_{(0)} - 2 K_{(2)}  \right\} \right] \right) \\ 
    & = \frac{3}{32\pi G} \int_{\mathscr{I}^+} \sqrt{g^+_{(0)}} g_{(3)}^{+ \; ij} \delta g^+_{(0)ij}\,,
    \end{split}
\end{align}
and thus using the tracelessness (\ref{eq: em_traceless}) and conservation (\ref{eq: em_conserved}) of the energy-momentum tensor, we conclude that the variational problem is well-posed when
\begin{equation} \label{eq: well_posed_bcs}
     \delta g^+_{(0)ij} = 2  \sigma(\tilde{x}) g^+_{(0)ij} + \mathcal{L}_{\zeta} g^+_{(0)ij} = 2\left(\sigma g^+_{(0)ij} + \nabla^+_{(0)(i} \zeta_{j)}\right)\,,
\end{equation}
{\it i.e.} for transformations that act on the representative of the conformal class via a combination of a Weyl rescaling parameterised by $ \sigma$ and a boundary diffeomorphism in the direction of the generic boundary vector field $\zeta$.

\subsubsection{Comparison with AlAdS variational problem}
The above analysis demonstrates that the addition of the boundary terms consisting of the Gibbons-Hawking term and the holographic counterterms, together with imposing the boundary conditions (\ref{eq: well_posed_bcs}), is sufficient to provide a well-posed variational problem. This is in agreement with previous work in AlAdS \cite{Papadimitriou:2005ii} as well as some recent literature concerning AldS \cite{Compere:2020lrt, Kolanowski:2021hwo}. We note that we are considering more explicitly the role of the boundary diffeomorphisms $\zeta$ than 
in \cite{Papadimitriou:2005ii}.
It was shown in \cite{Papadimitriou:2005ii} that for AlAdS spacetimes the most general vector fields that preserve a conformal class
are \textit{asymptotic conformal Killing vectors}, bulk vector fields which asymptote to \textit{conformal Killing vectors} of $g_{(0)}$. We will see shortly  
that a similar consideration is intimately related to the existence of a conserved charge
at a given cross-section of $\mathscr{I}^+$ in AldS spacetime.

\subsection{Invariance of the renormalised action and asymptotic symmetries} \label{sec: inv_off_shell}

Now that we have determined the variations at $\mathscr{I}^+$ which lead to a well-posed variational problem \eqref{eq: well_posed_bcs}, it remains to identify the asymptotic symmetry group (ASG) at $\mathscr{I}^+$ of AldS spacetimes. The ASG is generated by vector fields $\xi$ which leave both the off-shell (renormalised) action \eqref{eq: S_ren} and a choice of boundary conditions invariant.

We first determine the transformations which preserve the gauge (\ref{eq: metric_ADM}). The functional form of the leading order metric near the conformal boundary is 
\begin{equation} \label{eq: BCs_functional}
    \gamma_{ij}(\tau, x) \sim e^{2\tau} g^+_{(0)ij} (x)\,,
\end{equation}
and we do not impose any condition on $g^+_{(0)ij}$ for now. We begin with the transformations of the bulk metric components which are given by Lie derivatives 
\begin{align} \label{eq: Lie_metric}
\begin{split}
\delta_{\xi} g_{\tau \tau} & = \mathcal{L}_{\xi} g_{\tau \tau}  = - 2 \dot{\xi}^{\tau}\,, \\
\delta_{\xi} g_{\tau i} & = \mathcal{L}_{\xi} g_{\tau i}  = \gamma_{ij} (\dot{\xi}^j-\partial^j \xi^{\tau})\,, \\
\delta_{\xi} g_{ij} & = \mathcal{L}_{\xi} g_{ij} = L_{\xi} \gamma_{ij} + 2K_{ij} \xi^{\tau} \sim L_{\xi} \gamma_{ij} + 2\gamma_{ij} \xi^{\tau}\,,
\end{split} 
\end{align}
where $L_{\xi}$ denotes the Lie derivative along the directions spanned by $\Sigma_{\tau}$. Useful formulae for the transformations of the induced quantities ($\gamma_{ij}, K_{ij}$ {\it etc}.) are given in Appendix \ref{app: Variations}.  The vectors which preserve the gauge fixing up to next-to-normalisable order satisfy 
\begin{equation}
    \mathcal{L}_{\xi} g_{\tau \tau} = \mathcal{L}_{\xi} g_{\tau i} = \mathcal{O}(e^{-3\tau})\,,
\end{equation}
and thus we can integrate this condition using (\ref{eq: Lie_metric}) and find 
\begin{align}
\begin{split} \label{eq: gauge_preservation}
    \xi^{\tau} & = \sigma_{\xi}(x) + \mathcal{O}(e^{-3\tau})\,, \\
    \xi^{i} & = \xi^i_{(0)}(x) - \partial_j \sigma_{\xi}(x) \int^{\infty}_{\tau} \gamma^{ij}(\tau', x) \, d\tau' + \mathcal{O}(e^{-5\tau})\,,
    \end{split}
\end{align}
as the general solution. We note as in \cite{Papadimitriou:2005ii} that the transformation with $\xi_{(0)}=0$ is precisely the `Penrose-Brown-Henneaux' (PBH) transformation \cite{penrose_rindler_1986, Brown:1986nw} which performs a Weyl rescaling of the conformal boundary.

We now determine the subset of (\ref{eq: gauge_preservation}) that preserves the renormalised (off-shell) action, including boundary terms.  
We recall the renormalised action takes the form\footnote{The limit of $\tau_0 \rightarrow \infty$ is implicit in the definition of $S_{\text{ren}}$.} 
\begin{equation}
S_{\text{ren}} =  S_{\text{bulk}} + S_{\text{GHY}} + S_{\text{ct}}\,,
\end{equation}
where
\begin{align}
    S_{\text{bulk}} & = \int_{M_{\tau_0}} \mathbf{L}_{\text{bulk}} = \frac{1}{16 \pi G} \int_{M_{\tau_0}} d^4x \, \sqrt{-g} 
    (R - 6)\,, \\ 
    S_{\text{GHY}} & =  - \frac{1}{8\pi G} \int_{\Sigma_{\tau_0}} d^3x \sqrt{\gamma} K\,, \\
    S_{\text{ct}} & = \frac{1}{8\pi G}  \int_{\Sigma_{\tau_0}} d^3x  \sqrt{\gamma} \sum_{n=0}^2(K_{(n)} - \lambda_{(n)}) =  \frac{1}{8 \pi G}  \int_{\Sigma_{\tau_0}} d^3x  \sqrt{\gamma} \left( 2 - \frac{R[\gamma]}{2}\right)\,,
\end{align}
and now we want to study $\delta_{\xi} S_{\text{ren}}$, which we can do term by term. Starting with  $S_{\text{bulk}}$, we need to consider 
\begin{equation}
    \delta_{\xi} \mathbf{L}_{\text{bulk}} = \mathcal{L}_{\xi} \mathbf{L}_{\text{bulk}} = (i_{\xi} d+ d i_{\xi}) \mathbf{L}_{\text{bulk}} = d i_{\xi} \mathbf{L}_{\text{bulk}}\,,
\end{equation}
and thus
\begin{equation} \label{eq: bulk_variation_xi}
    \delta_{\xi} S_{\text{bulk}} = \int_{M_{\tau_0}} d i_{\xi} \mathbf{L}_{\text{bulk}} = \int_{\Sigma_{\tau_0}} i_{\xi} \mathbf{L}_{\text{bulk}} = \frac{1}{16 \pi G} \int_{\Sigma_{\tau_0}} d^3 x \, \xi^{\tau} \sqrt{\gamma} (R - 6)\,.
\end{equation} 
In order to combine with the second term, we rewrite the bulk Ricci scalar as \cite{Wald:1984rg} 
\begin{equation} \label{eq: Ricci_induced}
    R = R[\gamma] + K_{ij} K^{ij} - K^2 + \frac{2}{\sqrt{\gamma}} \partial_{\tau}\left( \sqrt{\gamma} K \right)\,,
\end{equation}
which then gives 
\begin{equation} \label{eq: bulk_variation_final}
     \delta_{\xi} S_{\text{bulk}} =  \int_{\Sigma_{\tau_0}} d^3 x \, \xi^{\tau} \sqrt{\gamma} \left( R[\gamma] + K_{ij} K^{ij} - K^2 + \frac{2}{\sqrt{\gamma}} \partial_{\tau}\left( \sqrt{\gamma} K \right) -6 \right)\,.
\end{equation}

Moving on to the second term, $\delta_{\xi} S_{\text{GHY}}$, an explicit computation shows 
\begin{equation} \label{eq: variation_GHY}
    \delta_{\xi} S_{\text{GHY}} = - \frac{1}{8 \pi G} \int_{\Sigma_{\tau_0}} d^3x \, \delta_{\xi} \left( \sqrt{\gamma} K \right) = - \frac{1}{8 \pi G} \int_{\Sigma_{\tau_0}} d^3x \, \xi^{\tau} \partial_{\tau} \left( \sqrt{\gamma} K  \right)\,,
\end{equation}
where we have made use of (\ref{eq: Variation_det_K}).

The third term is $\delta_{\xi} S_{\text{ct}}$ which we can write as 
\begin{equation}
    \delta_{\xi} S_{\text{ct}} = \frac{1}{8\pi G}  \int_{\Sigma_{\tau_0}} d^3x \sum_{n=0}^2 \delta_{\xi} \left( \sqrt{\gamma} (K_{(n)} - \lambda_{(n)}) \right) = - \int_{\Sigma_{\tau_0}} d^3x\, \sqrt{\gamma} \sum_{n=0}^2 \hat{\pi}_{(n)}^{ij} \delta_{\xi} \gamma_{ij}\,, 
\end{equation}
where we have made use of (\ref{eq: momenta_variation_no_integral}) and followed \cite{Papadimitriou:2005ii} in putting a hat on the counterterm momenta. This notation indicates they are defined as predetermined local functionals of the off-shell induced fields rather than determined by the asymptotic behaviour of the on-shell fields. We then apply the transformation (\ref{eq: Lie_metric}) and make use of the initial value constraint (\ref{eq: Constraint_2})\footnote{Even though we are working off-shell, we can still apply this constraint to the counterterms as they satisfy this by construction.} to write this term as 
\begin{equation} \label{eq: ct_variation}
     \delta_{\xi} S_{\text{ct}} = - 2 \int_{\Sigma_{\tau_0}} d^3x\, \sqrt{\gamma} \sum_{n=0}^2 \hat{\pi}_{(n)}^{ij} K_{ij}.
\end{equation}
We can now combine equations (\ref{eq: bulk_variation_final}), (\ref{eq: variation_GHY}) and (\ref{eq: ct_variation}) and arrive at
\begin{equation} \label{eq:bdry}
    \delta_{\xi} S_{\text{ren}} = \frac{1}{16 \pi G} \int_{\Sigma_{\tau_0}} d^3 x \, \xi^{\tau} \sqrt{\gamma} \left( R[\gamma] + K_{ij} K^{ij} - K^2  -6 - 32 \pi G \sum_{n=0}^2 \hat{\pi}_{(n)}^{ij} K_{ij} \right)\,,
\end{equation}
which is our final expression for the variation of the renormalised action. The fact that this is a boundary term reflects the fact that the bulk theory is diffeomorphism invariant, and that we looked for invariance under (a class of) diffeomorphisms. Asymptotic symmetries are the transformations for which (\ref{eq:bdry}) vanishes.

Using (\ref{eq: BCs_functional}), we see that the apparent leading divergence at $\mathcal{O}(e^{3\tau})$ in fact vanishes and thus the integrand above has a leading order divergence at $\mathcal{O}(e^{2\tau})$. We thus conclude that in order to preserve the renormalised action \textit{off-shell} we require the falloff on $\xi^{\tau}$ to be 
\begin{equation} \label{eq: sigma=0}
    \xi^{\tau} = \mathcal{O}(e^{-3\tau}) \iff \sigma_{\xi}(x) = 0\,,
\end{equation}
and thus our ASG is generated by vectors of the form 
\begin{align}
\begin{split} \label{eq: xi_ASG}
    \xi^{\tau} & = \mathcal{O}(e^{-3\tau})\,, \\
    \xi^{i} & = \xi^i_{(0)}(x) + \mathcal{O}(e^{-5\tau})\,,
    \end{split}
\end{align}
where $\xi_{(0)}$ is an arbitrary vector field. We note that for \textit{on-shell} configurations, the renormalised action is preserved even when $\sigma_{\xi} \neq 0$. This can be seen using the Hamiltonian constraint (\ref{eq: Einstein_ADM})
\begin{align}
\begin{split}
    \delta_{\xi} S_{\text{ren}} & \approx \frac{1}{8\pi G} \int_{\Sigma_{\tau_0}} d^3 x \, \xi^{\tau} \sqrt{\gamma} \left( K_{ij}K^{ij} - K^2 - 16\pi G \sum_{n=0}^2 \pi_{(n)}^{ij} K_{ij} \right) \\
    & = 2 \int_{\Sigma_{\tau_0}} d^3 x \, \xi^{\tau} \sqrt{\gamma} \left( \pi^{ij} K_{ij} - \sum_{n=0}^2 \pi_{(n)}^{ij} K_{ij} \right) \\
    & = 2 \int_{\Sigma_{\tau_0}} d^3 x \, \xi^{\tau} \sqrt{\gamma}  \pi_{(3)}^{ij} K_{ij} = 0\,,
    \end{split}
\end{align}
where in the final equality we used equations (\ref{eq: K_eigenfunctions}), 
(\ref{eq: em_traceless}) and (\ref{eq: EM_future}). 
This reflects the fact that \eqref{eq: well_posed_bcs} are associated with a well-posed variational problem.

We thus found that the (off-shell) action is invariant  under the transformations (\ref{eq: xi_ASG}). Then by Noether's theorem there should exist a corresponding charge that generates these transformations. The transformations (\ref{eq: xi_ASG}) are local transformations and, as such, their associated charges should be zero, and we will confirm this in the next subsection.  Non-trivial conserved charges appear when we impose that the {\it conformal class is fixed}. In deriving the charges by various alternative methods in the forthcoming sections, we will make use of the boundary condition of preservation of the conformal class
\begin{equation} \label{eq: conformal_dirichlet}
    \delta \gamma_{ij}  \propto \gamma_{ij} \; \;\; \text{on} \; \mathscr{I}^+ \,,
\end{equation}
which forces the following relation on the vector fields $\xi$
\begin{align}
\begin{split} \label{eq: xi_ASG_CKV}
    \xi^{\tau} & = \mathcal{O}(e^{-3\tau}), \\
    \xi^{i} & = \xi^i_{(0)}(x) + \mathcal{O}(e^{-5\tau}), \qquad L_{\xi_{(0)}} \gamma_{ij} = \frac{2}{3} \gamma_{ij} D_k \xi^k \; \;\; \text{on} \; \mathscr{I}^+ \,,
    \end{split}
\end{align}
{\it i.e} the subset of the boundary diffeomorphisms (\ref{eq: xi_ASG}) where $\xi_{(0)}$ is a conformal Killing vector (CKV) of $g_{(0)}^+$. We note that these conditions are precisely the AldS analogue of the conditions in Appendix B of \cite{Papadimitriou:2005ii} which define an \textit{asymptotic conformal Killing vector} (ACKV). Not all spacetimes admit such vectors, and some spacetimes may only admit such vectors over a portion of  $\mathscr{I}^+$. In such cases (as we will see) one can define gravitational charges that are conserved at these portions and satisfy balance laws outside these portions where the change in the charge is balanced by the net flux of the corresponding current density. These balance laws follow from the invariance of the action under the more general transformations  (\ref{eq: xi_ASG}) and the vanishing of the charges that generate them.

\subsection{Gravitational charges at \texorpdfstring{$\mathscr{I}^+$}{Scrim}} \label{sec: grav_charges}

Having established that the action has an invariance, one may proceed to derive the conserved charges associated with the asymptotic symmetries using Noether's theorem(s), following analogous steps to the derivation of conserved charges for a QFT on a fixed background.  The reason we are able to carry out this computation in complete generality is because we have full control over the asymptotic form of the gravitational field (and this is the case because the cosmological constant is non-zero). We provide two derivations of the charges associated with asymptotic conformal Killing vectors (\ref{eq: xi_ASG_CKV}), following closely the analogous derivations in AlAdS presented in \cite{Papadimitriou:2005ii}, and we will also show that these charges satisfy balance laws when $\xi$ is a general vector. We will then show that these charges can also be obtained as Wald Hamiltonians using the covariant phase techniques discussed in Section \ref{sec: covariant}.

\subsubsection{Gravitational charges from Noether's first theorem} \label{sec: Noeth1}

Our starting point is the theory with a conformal class kept fixed, and we 
will derive the corresponding gravitational charge formulae using Noether's first theorem. 
To apply Noether's method, we consider vector fields of the form (\ref{eq: xi_ASG_CKV}) and field variations 
\begin{equation} \label{eq: variation_Noether_1}
    \delta_f \psi = f(\tau, x) \mathcal{L}_{\xi}\psi\,,
\end{equation}
where $f$ is an arbitrary function. As usual with Noether's method, such a variation does not generically preserve the renormalised action (this is the case only when $f$ is a constant), although it does preserve the boundary conditions \eqref{eq: conformal_dirichlet}. Following \cite{Papadimitriou:2005ii}, we will denote with a bar an object that is defined on $\Sigma_{\tau}$, for example $\bar{f}$ will denote the pullback of $f$ and $\bar{d}$ the boundary exterior derivative. As shown in (\ref{eq: well_posed_variational}), the general variation of the action takes the form 
\begin{equation}
    \delta S_{\text{ren}} = \int_{M_{\tau_0}} \mathbf{E} \delta \psi + \int_{\Sigma_{\tau_0}} d^3 x \, \sqrt{\gamma} \pi_{(3)}^{ij} \delta \gamma_{ij}\,,
\end{equation}
where we have now included the $\mathbf{E}$ term as we work off-shell. Applying the variation (\ref{eq: variation_Noether_1}) to this we obtain 
\begin{align}
\begin{split} \label{eq: delta_f_S}
    \delta_f S_{\text{ren}} & = \int_{M_{\tau_0}} f \mathbf{E} \mathcal{L}_{\xi} \psi + \int_{\Sigma_{\tau_0}} d^3 x \, \sqrt{\gamma} f \pi_{(3)}^{ij} L_{\xi} \gamma_{ij} \\
    & = - \int_{M_{\tau_0}} f d (\bm{\Theta}(\psi, \mathcal{L}_{\xi}\psi)- i_{\xi} \mathbf{L}) + \int_{\Sigma_{\tau_0}} d^3 x \, \sqrt{\gamma} f \pi_{(3)}^{ij} L_{\xi} \gamma_{ij}  \\
    & = \int_{M_{\tau_0}} d f \wedge \mathbf{J}[\xi] - \int_{\Sigma_{\tau_0}} f ( \bm{\Theta}(\psi, \mathcal{L}_{\xi}\psi)- i_{\xi} \mathbf{L} ) + \int_{\Sigma_{\tau_0}} d^3 x \, \sqrt{\gamma} f \pi_{(3)}^{ij} L_{\xi} \gamma_{ij} \,,
    \end{split}
\end{align}
where we used \eqref{eq: Lagrangian_variation} in moving to the second line and integration by parts together with the definition \eqref{eq: Noether_current} in the third line. We want to extract the full Noether current by manipulating the above expression in order to show that the integrand is proportional to the derivative of $f$. Clearly the bulk term above is already of the correct form, but the boundary terms require some more work.

In order to correctly manipulate the boundary terms above, we begin by noting from (\ref{eq: well_posed_variational}) that the pullback of the symplectic potential takes the form 
\begin{equation} \label{eq: theta_pullback}
\left.\bm{\Theta} \right|_{\Sigma_{\tau_0}} = \delta \mathbf{B} + \sqrt{\gamma} \pi^{ij}_{(3)} \delta \gamma_{ij} \, d\mu + d \mathbf{Y}\,, 
\end{equation}
where 
\begin{equation} \label{eq: B_def}
 \mathbf{B} = - \mathbf{L}_{\text{GHY}} - \mathbf{L}_{\text{ct}}\,, 
\end{equation} 
and $\mathbf{Y}$ is a $2$-form which is left undetermined in the variational problem as the integral of an exact form over a manifold without boundary (in this case $\Sigma_{\tau_0}$) is zero. Although $\mathbf{Y}$ does not affect the variational problem, it will be important in the final expression for the charges. The explicit expression for this ``corner term'' has appeared in several related works concerning charges in Al(A)dS$_4$ spacetimes \cite{Compere:2008us, Anninos:2010zf, Compere:2020lrt, Kim:2023ncn} and is often referred to as the \textit{counterterm symplectic potential}, given explicitly via
\begin{equation}
    d \mathbf Y = d \bm{\Theta}_{\text{ct}}( \gamma, \delta \gamma ) \equiv \delta \mathbf{L}_{\text{ct}} - \delta \gamma_{ij} \frac{\delta \mathbf{L}_{\text{ct}}}{\delta \gamma_{ij}}\,,
\end{equation}
and for our counterterm action (\ref{eq: S_ct}) we have
\begin{equation}
    \bm{\Theta}_{\text{ct}}(\gamma, \delta \gamma) = - \bm{\Theta}^{(3)}_{\text{EH}}(\gamma, \delta \gamma) =  \frac{1}{16 \pi G}\cdot \frac{1}{2!} \varepsilon_{i j k} \left(D_{l} \delta \gamma^{l i} - D^{i} (\gamma_{ l m} \delta \gamma^{l m} ) \right) dx^{j} \wedge dx^{k}\,,
\end{equation}
a result which we note is special to $d=4$ bulk dimensions.\footnote{In higher dimensions the counterterm action would be supplemented by higher curvature terms \cite{deHaro:2000vlm} and it would be interesting to examine the analogous modifications to the symplectic potential.} 
In what follows, we will find it more convenient to work with 
\begin{equation} \label{eq: Shifted_theta}
    \left.\bm{\Theta} \right|_{\Sigma_{\tau_0}} - d  \bm{\Theta}_{\text{ct}}(\gamma, \delta \gamma) =  \delta \mathbf{B} +\sqrt{\gamma} \pi^{ij}_{(3)} \delta \gamma_{ij} \, d\mu\,,
\end{equation}
which corresponds to utilising the inherent ambiguity present in $\bm{\Theta}$ in always being able to add an exact form \cite{Iyer:1994ys}. Replacing the generic variation $\delta$ with $\delta_{\xi}$ where $\xi$ is a vector in our asymptotic symmetry group (\ref{eq: xi_ASG_CKV}) gives 
\begin{equation} \label{eq: variation_B_CKV}
     \left.\bm{\Theta} (\psi, \mathcal{L}_{\xi} \psi)\right|_{\Sigma_{\tau_0}} - d  \bm{\Theta}_{\text{ct}}(\psi, \mathcal{L}_{\xi} \psi) = \delta_{\xi} \mathbf{B} + \sqrt{\gamma} \pi^{ij}_{(3)} L_{\xi} \gamma_{ij} \, d\mu\,.
\end{equation}

With this established, we can now perform the following manipulations of the variation of the action \eqref{eq: delta_f_S}
\begin{align} \label{eq: delta_f_S_int}
    \begin{split}
           \delta_f S_{\text{ren}} & = \int_{M_{\tau_0}} d f \wedge  \mathbf{J} [\xi] - \int_{\Sigma_{\tau_0}} f [ \bm{\Theta}(\psi, \mathcal{L}_{\xi} \psi) - i_{\xi} \mathbf{L} ] +  \int_{\Sigma_{\tau_0}} d^3 x \, \sqrt{\gamma} f \pi_{(3)}^{ij} L_{\xi} \gamma_{ij} \\
           & = \int_{M_{\tau_0}} d f \wedge  \mathbf{J} [\xi] - \int_{\Sigma_{\tau_0}} f \bm{\Theta}(\psi, \mathcal{L}_{\xi} \psi) +  \int_{\Sigma_{\tau_0}} d^3 x \, \sqrt{\gamma} f \pi_{(3)}^{ij} L_{\xi} \gamma_{ij} \\
           & = \int_{M_{\tau_0}} d f \wedge  \mathbf{J} [\xi] - \int_{\Sigma_{\tau_0}} f [\delta_{\xi} \mathbf{B} + d \bm{\Theta}_{\text{ct}}(\psi, \mathcal{L}_{\xi} \psi)] \\
           & = \int_{M_{\tau_0}} d f \wedge \left\{ \mathbf{J} [\xi] - d \bm{\Theta}_{\text{ct}}(\psi, \mathcal{L}_{\xi} \psi)\right\} -  \int_{\Sigma_{\tau_0}} f \delta_{\xi} \mathbf{B}\,,
    \end{split}
\end{align}
where we made use of the fact that $\xi$ as defined in (\ref{eq: xi_ASG_CKV}) is tangential to $\Sigma_{\tau_0}$ in going from the first to second line, equation (\ref{eq: variation_B_CKV}) in the second to third line and Stokes' theorem in the final line. We note that 
\begin{equation}
    \mathbf{B} = -\mathbf{L}_{\text{GHY}} - \mathbf{L}_{\text{ct}} = \frac{1}{8\pi G} \cdot \frac{1}{3!} \left[ K -  \sum_{n=0}^2 ( K_{(n)} - \lambda_{(n)})  \right] \epsilon_{ijk} dx^i \wedge dx^j \wedge dx^k\,,
\end{equation}
which is covariant with respect to boundary diffeomorphisms (of which $\xi$ is) and thus 
\begin{equation}
    \delta_{\xi} \mathbf{B} = L_{\xi} \mathbf{B} = \bar{d} i_{\xi} \mathbf{B}\,,
\end{equation}
where we used $L_{\xi} = i_{\xi} \bar{d} + \bar{d} i_{\xi}$ and the fact that $\mathbf{B}$ is a top-form of the boundary. Using this, we can further manipulate \eqref{eq: delta_f_S_int} as 
\begin{align}
    \begin{split}
          \delta_f S_{\text{ren}} & = \int_{M_{\tau_0}} d f \wedge \left\{ \mathbf{J} [\xi] - d \bm{\Theta}_{\text{ct}}(\psi, \mathcal{L}_{\xi} \psi)\right\} -  \int_{\Sigma_{\tau_0}} f \bar{d} i_{\xi} \mathbf{B} \\
          &= \int_{M_{\tau_0}} d f \wedge \left\{ \mathbf{J} [\xi] - d \bm{\Theta}_{\text{ct}}(\psi, \mathcal{L}_{\xi} \psi)\right\} +  \int_{\Sigma_{\tau_0}} \bar{d} f \wedge i_{\xi} \mathbf{B} \\ 
          &= \int_{M_{\tau_0}} d f \wedge \left\{ \mathbf{J} [\xi] - d \bm{\Theta}_{\text{ct}}(\psi, \mathcal{L}_{\xi} \psi)\right\} +  \int_{M_{\tau_0}} \rho(\Sigma_{\tau_0}) \wedge d f \wedge i_{\xi} \mathbf{B} \\ 
          & = \int_{M_{\tau_0}} d f \wedge \left\{ \mathbf{J} [\xi] - d \bm{\Theta}_{\text{ct}}(\psi, \mathcal{L}_{\xi} \psi) - \rho(\Sigma_{\tau_0}) \wedge i_{\xi} \mathbf{B} \right\} \,,
    \end{split}
\end{align}
where we introduced the \textit{Poincar\'e dual} of $\Sigma_{\tau_0}$ in $M_{\tau_0}$, $\rho(\Sigma_{\tau_0})$, a 1-form with $\delta$-function support on $\Sigma_{\tau_{0}}$. We have thus established that the variation of the off-shell renormalised action is proportional to the derivative of $f$.

In order to construct the current, we now follow the usual argument of Noether's theorem in noticing that the on-shell variation of the renormalised action is
\begin{equation}
\delta_f S_{\text{ren}} \approx 0\,,
\end{equation}
which can be seen from the first line of \eqref{eq: delta_f_S} using $\mathbf{E} = 0$ together with the fact that $\xi$ is an ACKV \eqref{eq: xi_ASG_CKV} and $\pi_{(3)}^{ij}$ is traceless \eqref{eq: em_traceless}. This allows us to identify the Noether current as
\begin{equation} \label{eq: Noether_current_full}
    \tilde{\mathbf{J}}[\xi] \equiv \mathbf{J} [\xi] - d \bm{\Theta}_{\text{ct}}(\psi, \mathcal{L}_{\xi} \psi) - \rho(\Sigma_{\tau_0}) \wedge i_{\xi} \mathbf{B} \approx d ( \mathbf{Q}[\xi] - \bm{\Theta}_{\text{ct}}(\psi, \mathcal{L}_{\xi} \psi)) -  \rho(\Sigma_{\tau_0}) \wedge i_{\xi} \mathbf{B}\,,
\end{equation}
where we used \eqref{eq: Q_def} in the on-shell evaluation. 

\begin{figure}
    \centering
    \includegraphics{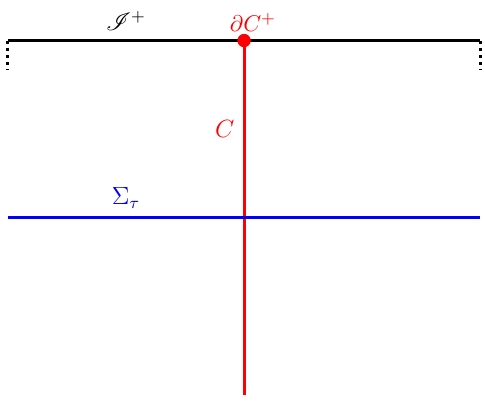}
    \caption{Penrose diagram showing the setup which we use to construct the charges for AldS spacetimes. The timelike hypersurface $C$ shown in red is the hypersurface which we integrate the Noether current $\tilde{\mathbf{J}}$ over in equation (\ref{eq: Charge_noether_1}). The charge will be defined as an integral over the surface $\partial C^+ = C \cap \mathscr{I}^+$. For comparison, we also include a spacelike ``instant of time'' hypersurface $\Sigma_{\tau}$, which is compact and does not intersect $\mathscr{I}^+$.}
    \label{fig: ds_timelike_slice}
\end{figure}

In order to construct the Noether charge, we will need to integrate the current \eqref{eq: Noether_current_full} over a suitable hypersurface $C$ which intersects a cross-section of the conformal boundary. The natural choice of $C$ made in both asymptotically flat \cite{Wald:1993nt,Iyer:1994ys,Iyer:1995kg, Wald:1999wa} and AlAdS \cite{Papadimitriou:2005ii, Compere:2020lrt} is a spacelike hypersurface which represents an ``instant of time'' and $\partial C$ is the intersection of $C$ with the conformal boundary. In the case of AldS spacetimes, ``instants of time'' are now compact manifolds and thus do not serve as suitable hypersurfaces due to their lack of boundary. Due to this, in AldS spacetimes we will take $C$ to be a \textit{timelike} slice in spacetime, which intersects the conformal boundary as shown in Figure \ref{fig: ds_timelike_slice}.

We note that different choices of hypersurfaces $C$ have appeared in the literature: our choice of timelike $C$ is in keeping with the work of \cite{Compere:2020lrt, Kolanowski:2021hwo}, whereas   \cite{Ashtekar:2014zfa, Ashtekar:2015lla, Ashtekar:2015lxa, Anninos:2010zf, PremaBalakrishnan:2019jvz} utilise \textit{spacelike} hypersurfaces in various contexts. We note that everywhere spacelike hypersurfaces are not possible
%consistent 
for charges defined at cross-sections of $\mathscr{I}^+$ as such a hypersurface must necessarily be compact and thus cannot intersect the conformal boundary in a non-trivial cross-section. This tells us that at least somewhere on the hypersurface, the causal character must change from spacelike. Another possible choice used in \cite{Chrusciel:2020rlz, Chrusciel:2021ttc, Chrusciel:2023umn} is to use \textit{null cones} which intersect $\mathscr{I}^+$. We shall see that for one-ended hypersurfaces $C$ the charges will only depend on $\partial C$ and thus the bulk causal nature of $C$ will be irrelevant. We will show in Section \ref{sec: past_contribution} that our choice of timelike $C$ naturally allows one to include the contribution from $\mathscr{I}^-$. For now, we restrict consideration to the one-ended case and define the Noether charge as 
\begin{equation} \label{eq: Charge_noether_1}
    Q_{\xi} = \int_C  \tilde{\mathbf{J}}[\xi] \approx \int_{\partial C}  \mathbf{Q}[\xi] - \bm{\Theta}_{\text{ct}}(g, \mathcal{L}_{\xi} g) -  i_{\xi} \mathbf{B}\,. 
\end{equation}

\begin{figure}[H]
    \centering
    \includegraphics{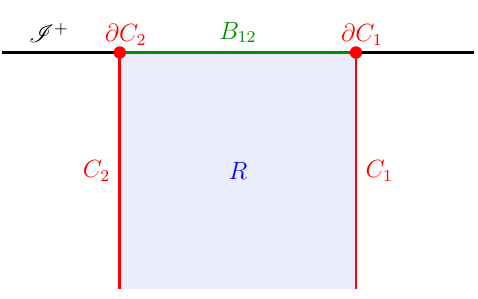}
    \caption{Penrose diagram displaying an AldS$_4$ spacetime with radially invariant charges. The charges associated with the timelike slices $C_{1,2}$ are equivalent.}
    \label{fig: ds_no_flux}
\end{figure}

It remains to show that the charge is invariant along $\mathscr{I}^+$, a notion we will refer to as \textit{radial invariance}. To show this, we first use Stokes' theorem to re-write the difference of charges as 
\begin{align}
    \begin{split}
        \left. Q_{\xi} \right |_{C_2} -  \left. Q_{\xi} \right |_{C_1} = \int_{C_2}  \tilde{\mathbf{J}}[\xi] - \int_{C_1}  \tilde{\mathbf{J}}[\xi] = \int_{R} d  \tilde{\mathbf{J}}[\xi] - \int_{B_{12} \subset \mathscr{I}^+} \tilde{\mathbf{J}}[\xi]\,,
    \end{split}
\end{align}
where the setup is depicted in Figure \ref{fig: ds_no_flux}. Taking this expression on-shell we find
\begin{align}
    \begin{split} \label{eq: conservation_Noether_1}
         \left. Q_{\xi} \right |_{C_2} -  \left. Q_{\xi} \right |_{C_1} & \approx - \int_{\partial R}   \rho(\Sigma_{\tau_0}) \wedge i_{\xi} \mathbf{B}  - \int_{B_{12}} \mathbf{J}[\xi]  - d \bm{\Theta}_{\text{ct}}(\psi, \mathcal{L}_{\xi} \psi) - \rho(\Sigma_{\tau_0}) \wedge i_{\xi} \mathbf{B} \\
         & = \int_{C_1}   \rho(\Sigma_{\tau_0}) \wedge i_{\xi} \mathbf{B} - \int_{C_2}   \rho(\Sigma_{\tau_0}) \wedge i_{\xi} \mathbf{B}  -  \int_{B_{12}} \mathbf{J}[\xi]  - d \bm{\Theta}_{\text{ct}}(\psi, \mathcal{L}_{\xi} \psi) \\
         & =  \int_{\partial C_1} i_{\xi} \mathbf{B} -  \int_{\partial C_2} i_{\xi} \mathbf{B} - \int_{B_{12}} \mathbf{J}[\xi]  - d \bm{\Theta}_{\text{ct}}(\psi, \mathcal{L}_{\xi} \psi) \\
         & = - \int_{B_{12}} \mathbf{J}[\xi]  - d \bm{\Theta}_{\text{ct}}(\psi, \mathcal{L}_{\xi} \psi) - \bar{d} i_{\xi} \mathbf{B} \\
             & = - \frac{3}{32 \pi G} \int_{B_{12}} d^3 x \sqrt{g^+_{(0)}} g_{(3)}^{+\; ij} \delta_{\xi} g^+_{(0)ij} = 0\,,
         \end{split}
\end{align}
where in the last equality we used the fact that $\xi$ is ACKV, so $\delta_{\xi} g^+_{(0)ij} \propto  g^+_{(0)ij}$ and 
the final expression vanishes since $g_{(3)}^{+\; ij}$ is traceless on-shell. This demonstrates that the Noether charges are radially invariant.  

We now relax the condition that $\xi$ is an ACKV and only assume 
\eqref{eq: xi_ASG}. One may still define the charges using \eqref{eq: Charge_noether_1} but this charge is no longer conserved - it rather satisfies a flux-balance law,
\begin{equation} \label{flux-balance}
   \left. Q_{\xi} \right |_{C_2} -  \left. Q_{\xi} \right |_{C_1} = - \int_{B^{+}_{12}} \mathbf{F}^+_{\xi_{(0)}}\,,
\end{equation}
where 
\begin{equation} \label{flux_def}
   \mathbf{F}^+_{\xi_{(0)}} = - \sqrt{g^+_{(0)}} T^{ij} \nabla^+_i \xi^{(0)}_{j} \, d\mu\,,
\end{equation}
is the flux of the current density associated with $Q_{\xi}$. 
This balance equation follows from the same steps given in \eqref{eq: conservation_Noether_1}, except that now the expression in the last step does not vanish but rather defines the flux.
In writing the last term in \eqref{eq: conservation_Noether_1} as in \eqref{flux_def} we used \eqref{eq: EM_future}.
In the next subsection we will give another perspective around this flux-balance law, and we also present another derivation later using covariant phase space methods.

Note that \eqref{flux-balance} implies that the gravitational charge is conserved if and only if the flux vanishes. If $\xi$ is an ACKV, $\mathbf{F}^+_{\xi_{(0)}}=0$ point-wise and we recover the conservation law we derived earlier. However, as the r.h.s. of \eqref{flux-balance} involves the integral of  $\mathbf{F}^+_{\xi_{(0)}}$, it would suffice for $\mathbf{F}^+_{\xi_{(0)}}$ to be exact over the compact directions of $B_{12}$ in order for $Q_{\xi}$ to be conserved. We will discuss examples of  such more general conservation laws in Section  \ref{sec: RTdS}.

\subsubsection{Gravitational charges from Noether's second theorem} \label{sec: Noet2}

Now we will consider a second derivation of the gravitational charges using Noether's second theorem. As in the previous subsection, we start by considering the case of keeping the conformal class fixed. In order to derive the charges via this approach we consider vector fields of the form (\ref{eq: xi_ASG_CKV}) and variations
\begin{equation} \label{eq: h_variation}
    \delta_h \psi = \mathcal{L}_{h \xi} \psi\,,
\end{equation}
where $h=h(x)$ is an arbitrary function on the boundary. We note that (\ref{eq: h_variation}) will preserve the renormalised action off-shell 
\begin{equation}
   0 =  \delta_h S_{\text{ren}} = \int_{M_{\tau_0}} \mathbf{E} \delta_{h} \psi + \int_{\Sigma_{\tau_0}} d^3 x \, \sqrt{\gamma} \pi_{(3)}^{ij} \delta_h \gamma_{ij}\,,
\end{equation}
since $h \xi$ is of the form \eqref{eq: xi_ASG}, but will not preserve the boundary conditions \eqref{eq: conformal_dirichlet}. In order to derive the charges, we note that on-shell we have
\begin{equation} \label{2ndNoether}
    0 = \int_{\Sigma_{\tau_0}} d^3 x \, \sqrt{\gamma} \pi_{(3)}^{ij} \delta_h \gamma_{ij}  = \int_{\Sigma_{\tau_0}} d^3 x \, \sqrt{\gamma} \pi_{(3)}^{ij} \left( h L_{\xi} \gamma_{ij} + 2 \xi_i D_j h \right)\,,
\end{equation}
and now performing integration by parts on the second term and using the fact that $h$ is an arbitrary function we have
\begin{align}
    -2 D_i \left( \pi_{(3)}^{ij} \xi_j\right) = -\pi_{(3)}^{ij}  L_{\xi} \gamma_{ij}  = 0\,,
\end{align}
where in the second equality we used the condition in (\ref{eq: xi_ASG_CKV}) that $\xi_{(0)}$ is a CKV and $\pi_{(3)}^{ij}$ is traceless. We thus reach the conclusion that 
\begin{equation} \label{eq: Q_noether_2}
    \mathcal{Q}_{\xi} = -\int_{\Sigma_{\tau_0} \cap C} \pi_{(3)j}^{\phantom{(3)}i} \xi^j_{(0)} \epsilon_{i k l} \,dx^k \wedge dx^l\,,
\end{equation}
is a radially invariant charge associated with a boundary conformal Killing vector $\xi_{(0)}$. 

The expression given above seems at first to be different to the charge expression (\ref{eq: Charge_noether_1}) derived using Noether's first theorem. Such a discrepancy parallels the AlAdS case studied in \cite{Papadimitriou:2005ii}, where the charges derived using each method were shown to be equivalent assuming some mild symmetry for the spacetimes, most importantly the stationarity of the AlAdS conformal boundary. As our analysis in the AldS case has followed that work closely, one might now expect a similar result to hold. As it turns out, the ``corner improvement" we introduced using the counterterm symplectic potential $\mathbf{\Theta}_{\text{ct}}$ is sufficient to remove the requirement to impose \textit{any} symmetry (beyond the existence of ACKVs) on $\mathscr{I}^+$ in order to have equivalence of (\ref{eq: Charge_noether_1}) and (\ref{eq: Q_noether_2}). This will become apparent after we present a third derivation in the next subsection using covariant phase space methods, but first we will discuss what happens if we relax the condition that $\xi$ is a ACKV.

We now consider the case $\xi$ satisfies the more general fall-off condition \eqref{eq: xi_ASG}.  Equation \eqref{2ndNoether} still holds and we conclude,
\begin{equation} \label{eq: Noether2}
  -2 D_i \left( \pi_{(3)}^{ij} \xi_j\right) = - \pi_{(3)}^{ij}  L_{\xi} \gamma_{ij}\, ,   
\end{equation}
but the r.h.s. is no longer zero. Integrating over $B_{12}$ in the boundary (see Figure \ref{fig: ds_no_flux}) and using the fact that the l.h.s. is a total derivative we obtain the flux-balance law in \eqref{flux-balance}.

\subsubsection{Wald Hamiltonians at \texorpdfstring{$\mathscr{I}^+$}{Scrim}} \label{sec: charges}

In this section we derive the expressions for the Wald Hamiltonians (\ref{eq: Hamiltonian_def}) on AldS$_4$ spacetimes associated to \textit{one-ended} hypersurfaces whose end is a cross-section of $\mathscr{I}^+$. To do this, we will first determine the subset of the well-posed boundary conditions for which the integrability condition (\ref{eq: H_existence}) is satisfied. We will compute the spacetime vector fields which preserve such boundary conditions and use these to provide an alternative definition of the ASG at $\mathscr{I}^+$ \eqref{eq: xi_ASG_CKV}. We will derive the explicit expression for the Hamiltonians and show that these are radially invariant quantities along $\mathscr{I}^+$. In this process we will demonstrate that they are equivalent to both (\ref{eq: Charge_noether_1}) and (\ref{eq: Q_noether_2}).

We will then consider the case of non-existence of the Hamiltonians {\it i.e.} when (\ref{eq: H_existence}) is not satisfied. This will be the relevant case in order to describe AldS$_4$ spacetimes with non-trivial gravitational radiation. In order to construct the charges in such a setting, we will construct the modified Hamiltonians (\ref{eq: mod_H}) and give a formula for the radial flux between neighbouring charges along $\mathscr{I}^+$. The results of this section reproduce several results of the prior literature \cite{Anninos:2010zf, Compere:2020lrt, Kolanowski:2021hwo} although using the Hamiltonian approach to the holographic renormalisation provides a fresh perspective on several aspects of the derivation. 

\paragraph{Existence and asymptotic symmetries} \label{sec: existence and asymptotic symmetries}

It was shown in \cite{Iyer:1995kg} that the existence of a well-posed variational problem is intimately related to the integrability of Hamilton's equations on phase space (\ref{eq: Hamiltonian_def}). In this section we will determine the subset of the boundary conditions (\ref{eq: well_posed_bcs}) which satisfy the integrability condition (\ref{eq: H_existence}) and thus give rise to a Wald Hamiltonian of the form (\ref{eq: Hamiltonian_integrable}). We will then compute the generic expression for these Hamiltonians at $\mathscr{I}^+$.

Applying the well-posed boundary conditions (\ref{eq: well_posed_bcs}) to (\ref{eq: Shifted_theta}) we find 
\begin{equation} \label{eq: theta_pullback_well_posed}
     \left.\bm{\Theta} \right|_{\mathscr{I^+}} - d  \bm{\Theta}_{\text{ct}} = \delta \mathbf{B} + d(* \bm{X} )\,,
\end{equation}
where
\begin{equation}
  X_i = -\langle T_{ij} \rangle^{\mathscr{I^+}}\zeta^j\,.
  \end{equation}
This structure is the same as that found in equation (11) of \cite{Iyer:1995kg}, which goes on to prove that the integrability condition (\ref{eq: Hamiltonian_integrable}) is satisfied when $ * \bm{X} = 0 $,
although this is not the most generic choice which allows for existence of a Hamiltonian. In order to determine the latter, we begin by evaluating the symplectic current (\ref{eq: omega_def}) pulled back to $\mathscr{I}^+$ using \eqref{eq: Shifted_theta}
\begin{align}
\begin{split} \label{eq: pullback_omega}
    \left. \bm{\omega} \right|_{\mathscr{I}^+} - d \bm{\omega}_{\text{ct}}  & = \left( -\frac{1}{2} \delta_1 \left( \sqrt{g^+_{(0)}} T^{ij} \right) \delta_2 (g^+_{(0)ij}) - ( 1 \leftrightarrow 2) \right) d\mu \\
    & = \left( - \delta_1 \left( \sqrt{g^+_{(0)}} T^{ij} \nabla^+_{(0)i} \zeta_{2j}  \right)  - ( 1 \leftrightarrow 2) \right)  d\mu\,,
    \end{split}
 \end{align}
which we see is generically non-zero for boundary diffeomorphisms generated by $\zeta_{1,2}$. In order to satisfy the existence criterion (\ref{eq: H_existence}) at the level of the integrand vanishing (when $\xi$ is not tangential to $C \cap \mathscr{I}^+$)  we must have 
\begin{equation} \label{eq: conformal_Killing}
    \nabla^+_{(0)(i} \zeta_{j)} = \frac{1}{3} g^+_{(0)ij} \nabla^+_{(0)k} \zeta^{k}\,,
\end{equation}
{\it i.e.}  $\zeta$ must be a conformal Killing vector of $g^+_{(0)}$. Such a requirement means that (\ref{eq: well_posed_bcs}) becomes 
\begin{equation} \label{eq: Existence_bcs}
     \delta g^+_{(0)ij} = 2\left( \sigma +  \frac{1}{3}  \nabla^+_{(0)k} \zeta^{k} \right) g^+_{(0)ij}  = 2  \tilde{\sigma}  g^+_{(0)ij} \,,
\end{equation}
hence the boundary conditions which lead to both a well-posed variational problem and existence of a Hamiltonian are the preservation of the conformal class at $\mathscr{I}^+$. We also note that due to (\ref{eq: conformal_Killing}), the pullback of the symplectic potential (\ref{eq: theta_pullback}) becomes 
\begin{equation} \label{eq: pullback_theta_ckv}
    \left.\bm{\Theta} \right|_{\mathscr{I^+}} - d  \bm{\Theta}_{\text{ct}}(\gamma, \delta \gamma)  = \delta \mathbf{B}\,.
\end{equation}

Finally, we want to determine which vectors in (\ref{eq: gauge_preservation}) respect this existence criterion.  The action of these vector fields on the representative of the conformal class at $\mathscr{I}^+$ can be determined by extracting the $\mathcal{O}(e^{2\tau})= \mathcal{O}(\rho_+^{-2})$ term of the transformation of the full metric tensor $\delta_{\xi} g_{ij}= \mathcal{L}_{\xi} g_{ij}$ and reads
\begin{equation} \label{eq: g_0_trans}
    \delta_{\xi} g^+_{(0)ij} = \mathcal{L}_{\xi_{(0)}^k} g^+_{(0)ij} + 2 \sigma_{\xi} g^+_{(0) ij}\,,
    \end{equation}
which we notice precisely matches the general transformation rule which ensures a well-posed variational problem (\ref{eq: well_posed_bcs}). In order to define a Hamiltonian, the argument given in equation (\ref{eq: conformal_Killing}) ensures that $\xi_{(0)}^i$ must be a conformal Killing vector of $g^+_{(0)ij}$
\begin{equation} \label{eq: conformal_Killing_xi}
     \nabla^+_{(0)(i} \xi^{(0)}_{j)} = \frac{1}{3} g^+_{(0)ij} \nabla^+_{(0)k} \xi_{(0)}^{k}\,,
\end{equation} 
and we also note from \eqref{eq: gauge_preservation} that  
\begin{equation} \label{eq: sigma_xi_zero}
    \sigma_{\xi} =0\,,
\end{equation}
ensures that we use field independent vector fields $\delta \xi =0$ as considered in \cite{Wald:1993nt, Iyer:1994ys, Iyer:1995kg, Wald:1999wa}. We have thus arrived at the ASG of ACKVs as defined in (\ref{eq: xi_ASG_CKV}), although we note that this time we derived the group by demanding existence of a Hamiltonian associated to field independent vectors $\xi$, rather than by invariance of the renormalised off-shell action \eqref{eq: S_ren} and the choice of boundary conditions \eqref{eq: conformal_dirichlet}. 

If we instead attempted to use vectors with $\sigma_{\xi} \neq 0$ as our asymptotic symmetries, then the appearance of the inverse metric in the $\xi^i$ component of (\ref{eq: gauge_preservation}) would force one to consider ``field dependent" symmetries where $\delta \xi \neq 0$ and the charge algebra becomes soft. Such a formalism is a topic of some study, (see {\it e.g.} related formulae in \cite{Compere:2019bua, Ciambelli:2024vhy}) but we do not consider such cases here.

\paragraph{Radially invariant Hamiltonians}

With the asymptotic symmetries established, we are now ready to give an explicit expression for the Hamiltonians (\ref{eq: Hamiltonian_def}) of AldS
spacetimes associated with the vector fields $\xi$ in \eqref{eq: gauge_preservation} satisfying properties \eqref{eq: conformal_Killing_xi} and \eqref{eq: sigma_xi_zero}. For convenience, we recall Hamilton's equations of motion on phase space  
\begin{equation} \label{eq: Hamiltonian_recall}
\delta H_{\xi} = \Omega_{C} (\psi, \delta \psi, \mathcal{L}_{\xi} \psi)= \int_{C} \boldsymbol{\omega} (\psi, \delta \psi, \mathcal{L}_{\xi} \psi) = \int_{\partial C} \delta \mathbf{Q}[\xi] - i_{\xi} \bm{\Theta}(\psi, \delta \psi)\,,
\end{equation}
where $C$ is again a timelike slice as displayed in Figure \ref{fig: ds_timelike_slice}. We recall that the equation above has been written without any consideration of the ambiguities present in the definition of $\bm{\Theta}$. Since we employ the shift (\ref{eq: Shifted_theta}), the variation of our Hamiltonian is given by \cite{Iyer:1994ys}
\begin{align}
\begin{split} \label{eq: Hamiltonian_modified_exp}
    \delta H_{\xi} & = \int_{C} \boldsymbol{\omega} (\psi, \delta \psi, \mathcal{L}_{\xi} \psi) - d \boldsymbol{\omega}_{\text{ct}} (\psi, \delta \psi, \mathcal{L}_{\xi} \psi) \\
    &= \int_{\partial C} \delta \mathbf{Q}[\xi] - i_{\xi} \bm{\Theta}(\psi, \delta \psi) - \bm{\omega}_{\text{ct}}(\psi, \delta \psi, \mathcal{L}_{\xi} \psi) \\
    & =  \int_{\partial C} \delta \mathbf{Q}[\xi] - i_{\xi} \bm{\Theta}(\psi, \delta \psi) - \delta \bm{\Theta}_{\text{ct}} (\mathcal{L}_{\xi} \psi, \psi) + \mathcal{L}_{\xi} \bm{\Theta}_{\text{ct}}(\delta \psi, \psi) \\
     & = \int_{\partial C} \delta \mathbf{Q}[\xi] - i_{\xi} \bm{\Theta}(\psi, \delta \psi) - \delta \bm{\Theta}_{\text{ct}} (\mathcal{L}_{\xi} \psi, \psi) + i_{\xi} d \bm{\Theta}_{\text{ct}}(\delta \psi, \psi) + d i_{\xi}  \bm{\Theta}_{\text{ct}}(\delta \psi, \psi) \\ 
     & = \int_{\partial C} \delta \left[ \mathbf{Q}[\xi] - \bm{\Theta}_{\text{ct}} (\mathcal{L}_{\xi} \psi, \psi)\right]   - i_{\xi} \left[ \bm{\Theta}(\psi, \delta \psi) - d \bm{\Theta}_{\text{ct}}(\delta \psi, \psi) \right]\,,
    \end{split}
\end{align}
where in the first line we used Stokes' theorem, in the second the definition of the symplectic current, \eqref{eq: omega_def}, in the third, ${\cal L}_\xi= i_\xi d+ d i_\xi$, and in final line we packaged the terms together in order to demonstrate how the Noether charge and the symplectic potential transform respectively. Notice that we have ignored the contribution of the $d$-exact term as we assume that $\partial (\partial C) = \emptyset$. This term would have to be included if one considers a spacetime whose cross-sections contain non-trivial boundary, for example the C-metric solutions \cite{Kinnersley:1970zw} where the contribution of such terms was considered in the AlAdS case \cite{Kim:2023ncn}.

With the hypersurface $C$ and the vector $\xi$ established, it remains to compute the explicit formulae for the charges (\ref{eq: Hamiltonian_modified_exp}). With the explicit form of Hamilton's equations on phase space now derived, we extract the Hamiltonian directly from integrating the right hand side of (\ref{eq: Hamiltonian_modified_exp}). To do this, we note the well-known result for the Noether charge of general relativity 
\begin{equation} \label{eq: Q_GR}
    \mathbf{Q}[\xi] = - * \mathbf{\Xi}[\xi]\,,
\end{equation}
where 
\begin{equation}
    \Xi_{\mu \nu} = \frac{1}{8\pi G} \nabla_{[\mu} \xi_{\nu]}\,.
\end{equation}
Now using the gauge (\ref{eq: metric_ADM}), the Christoffel symbols (\ref{eq: Christoffel_symbols}) and the asymptotic symmetry conditions (\ref{eq: xi_ASG_CKV}) for $\xi$ we find the relevant components are 
\begin{equation} \label{eq: Xi_t_i}
\nabla^{[i} \xi^{\tau]} = K^i_j \xi^j + \mathcal{O}(\rho_+^5)\,,
\end{equation}
and so
\begin{equation}
     \mathbf{Q}[\xi] = - \frac{1}{8 \pi G} \cdot \frac{1}{2!} \epsilon_{ij k \tau} \nabla^{[k} \xi^{\tau]} dx^i \wedge dx^j = \frac{1}{8\pi G} \cdot \frac{1}{2!}  K^k_l \xi^l \epsilon_{k i j} dx^i \wedge dx^j   + \mathcal{O}(\rho_+)\,,
\end{equation}
which gives us the first term in the integral (\ref{eq: Hamiltonian_modified_exp}). 

In order to derive the second term we note
\begin{equation} \label{eq: Theta_3_EH}
  -\left( *_{3} \bm{\Theta}_{\text{ct}} \right)^{i} = \left( *_{3} \bm{\Theta}^{(3)}_{\text{EH}} \right)^{i} = \frac{1}{16 \pi G}\left( D^i(\gamma_{jk} \delta \gamma^{jk} ) - D_j \delta \gamma^{ij}\right) = v^i_{(2)} = v^i + \mathcal{O}(\rho_+^5)\,, 
\end{equation}
where $v^i$ is defined in \eqref{eq: sym_pot} and the subscript indicates dilatation weight (see Section \ref{sec: Theory and holographic renormalisation}) and hence 
\begin{equation}
   \bm{\Theta}_{\text{ct}}(\gamma, \delta_{\xi} \gamma) = -  *_{3} *_{3} \bm{\Theta}^{(3)}_{\text{EH}}(\gamma, \delta_{\xi} \gamma) = - \frac{1}{2!}  v_{(2)}^k \epsilon_{kij} dx^i \wedge dx^j\,.
\end{equation}

For the third term we use (\ref{eq: pullback_theta_ckv}) to write 
\begin{equation}
    i_{\xi} \left[\bm{\Theta} - d\bm{\Theta_{\text{ct}}} \right] = \delta i_{\xi} \mathbf{B} = \frac{1}{16\pi G} \delta \left\{   \left[ K - \sum_{n=0}^2 ( K_{(n)} - \lambda_{(n)})  \right] \epsilon_{kij}  \right\} \xi^k  dx^i \wedge dx^j + \mathcal{O}(\rho_+)\,,
\end{equation}
and thus the Hamiltonian (\ref{eq: Hamiltonian_modified_exp}) takes the form 
\begin{align}
\begin{split} \label{eq: Corner_Hamiltonian}
    H_{\xi} & = \int_{\partial C} \mathbf{Q}[\xi] - \bm{\Theta}_{\text{ct}} (\mathcal{L}_{\xi} \psi, \psi)- i_{\xi} \mathbf{B}(\psi) \\
    & = - \frac{1}{8 \pi G} \cdot \frac{1}{2!} \int_{\partial C}  \left( K^k_l \xi^l + v_{(2)}^k  - \left\{  K - \sum_{n=0}^2 ( K_{(n)} - \lambda_{(n)}) \right\} \xi^k\right) \epsilon_{kij} dx^i \wedge dx^j  \\
    & = - \int_{\partial C} \left( \pi^{\phantom{(3)} k}_{(3)l} \xi^l + \frac{1}{16 \pi G}\left( v_{(2)}^k +  \sum_{n=0}^2 ( K^{\phantom{(n)}k}_{(n)l} - \lambda_{(n)} \delta^k_l) \xi^l \right)  \right) \epsilon_{kij} dx^i \wedge dx^j\,, 
    \end{split} 
\end{align}
where the extra minus sign which appears in front of the charge is due to the fact that the orientation of $\partial C$ when embedded in $C$ is opposite that of the ``natural'' orientation on surfaces in AldS$_4$ spacetimes and we employed the definition of the conjugate momenta (\ref{eq: momenta}) in going to the final line. The na\"ive conclusion from counting dilatation weights of these terms is that the first term is finite and the second is divergent at $\mathscr{I}^+$. In fact, the second term vanishes, a proof of which is given in Appendix \ref{app: Hamiltonian_proof}. The inclusion of the counterterm symplectic potential is important in the vanishing of the second term and allows for a relaxation of assumption \textit{ii)} of Lemma 4.1 in \cite{Papadimitriou:2005ii}, where the term was not included and the additional assumption of the metric $\gamma_{ij}$ having an isometry direction orthogonal to the hypersurface $\partial C$ was required. Such a relaxation is only complete for $d=4$, and would be interesting to pursue in higher spacetime dimensions. Looking ahead to the next subsection, we also note that the main use of relaxaing this assumption is in being able to extend the notion of a phase space Hamiltonian to the cases where $\slashed{\exists}$ ACKVs, as many of the known solutions which possess ACKVs ({\it e.g.} Kerr-Newman-dS$_4$) already satisfy the assumptions of \cite{Papadimitriou:2005ii}. 

We summarise the main result as follows: for an AldS$_4$ spacetime, the Wald Hamiltonians corresponding to the ACKVs $\xi$ are given by
\begin{equation} \label{eq: Final_Hamiltonian}
    H_{\xi} = - \int_{\partial C}  \pi^{\phantom{(3)} k}_{(3)l} \xi^l \epsilon_{kij} dx^i \wedge dx^j  = \frac{1}{2} \int_{\partial C} T^{k}_{l} \xi_{(0)}^l \epsilon^{(0)}_{kij} dx^i \wedge dx^j\,.
\end{equation} 
where we used (\ref{eq: EM_future}) and the fact that the only terms which contribute are those at leading order. We also note that by equations (\ref{eq: Corner_Hamiltonian}) and (\ref{eq: Final_Hamiltonian}) the Wald Hamiltonians are equivalent to \textit{both} Noether charges $Q_{\xi}$ as defined in (\ref{eq: Charge_noether_1}) and $\mathcal{Q}_{\xi}$ defined in (\ref{eq: Q_noether_2}). This means that all three derivations of the charges ar $\mathscr{I}^+$ give the same value and are equivalent. 

The fact that the Hamiltonians are the same as the Noether charge also means they are radially invariant by equation (\ref{eq: conservation_Noether_1}). We can equivalently show this using (\ref{eq: H_difference}), where the difference between two charges defined at cross-sections $\partial C_1, \partial C_2$ is given by
\begin{align}
\begin{split} 
    \left. \delta H_{\xi} \right|_{\partial C_2} -   \left. \delta H_{\xi} \right|_{\partial C_1} & = - \int_{B_{12}} \boldsymbol{\omega} (\psi, \delta \psi, \mathcal{L}_{\xi} \psi)- d \bm{\omega}_{\text{ct}}  (\psi, \delta \psi, \mathcal{L}_{\xi} \psi) \\
    & = \delta \int_{B_{12}}  \left(\sqrt{g^+_{(0)}} T^{ij} \nabla^+_{(0)(j} \xi^{(0)}_{i)} \right) d\mu \\
    & = 0\,, \label{eq: conservation}
    \end{split}
\end{align}
where we used the conformal Killing equation (\ref{eq: conformal_Killing_xi}) in the final step.

A final important note concerning the charges derived using this technique is that they appear to suffer from ambiguities arising from the integration of Hamilton's equations on the phase space. If $H_{\xi}$ satisfies (\ref{eq: Hamiltonian_modified_exp}), then so does 
\begin{equation} \label{eq: H_ambiguity}
    H'_{\xi} = H_{\xi} + \int_{\partial C} H^i_j \xi^j \epsilon_{ikl} dx^k \wedge dx^l\,, \qquad\delta\left( \int_{\partial C} H^i_j \xi^j \epsilon_{ikl} dx^k \wedge dx^l \right) =0\,,
\end{equation}
where the components of $H^i_j$ are analytic functionals of the intrinsic quantities defined on the boundary. The requirement for the second term to be in the Kernel of $\delta$ enforces constraints on $H^i_j$ which depend on the boundary conditions. For the boundary conditions \eqref{eq: conformal_dirichlet} (equivalent to \eqref{eq: Existence_bcs} which give existence of a Hamiltonian), we have up to total derivatives
\begin{equation}
    \delta \left( H^i_j \sqrt{\gamma} \right) = \sqrt{\gamma} \delta H^i_j  +3 \tilde{\sigma} H^i_j \sqrt{\gamma} = 0 \iff \delta H^i_j = -3 \tilde{\sigma} H^i_j\,, \label{eq: amb_sc}
\end{equation}
{\it i.e.} $H^i_j$ has dilatation weight $3$. By the requirement of radial invariance of the charge $H_{\xi}'$ along $\mathscr{I}^+$ we also have 
\begin{equation}
    D_{i}\left( H^i_j \xi^j \right) = \xi^j D_i H^i_j + H^i_j D_i \xi^j = \xi^j D_i H^i_j + \frac{1}{3} H^i_i D_{k} \xi^k = 0\,,
\end{equation}
where we used the fact that $\xi$ is an ACKV of the boundary \eqref{eq: xi_ASG_CKV}. The solution to the above equation is 
\begin{equation} \label{eq: amb}
D_{i} H^i_j = H^i_i =0\,, 
\end{equation}
{\it i.e.} $H^i_j$ is traceless and conserved. In $d=4$ there is no quantity which satisfies this property other than $\pi_{(3)j}^{\phantom{(3)}i}$.\footnote{This is no longer be the case if we allow for parity odd conserved tensor: the Cotton-York tensor, $C_{ij} = \epsilon_i{}^{kl} D_k (R_{lj} - \frac{1}{4} R g_{lj})$ satisfies the conditions \eqref{eq: amb}. Such term would be produced if we include the Pontryagin term  
in the bulk action.}. In fact there is no local parity-even diffeomorphism covariant quantity constructed from a metric that satisfies \eqref{eq: amb_sc} even before we impose any additional conditions (such as  \eqref{eq: amb}). Note that $\pi_{(3)j}^{\phantom{(3)}i}$ is not a local function of $g_{(0)}$, and it already appears in the expression (\ref{eq: Final_Hamiltonian}) for the radially invariant Hamiltonian and thus the only ambiguity present is shifting the overall numerical factor multiplying the charge. The radially invariant Hamiltonian is thus defined unambiguously once one chooses a suitable normalisation of the ACKV $\xi$. 

\paragraph{Fluxes - Modified Hamiltonians} \label{sec: fluxes_one_ended}

We have constructed in equation (\ref{eq: Final_Hamiltonian}) the generic expression for the Wald Hamiltonian at $\mathscr{I}^+$ of an AldS$_4$ spacetime associated with an ACKV $\xi$. A shortcoming of this formula is that it does not apply to spacetimes which fail to admit the required asymptotic isometries at (at least some portion of) $\mathscr{I}^+$. The natural conclusion is that such spacetimes do not possess Wald Hamiltonians which are finite and radially invariant at $\mathscr{I}^+$ and must instead exhibit non-trivial gravitational radiation reaching $\mathscr{I}^+$. In order to describe the presence of this radiation, we will now construct the modified Hamiltonians following the procedure described in Section \ref{sec: Modified_H}. The formulae we derive will be applicable to a portion of $\mathscr{I}^+$ where $\nexists$ an ACKV, although we will be interested in cases where the spacetime does admit an ACKV in a particular portion or limit of $\mathscr{I}^+$, an explicit example of which we will discuss in Section \ref{sec: RTdS}.

We begin by noting that for a spacetime which does not possess an ACKV, one generically has 
\begin{equation}
    g_{(3)}^{+\; ij} \delta_{\xi} g^+_{(0)ij} = g_{(3)}^{+\; ij} \mathcal{L}_{\xi_{(0)}^k} g^+_{(0)ij} \neq 0\,,
\end{equation}
as there is no conformal Killing vector $\xi_{(0)}$. Recalling equations (\ref{eq: theta_pullback}) and (\ref{eq: conformal_Killing}), we note that a sufficient condition for existence of a Hamiltonian was that the boundary vector field $\zeta$ in the well-posed boundary conditions (\ref{eq: well_posed_bcs}) was a CKV although now such a vector field does not exist! 

\begin{figure}
    \centering
    \includegraphics{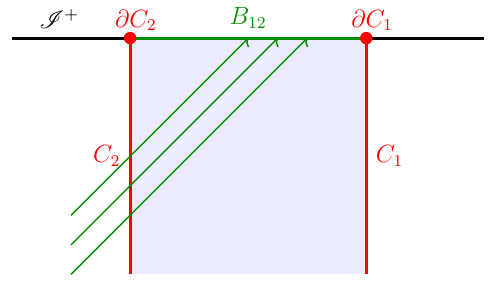}
    \caption{Penrose diagram displaying an AldS$_4$ spacetime with gravitational radiation reaching $\mathscr{I}^+$. Due to the radiation, the Hamiltonians associated with the timelike slices $C_{1,2}$ are not equivalent, with their difference given by the flux formula (\ref{eq: flux_ds}).}
    \label{fig: ds_flux}
\end{figure}

We revisit the pullback of the symplectic current to $\mathscr{I}^+$ (\ref{eq: pullback_omega})
\begin{equation}
 \bm{\omega} (\psi, \delta_1 \psi, \delta_2 \psi) - d \bm{\omega}_{\text{ct}}  (\psi, \delta_1 \psi, \delta_2 \psi)  = \left( - \delta_1 \left( \sqrt{g^+_{(0)}} T^{ij} \nabla^+_{(0)i} \zeta_{2j}  \right)  - ( 1 \leftrightarrow 2) \right)  d\mu\,,
\end{equation}
and we note that this is now generically non-zero and the existence criterion (\ref{eq: H_existence}) fails. Clearly we need to apply the modification procedure in order to define a Wald Hamiltonian. In order to identify a suitable candidate for the modification term in (\ref{eq: mod_H}), we note that the pullback of the symplectic potential is now given by
\begin{equation}
   \left.\bm{\Theta} \right|_{\mathscr{I^+}} - d  \bm{\Theta}_{\text{ct}}(\gamma, \delta \gamma) = \delta \mathbf{B} - \frac{1}{2} \sqrt{g^+_{(0)}} T^{ij} \delta g^+_{(0)ij} d\mu\,,
\end{equation}
and thus the pullback of the symplectic current is
\begin{equation}
   \bm{\omega} (\psi, \delta_1 \psi, \delta_2 \psi) - d \bm{\omega}_{\text{ct}}  (\psi, \delta_1 \psi, \delta_2 \psi)  = - \frac{1}{2} \delta_1 \left(  \sqrt{g^+_{(0)}} T^{ij}  \delta_2 g^+_{(0)ij}\right) - (1 \leftrightarrow 2)\,,
\end{equation}
allowing us to identify a candidate for the modification term as
\begin{equation} \label{eq: modification_term_FG}
    \bm{\theta} (\psi, \delta \psi) = - \frac{1}{2} \sqrt{g^+_{(0)}} T^{ij} \delta g^+_{(0)ij} d\mu + \delta \mathbf{W}(\psi)\,,
\end{equation}
where we have included the ambiguity term $\mathbf{W}$ which can be fixed with some physical criteria. The criteria discussed in \cite{Kolanowski:2021hwo} of locality, conformal invariance and analyticity uniquely fix 
\begin{equation} \label{eq: flux_ambiguity}
    \mathbf{W} = 0\,, 
\end{equation}
and thus with this choice one finds the modified Hamiltonian as defined in (\ref{eq: mod_H}) is 
\begin{align}
\begin{split} \label{eq: Mod_H}
\delta \mathcal{H}_{\xi} & =  \int_{\partial C} \delta \left[ \mathbf{Q}[\xi] - \bm{\Theta}_{\text{ct}} (\psi, \mathcal{L}_{\xi} \psi)\right]   - i_{\xi} \left[ \bm{\Theta}(\psi, \delta \psi) - d \bm{\Theta}_{\text{ct}}(\psi, \delta \psi) - \bm{\theta}(\psi, \delta \psi) \right]   \\
& = \delta \int_{\partial C}   \mathbf{Q}[\xi] - \bm{\Theta}_{\text{ct}} (\psi, \mathcal{L}_{\xi} \psi)  - i_{\xi} \mathbf{B} \,,
\end{split}
\end{align}
which is in agreement with (\ref{eq: Corner_Hamiltonian}). We note that the flux between neighbouring hypersurfaces (\ref{eq: flux}) is given by 
\begin{equation} \label{eq: flux_ds}
    \left.  \mathcal{H}_{\xi} \right|_{\partial C_2} -   \left.  \mathcal{H}_{\xi} \right|_{\partial C_1} =  - \int_{B_{12}} \bm{\theta} (\psi, \mathcal{L}_{\xi} \psi) = \int_{B_{12}} \sqrt{g^+_{(0)}} T^{ij} \nabla^+_{(0)(i} \xi^{(0)}_{j)} \,d\mu \,,
\end{equation}
which is again clearly non-zero due to the absence of CKVs. A Penrose diagram depicting this flux is shown in Figure \ref{fig: ds_flux}.

For the modified Hamiltonian (\ref{eq: Mod_H}) there is again a possibility of an ambiguous term arising in the kernel of $\delta$. This equation has to be examined separately to the radially invariant case as the boundary conditions are different. The possible ambiguity is of the form 
\begin{equation}
    \mathcal{H}'_{\xi} = \mathcal{H}_{\xi} + \int_{\partial C} \mathcal{H}^i_j \xi^j \epsilon_{ikl} dx^k \wedge dx^l\,, \qquad\delta\left( \int_{\partial C} \mathcal{H}^i_j \xi^j \epsilon_{ikl} dx^k \wedge dx^l \right) =0\,,
\end{equation}
where now we do not fix the conformal class at $\mathscr{I}$ and $\xi$ is no longer an ACKV (as the boundary metric does not admit such isometries). We only require that $\xi$ satisfies the fall-off conditions \eqref{eq: xi_ASG}, and thus they act on the boundary as local diffeomorphisms. This means that any local diffeomorphism covariant tensor $\mathcal{H}^i_j$ provides such ambiguity. Imposing finiteness, the tensor should have
dilatation weight $3$ (dilatation weight $(d-1)$ in general dimension), and as we already remarked earlier (below \eqref{eq: amb}) there is no such  (parity even) term in pure gravity when $d=4$. We also refer to \cite{Chandrasekaran:2021vyu, Kolanowski:2021hwo} for complementary discussion about this point.

\paragraph{Comment on different approaches to defining charges}

In the previous sections we have utilised three different approaches to derive the charges defined at cross-sections of $\mathscr{I}^+$ in an AldS$_4$ spacetime. We now summarise the main points and advantages/drawbacks of each method:

\vspace{5mm}

\noindent \textit{Noether's first theorem}

\vspace{5mm}

This method is the most direct analogue of derivation of conserved charges in field theory. 
The method relies on considering field variations which do not preserve the off-shell action but do preserve the boundary conditions (\ref{eq: xi_ASG_CKV}). This approach allows for a derivation of radially invariant charges along $\mathscr{I}^+$. Relaxing the boundary conditions to \eqref{eq: xi_ASG} yields a flux-balance equations and a flux formula.

\vspace{5mm}

\noindent \textit{Noether's second theorem} 

\vspace{5mm}

This method relies on considering field variations which do preserve the action but do not preserve the boundary conditions (\ref{eq: xi_ASG_CKV}). The charges derived in this approach (\ref{eq: Q_noether_2}) are equivalent to those of the first method (\ref{eq: Charge_noether_1}) and are thus also radially invariant along $\mathscr{I}^+$.  If we relax the boundary conditions to \eqref{eq: xi_ASG} then the transforms are represent the local invariance of the off-shell action. This is the formulation that provides the fastest route to the final formulae, including the flux-balance law.

\vspace{5mm}

\noindent \textit{Wald Hamiltonians} 

\vspace{5mm}

In order to satisfy existence of the Hamiltonians as introduced in \cite{Wald:1993nt, Iyer:1994ys, Iyer:1995kg} we showed in equation (\ref{eq: conformal_Killing_xi}) that the spacetime must admit ACKVs at $\mathscr{I}^+$. The expression for the charges associated with the ACKVs are given in (\ref{eq: Final_Hamiltonian}) and they are invariant along $\mathscr{I}^+$. 

In order to generalise the construction of Wald Hamiltonians to spacetimes without ACKVs at $\mathscr{I}^+$, we used the modification procedure first outlined in \cite{Wald:1999wa}. This procedure results in a charge expression (\ref{eq: Mod_H}) which is the same as (\ref{eq: Final_Hamiltonian}), although now there is a non-zero flux between neighbouring charges, with the outgoing flux given in equation (\ref{eq: flux_ds}). We note that in order to apply this procedure it was crucial to include the contributions arising from the ``corner term'' $\mathbf{\Theta}_{\text{ct}}$ introduced by variation of the counterterm action. Had we ignored this term as in the AlAdS case studied in \cite{Papadimitriou:2005ii} then our expression for the radially invariant Hamiltonian (\ref{eq: Final_Hamiltonian}) would still have been valid, although when applied to spacetimes without ACKVs one would find a divergent Hamiltonian as the proof of finiteness given in Appendix \ref{app: Hamiltonian_proof} would no longer apply. This divergence is precisely cured by the inclusion of the corner terms.

In this article we are primarily interested in the case of pure AldS$_4$ gravity, {\it i.e.} with no matter present, where we have just shown that nothing non-trivial appears in the Kernel of $\delta$ for both the radially invariant and modified Hamiltonians. However, such ambiguities are present in other situations ({\it e.g.} $d>4$, $d=4$ with matter), see the discussion in \cite{Papadimitriou:2005ii}. It would interesting to understand how to fix this ambiguity in general, see \cite{Chandrasekaran:2021vyu} for work in this direction.

In conclusion, all three approaches are equivalent, both when describing radially invariant quantities at $\mathscr{I}^+$ as well as producing a flux formula which describes the passage of gravitational waves in AldS$_4$ spacetimes. The flux-balance law comes from relaxing the boundary conditions from \eqref{eq: xi_ASG_CKV} to \eqref{eq: xi_ASG} for the Noether methods and gives the same result as the Wald-Zoupas modification procedure \cite{Wald:1999wa} for the covariant phase space Hamiltonians. For clarity, we will work in the language of the covariant phase space for the remainder of this article.

\subsection{Past contribution to the charges} \label{sec: past_contribution}

In the previous sections we considered the Wald Hamiltonians associated to one-ended timelike hypersurfaces $C$ which end at $\mathscr{I}^+$. Despite the global nature being different between AldS and AlAdS, the local structural form of the AldS$_4$ Hamiltonians inherits much from the analogous results in AlAdS$_4$ spacetimes \cite{Papadimitriou:2005ii}. In this section we will now discuss one of the important differences present in AldS vs AlAdS spacetimes: the causal structure of AldS$_4$ spacetimes indicates that one-ended hypersurfaces are not sufficient to fully describe the effects of gravitational radiation and one should instead consider \textit{two-ended} $C$ upon which to define the Hamiltonians. This will lead one to consider an additional contribution to the charge arising from the \textit{past end} of the timelike hypersurface $C$. 

In order to motivate this additional consideration, we consider the simple examples of pure (A)dS$_4$. In AdS$_4$, a partial Cauchy spacelike hypersurface $C$ has one boundary component, namely $\partial C = C \cap \mathscr{I} $, as the conformal boundary is connected. On the other hand, in dS$_4$, a timelike hypersurface $C$ necessarily has two boundary components $\partial C = (C \cap \mathscr{I}^+  ) \sqcup ( C \cap \mathscr{I}^- )$, as the conformal boundary consists of two disconnected components. As we thus need to consider two-ended $C$ even in pure dS$_4$, a consideration of the past ends of $C$ for all AldS$_4$ spacetimes is a necessary ingredient to fully understand their charges. 

In this section we will develop the technology for the past contributions to the charges, showing how such considerations are essential in determining \textit{conserved quantities} in AldS spacetimes, where  \textit{conserved} now means under {\it time evolution}. We will pay attention to the cases of the past end being both a cut of $\mathscr{I}^-$ as well as a generic interior spacetime hypersurface. The second case is important in understanding both quasi-local charges as well as the notion of a two ended $C$ when $\mathscr{I}^{-}$ is not present, for example in the Robinson-Trautman de Sitter spacetime \cite{Robinson:1960zzb, stephani_kramer_maccallum_hoenselaers_herlt_2003, Bicak:1995vc} which we will discuss in more detail in Section \ref{sec: RTdS}.

\subsubsection{Contribution at \texorpdfstring{$\mathscr{I}^-$}{Scrim}} 

\begin{figure}[H]
\begin{center}  
\includegraphics[width=0.5\linewidth]{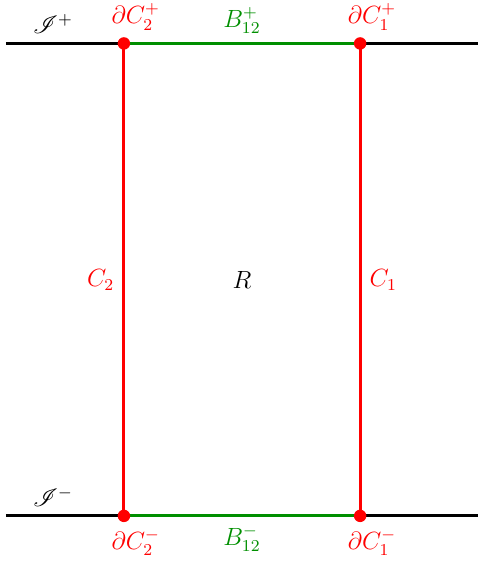}
\caption{Penrose diagram of an AldS spacetime where the timelike hypersurfaces $C_{1,2}$ stretch between $\mathscr{I}^+$ and $\mathscr{I}^-$} \label{fig: dS_both_boundaries}
\end{center}
\end{figure}

We begin with the case of the past end of $C$ being a cut of $\mathscr{I}^-$. As before, we want to investigate the differences between Hamiltonians associated with these timelike hypersurfaces, leading us to consider the scenario depicted in Figure \ref{fig: dS_both_boundaries}.

We recall that in neighbourhoods of both $\mathscr{I}^{\pm}$ one can use separate Starobinsky expansions \cite{Starobinsky:1982mr} to write the spacetime metric, {\it i.e.} near $\mathscr{I}^{\pm}$ one has 
\begin{equation}
ds^2 = \frac{\ell^2}{\rho_\pm^2} \left( - d\rho_\pm^2 + (g^{\pm}_{(0)ij} + \rho_\pm^2 g^{\pm}_{(2)ij} + \rho_\pm^3 g^{\pm}_{(3)ij} + \ldots) dx_\pm^i dx_\pm^j  \right)\,,
\end{equation}
where the coordinates are not required to overlap at all in their respective regions of validity. The power of this result is that one can apply many of the results that we have already proven at $\mathscr{I}^+$ directly to $\mathscr{I}^-$, namely that applying the procedure of holographic renormalisation at $\mathscr{I}^-$ and demanding asymptotic symmetries consisting of ACKVs at $\mathscr{I}^{-}$. At each boundary of the spacetime, we define the energy-momentum tensors $T_{ij}^{(\pm)}$ as 
\begin{equation}
T_{ij}^{\pm} = - \frac{3 \ell}{16\pi G} g^{(3)\pm}_{ij}\,.
\end{equation}

 In order to consider the charges depicted in Figure \ref{fig: dS_both_boundaries}, we require existence of a \textit{double ACKV}: a bulk vector field $\xi$ which satisfies $\xi |_{\mathscr{I}^{\pm}}= \xi^{\pm}_{(0)}$ where $\xi^{\pm}_{(0)}$ are conformal Killing vectors of $g^{\pm}_{(0)}$ respectively. In order to examine the past contribution to the charges, we consider the closed form 
 \begin{equation} \label{eq: omega_mod}
    \tilde{\boldsymbol{\omega}} (\psi, \delta \psi, \mathcal{L}_{\xi} \psi) =  \boldsymbol{\omega} (\psi, \delta \psi, \mathcal{L}_{\xi} \psi) - d \boldsymbol{\omega}_{\text{ct}} (\psi, \delta \psi, \mathcal{L}_{\xi} \psi)\,,
 \end{equation}
 where we note that $\boldsymbol{\omega}_{\text{ct}}$ recieves contributions from the counterterm actions that one adds on both $\mathscr{I}^{\pm}$. The precise form of the $\mathscr{I}^-$ term can be derived using the mappings in \eqref{eq: FG_scri+_to-} but will not be explicitly needed in our analysis. Employing Stokes' theorem
\begin{align}
0 & = \int_{R} d \tilde{\boldsymbol{\omega}} (g, \delta g, \mathcal{L}_{\xi} g) = \int_{\partial R} \tilde{\bm{\omega}} (g, \delta g, \mathcal{L}_{\xi} g)\,,
\end{align}
and taking $R$ to be as depicted in Figure \ref{fig: dS_both_boundaries}, we find
\begin{align}
 \int_{C_2} \tilde{\bm{\omega}} (g, \delta g, \mathcal{L}_{\xi} g) +  \int_{B_{12}^{+}} \tilde{\bm{\omega}} (g, \delta g, \mathcal{L}_{\xi} g) = \int_{C_1} \tilde{\bm{\omega}} (g, \delta g, \mathcal{L}_{\xi} g) + \int_{B_{12}^{-}} \tilde{\bm{\omega}} (g, \delta g, \mathcal{L}_{\xi} g)\,,
\end{align}
and now we note that $\tilde{\bm{\omega}}(g,\delta g, \mathcal{L}_{\xi} g) |_{B_{12}^{\pm}}= 0$ by virtue of $\xi$ being an ACKV of both $\mathscr{I}^{\pm}$. This simplifies the above result to 
\begin{equation}
 \int_{C_1} \tilde{\bm{\omega}} (g, \delta g, \mathcal{L}_{\xi} g) - \int_{C_2} \tilde{\bm{\omega}} (g, \delta g, \mathcal{L}_{\xi} g) = 0\,,
\end{equation}
which can be recast in terms of the Hamiltonians using \eqref{eq: Hamiltonian_modified_exp} and an integration over the phase space
\begin{equation} \label{eq: Hamiltonian_Conservation}
H_{\xi} |_{C_1} - H_{\xi} |_{C_2} = 0\,,
\end{equation}
\textit{i.e.} the Hamiltonians associated to the timelike slices are independent of the slice, just as in the one-ended case. As we consider two-ended slices, the Hamiltonians now take the form
\begin{align} \label{eq: H_two_ended}
    H_{\xi} |_C & = Q_{\xi}^+[C] - Q_{\xi}^-[C]\,,
\end{align}
with 
\begin{equation} \label{eq: Q_pm}
    Q_{\xi}^{\pm}[C] = \frac{1}{2} \int_{\partial C^{\pm}}T^{\pm k}_{\phantom{+}l} \xi_{\pm(0)}^l \epsilon^{(0)\pm}_{kij} dx^i \wedge dx^j =  \int_{\partial C^{\pm}} d\sigma_i^{\pm} T_{\pm}^{ij} \xi^{\pm}_{j(0)}\,,
\end{equation}
where we introduced $d\sigma_i^{\pm} = \frac{1}{2}\epsilon^{(0)\pm}_{ijk} dx^j \wedge dx^k$ and the derivation of the equation above is given in Appendix \ref{sec: scri_-}. Using \eqref{eq: H_two_ended}, we can express the slice independence property \eqref{eq: Hamiltonian_Conservation} as 
\begin{equation} \label{eq: charge_differences} 
\int_{\partial C_1^{+}} d\sigma_i^+ T_{+}^{ij} \xi^{+}_{j(0)} - \int_{\partial C_2^{+}} d\sigma^+_i T_{+}^{ij} \xi^{+}_{j(0)} = \int_{\partial C_1^{-}} d\sigma^-_i T_{-}^{ij} \xi^{-}_{j(0)} - \int_{\partial C_2^{-}} d\sigma_i^- T_{-}^{ij} \xi^{-}_{j(0)} \,,
\end{equation}
which tells us that the difference between the charges integrals at $\mathscr{I}^+$ is equal to the difference at $\mathscr{I}^-$, 
\begin{equation} \label{DeltaQ}
    \Delta Q_{\xi}^+ = \Delta Q_{\xi}^- \, ,
\end{equation}
where $\Delta Q_{\xi}^\pm = Q_{\xi}^{\pm}[C_2] - Q_{\xi}^{\pm}[C_1]$. We have thus shown that $\Delta Q_{\xi}$ is a conserved quantity under time evolution from $\mathscr{I}^-$ to $\mathscr{I}^+$.

We can in fact go further than this, and explicitly show that the differences at $\mathscr{I}^{\pm}$ are both identically zero (\textit{i.e.} each side of equation (\ref{eq: charge_differences}) vanishes identically). The proof is illustrated on $\mathscr{I}^+$ in \eqref{eq: conservation} and an identical proof follows at $\mathscr{I}^-$. We thus arrive at the result 
\begin{equation} \label{eq: Delta_Q_vanishing} 
Q_{\xi}^{+}[C_2] - Q_{\xi}^{+}[C_1] =  Q_{\xi}^{-}[C_2] - Q_{\xi}^{-}[C_1] = 0 \,,
\end{equation} 
which tells us that for spacetimes which possess double ACKVs $\xi$, not only is $\left. H_{\xi} \right|_{C_2} - \left. H_{\xi} \right|_{C_1}  = 0$, but the contributions vanish independently at each end. 

\subsubsection{Interior contribution} \label{sec: interior}

\begin{figure}[H]
\begin{center}  
\includegraphics[width=0.5\linewidth]{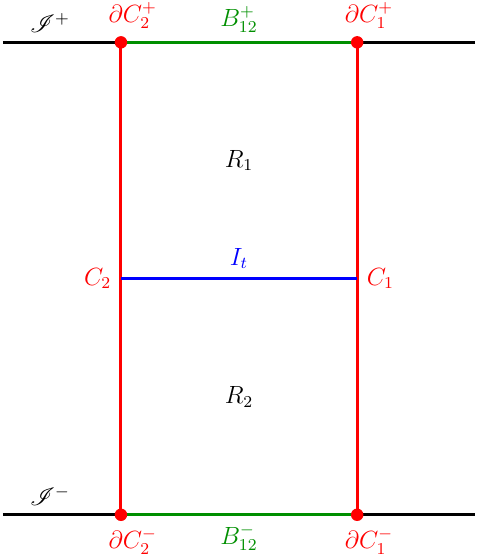}
\caption{Penrose diagram of an AldS spacetime where the timelike hypersurfaces $C$ stretch between $\mathscr{I}^+$ and $\mathscr{I}^-$, including an interior hypersurface $I_t$.} \label{fig: dS_both_boundaries_interior}
\end{center}
\end{figure}

We now consider a second type of past boundary, namely that of an interior spacelike hypersurface $I_t$, which we do not assume to be in either asymptotic region of the spacetime. We consider the scenario depicted in Figure \ref{fig: dS_both_boundaries_interior}
\noindent and we gauge fix the metric in the vicinity of the hypersurface $I_t$ to be 
\begin{equation}
    ds^2 = -dt^2 + \gamma^t_{ij} dx^i dx^j\,,
\end{equation}
over which we wish to perform similar integrals of the symplectic current to those considered in the previous subsection. We begin by considering the following integral 
\begin{align} \label{eq: dw_R2}
0 & = \int_{R_2} d \tilde{\bm{\omega}}(g, \delta g, \mathcal{L}_{\xi} g) = \int_{\partial R_2} \tilde{\bm{\omega}} (g, \delta g, \mathcal{L}_{\xi} g)\,,
\end{align}
where $\tilde{\bm{\omega}}$ is defined in \eqref{eq: omega_mod} and we again consider $\xi |_{\mathscr{I}^{-}}= \xi^{-}_{(0)}$ where $\xi^{-}_{(0)}$ is a conformal Killing vector of $g^{-}_{(0)}$. Performing Stokes' theorem again, we find 
\begin{equation}
- \int_{I_t} \tilde{\bm{\omega}} (g, \delta g, \mathcal{L}_{\xi} g) = \int_{C_2^\text{L}} \tilde{\bm{\omega}} (g, \delta g, \mathcal{L}_{\xi} g) - \int_{C_1^{\text{L}}} \tilde{\bm{\omega}} (g, \delta g, \mathcal{L}_{\xi} g) \,,
\end{equation}
where we have made use of $\tilde{\bm{\omega}}|_{\mathscr{I}^-}=0$ and $C_{1,2}^{\text{L}}$ denote the portions of $C_{1,2}$ to the past of $I_t$, bounding the region $R_2$ as shown in Figure \ref{fig: dS_both_boundaries_interior}. We note that this formula contains an interior contribution from the hypersurface $I_t$, which will be our main consideration in this section. As in the previous subsection, we use $\tilde{\bm{\omega}} = d\left[ \delta \mathbf{Q} - \bm{\omega}_{\text{ct}} - i_{\xi} \bm{\Theta}  \right]$ in order to rewrite the terms on the r.h.s. as 
\begin{align}
\begin{split} \label{eq: lower_box_midway}
-\int_{I_t} \tilde{\bm{\omega}} (g, \delta g, \mathcal{L}_{\xi} g) & = \int_{C_2^L \cap I_t} ( \delta \mathbf{Q}  - i_{\xi} \bm{\Theta} ) - \delta  \int_{C_2^L \cap \mathscr{I}^-} ( \mathbf{Q} - \bm{\Theta}_{\text{ct}} - i_{\xi} \mathbf{B} )  \\
& \phantom{aa} - \int_{C_1^L \cap I_t} ( \delta \mathbf{Q}  - i_{\xi} \bm{\Theta} ) + \delta  \int_{C_1^L \cap \mathscr{I}^-} ( \mathbf{Q}  - \bm{\Theta}_{\text{ct}} - i_{\xi} \mathbf{B} )  \\ 
& = \int_{C_2^L \cap I_t} ( \delta \mathbf{Q}  - i_{\xi} \bm{\Theta} ) - \int_{C_1^L \cap I_t} ( \delta \mathbf{Q}  - i_{\xi} \bm{\Theta} )\,,
\end{split}
\end{align}
where we have used the fact that the counterterms and boundary conditions give $\bm{\Theta}|_{\mathscr{I}^-} - d \bm{\Theta}_{\text{ct}}|_{\mathscr{I}^-} = \delta \mathbf{B}$, and the conservation of the charges under radial translation in $\mathscr{I}^-$.  
At this point we do not know if $\bm{\Theta}|_{I_t}   = \delta \mathbf{\tilde{B}}$, and thus we cannot even establish if Hamiltonians associated with $C_{1,2}^L$ even exist. Checking the existence criterion (\ref{eq: H_existence}) and using the formula \eqref{eq: theta_pullback_spacelike} for the pullback of the symplectic potential to a spacelike hypersurface, we find the Hamiltonians exist iff
\begin{equation} \label{eq: interior_existence}
\int_{C \cap I_t} i_{\xi} \tilde{\boldsymbol{\omega}} (g, \delta_1 g, \delta_2 g) = \int_{C \cap I_t} i_{\xi} \left[ \left(\delta_1 \pi^{ij} \wedge \delta_2 \gamma_{ij} \right)  d\mu \right]  =  0\,,
\end{equation}
which must be enforced by the behaviour of the bulk fields 
at $I_t$. We note that we can no longer use the asymptotic expansion of the fields as $I_t$ is assumed to be a generic hypersurface in the interior of the spacetime. In fact, we are not free to prescribe boundary conditions at $I_t$ as such conditions should be determined dynamically as a result of the boundary conditions at $\mathscr{I}^{\pm}$. 

As a concrete case we consider $I_t$ to lie in the future domain of dependence of $\mathscr{I}^-$. By causality, there can be no flux through $I_t$ and thus the Hamiltonians associated with $C_{1,2}^L$ must exist. This means that \cite{Iyer:1995kg}
\begin{equation} \label{eq: theta_pullback_bulk}
    \bm{\Theta}|_{I_t} = \delta \mathbf{\tilde{B}}\,,
\end{equation}
where we no longer need to consider the contribution of additional exact terms as in \eqref{eq: theta_pullback_well_posed} due to the fact that we are now working in the deep interior of spacetime and thus do not have to add the counterterms in order to deal with the boundary non-normalisable modes. Such a result means that the Hamiltonians take the form
\begin{equation} \label{eq: H_t_difference}
    H_{\xi} |_{C^L} = Q^t_{\xi}[C] - Q_{\xi}^-[C]\,,
\end{equation}
 where 
 \begin{equation} \label{eq: Interior_charge}
  Q^t_{\xi}[C] = \int_{C \cap I_t} \mathbf{Q}[\xi]  - i_{\xi}\mathbf{\tilde{B}}\,.
 \end{equation}
We will conclude this subsection by giving two examples of 
boundary conditions at $I_t$, which respect \eqref{eq: theta_pullback_bulk}.

\noindent \textbf{Dirichlet}: Our first example of such boundary conditions are Dirichlet boundary conditions, considered most notably in \cite{Brown:1992br}
\begin{equation} \label{eq: Dirichlet_BC}
    \delta \gamma_{ij} |_{I_t} = 0\,,
\end{equation}
for which equation \eqref{eq: theta_pullback_spacelike} becomes 
\begin{equation}
\bm{\Theta} = \frac{1}{8\pi G} \delta (K \sqrt{\gamma})  d\mu \implies \mathbf{\tilde{B}} = \frac{1}{8\pi G} K \sqrt{\gamma}  d\mu\,.
\end{equation}
In order to derive the charge, we note that the Dirichlet boundary conditions are preserved by a bulk vector field $\xi$ whose limit to $I_t$ satisfies 
\begin{equation} \label{eq: I_t_vector}
    \xi^{t} =0\,, \qquad \xi^i = \xi^i(x)\,, \qquad D_{(i} \xi_{j)} = 0\,,
\end{equation}
{\it i.e.} the limit to $I_t$ is a \textit{Killing vector} of $I_t$. Using \eqref{eq: Interior_charge} and the expression \eqref{eq: Q_GR} for the Noether charge 2-form, we can immediately compute the future contribution to $H_{\xi}|_{C^L}$
\begin{equation} \label{eq: Q_t_dirichlet}
   Q^t_\xi[C] =  \int_{C \cap I_t}  \mathbf{Q}[\xi] - i_{\xi} \mathbf{\tilde{B}} = - \int_{I_t \cap C} \pi^{k}_{l} \xi^l \varepsilon_{k i j} \, dx^i \wedge dx^j\,.
\end{equation}
We note that an alternative route to deriving this charge follows from computation of the pullback of $\bm{\omega}$ to $I_t$ (derivation in appendix \ref{app: pullback_omega})
\begin{equation} \label{eq: pullback_omega_2}
\bm{\omega}(g, \delta g, \mathcal{L}_{\xi} g) = \left[ 2 \delta \left(  D_i (\pi^{ij} \xi_j) \right) - D_i \left( \xi^i \pi^{jk} \delta \gamma_{jk} \right) \right] d\mu\,,
\end{equation}
and using the usual definition of the Hamiltonian \eqref{eq: Hamiltonian_def} we again reach \eqref{eq: Q_t_dirichlet}. 

Finally, we also note that the full Hamiltonian given in \eqref{eq: H_t_difference} is independently radially invariant at each end. The proof at $\mathscr{I}^{\pm}$ is given in \eqref{eq: conservation} and thus we will only need to check the radial invariance at $I_t$. Explicitly, we have
\begin{align}
\begin{split}
   \delta Q_{\xi}^t[C_2] - \delta Q_{\xi}^t[C_1] & = - \int_{I^{12}_{t}} \bm{\omega}(g, \delta g, \mathcal{L}_{\xi} g) = - \delta \int_{I^{12}_{t}} D_i (\pi^{ij} \xi_j) \, d\mu = 0\,,
\end{split}
\end{align}
where radial invariance is ensured by the momentum constraint \eqref{eq: Constraint_2} and the Killing equation \eqref{eq: I_t_vector}. 

\noindent \textbf{Neumann}: This is an alternative boundary condition which has been studied in the context of general relativity {\it e.g.} \cite{Compere:2008us, Krishnan:2016mcj}. In contrast to \eqref{eq: Dirichlet_BC}, this boundary condition is 
\begin{equation} \label{eq: Neumann}
    \delta \pi^{ij} |_{I_t} = 0\,,
\end{equation}
 for which equation \eqref{eq: theta_pullback_spacelike} becomes\footnote{Note that $\tilde{\mathbf{B}}=0$ is equivalent to the statement that the bare Eintein-Hilbert action is sufficient for a well-posed variational problem {\it i.e.} one does not have to add the Gibbons-Hawking or other boundary terms to the action. This phenomenon was observed in \cite{Krishnan:2016mcj} for general relativity with Neumann boundary conditions.} 
\begin{equation}
\bm{\Theta} = - \gamma_{ij} \delta \pi^{ij} = 0 \implies \mathbf{\tilde{B}} = 0\,,
\end{equation}
and thus the charge is
\begin{equation} \label{eq: Q_t_neumann}
   Q^t_\xi[C] =  \int_{C \cap I_t}  \mathbf{Q}[\xi]  = - \int_{I_t \cap C} K^{k}_{l} \xi^l \varepsilon_{k i j} \, dx^i \wedge dx^j\,,
\end{equation}    
which is also radially invariant using 
\begin{equation}
     \delta Q_{\xi}^t[C_2] - \delta Q_{\xi}^t[C_1]  = - \int_{I^{12}_{t}} \bm{\omega}(g, \delta g, \mathcal{L}_{\xi} g) = - \int_{I^{12}_t} \left(\delta \pi^{ij} \wedge \delta_{\xi} \gamma_{ij} \right)  d\mu = 0 \,, 
\end{equation}
via the Neumann boundary condition \eqref{eq: Neumann}.

We note that the imposition of Dirichlet/Neumann boundary conditions \eqref{eq: Dirichlet_BC}/\eqref{eq: Neumann} at the interior hypersurface are simply two examples of boundary conditions which define a Hamiltonian. More generally, the boundary conditions on the interior hypersurface which lead to existence of the Hamiltonian $H_{\xi}|_C^L$ will be any that satisfy equation \eqref{eq: interior_existence}. We note that when equation \eqref{eq: interior_existence} is satisfied, in analogy with \eqref{DeltaQ} we will always have
\begin{equation}
    \Delta Q_{\xi}^t = \Delta Q_{\xi}^- \,,
\end{equation}
where $\Delta Q^{t,-}_{\xi} = Q^{t,-}_{\xi}[C_2] -Q^{t,-}_{\xi}[C_1]$,
and thus the difference between neighbouring spacelike separated charges is a conserved quantity under time evolution from $\mathscr{I}^-$ to $I_t$. We will see in the next subsection how the violation of the existence property \eqref{eq: interior_existence} relates to the presence of gravitational flux through the hypersurface $I_t$.

\subsection{Past contribution to the fluxes} \label{sec: fluxes}

The two-ended cases discussed thus far have assumed existence of ACKVs, resulting in radial invariance of charges at $\mathscr{I}^{\pm}$. In this section we will relax this assumption, and derive the expressions for the modified Hamiltonians by allowing for fluxes at both components of the conformal boundary as well as an interior surface of the spacetime. The one-ended case of fluxes (usually taken as $\mathscr{I}^+$) has previously been studied in works such as \cite{Anninos:2010zf, Compere:2019bua, Compere:2020lrt, Fiorucci:2020xto, Kolanowski:2021hwo} as was outlined in Section \ref{sec: fluxes_one_ended}. In this section we extend such results to generic two-ended hypersurfaces. We first discuss the case of flux at both boundaries $\mathscr{I}^{\pm}$ and then we will discuss the case of flux through an interior hypersurface. 

\subsubsection{Contribution at \texorpdfstring{$\mathscr{I}^-$}{Scrimfluxes}} \label{sec: fluxes_scriminus}

\begin{figure}[H]
\begin{center}  
\includegraphics[width=0.5\linewidth]{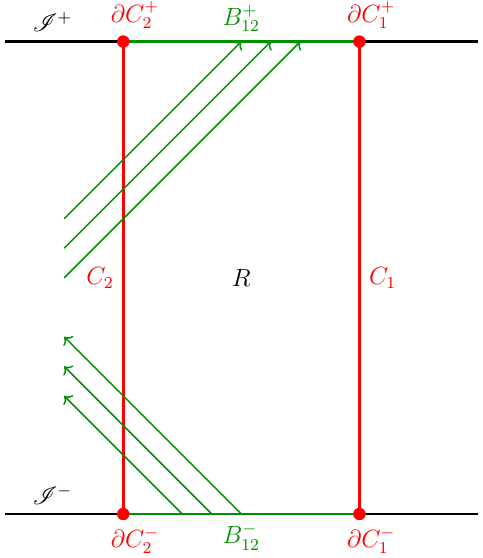}
\caption{Penrose diagram of an AldS spacetime with flux due to gravitational radiation passing through both boundaries $\mathscr{I}^{\pm}$.} \label{fig: dS_both_boundaries_fluxes}
\end{center}
\end{figure}

We begin by considering the setup for the passage of gravitational radiation as depicted in Figure \ref{fig: dS_both_boundaries_fluxes},
\noindent which we want to describe using the language of charges and fluxes. In order to do this, we utilise the Hamiltonian modification procedure described in \eqref{eq: mod_H} at both ends of $C$. From the one-ended modification procedure discussed in Section \ref{sec: fluxes_one_ended}, we see that the explicit charge expression \eqref{eq: Mod_H} is not altered by the modification procedure, but rather the flux formula \eqref{eq: flux_ds} is non-zero. Extending this argument to the two-ended case, we find that the modified Hamiltonian associated with a slice is now
\begin{align} \label{eq: H_mod_two_ended}
    \mathcal{H}_{\xi} |_C & = Q_{\xi}^+[C] - Q_{\xi}^-[C]\,,
\end{align}
with 
\begin{equation}
    Q_{\xi}^{\pm}[C] = \frac{1}{2} \int_{\partial C^{\pm}}T^{\pm k}_{\phantom{+}l} \xi_{\pm(0)}^l \epsilon^{(0)\pm}_{kij} \, dx^i \wedge dx^j =  \int_{\partial C^{\pm}} d\sigma_i^{\pm} \, T_{\pm}^{ij} \xi^{\pm}_{j(0)}\,,
\end{equation}
in agreement with equations \eqref{eq: H_two_ended} and \eqref{eq: Q_pm} as the charge expression does not change. As discussed, what changes are the flux formulae, with the difference between two modified Hamiltonians being given by 
\begin{equation}
    \mathcal{H}_{\xi} |_{C_2} - \mathcal{H}_{\xi} |_{C_1} =  Q_{\xi}^+[C_2] - Q_{\xi}^-[C_2] - Q_{\xi}^+[C_1] + Q_{\xi}^-[C_1]\,,
\end{equation}
where due to the non-zero flux at $\mathscr{I}^{\pm}$ we can no longer use $Q_{\xi}^{\pm}[C_2] - Q_{\xi}^{\pm}[C_1] =0$. Applying the flux formulae \eqref{eq: flux_ds} and recalling the definition of the flux form \eqref{eq: flux_form}
\begin{equation} \label{eq: fp_fluxes}
    Q_{\xi}^{\pm}[C_2] - Q_{\xi}^{\pm}[C_1] = - \int_{B^{\pm}_{12}} \mathbf{F}^{\pm}_{\xi_{(0)}^{\pm}} = \int_{B^{\pm}_{12}} \sqrt{g^{\pm}_{(0)}} T_{\pm}^{ij} \nabla^{\pm}_{(0)(i} \xi^{\pm}_{j)(0)} \, d\mu\, ,
\end{equation}
we find the overall difference between modified Hamiltonians is 
\begin{equation} \label{eq: mod_H_fluxes_two_ends}
     \mathcal{H}_{\xi} |_{C_2} - \mathcal{H}_{\xi} |_{C_1} = - \int_{B^{+}_{12}} \mathbf{F}^+_{\xi_{(0)}^+} + \int_{B^{-}_{12}} \mathbf{F}^-_{\xi_{(0)}^-}\,,
\end{equation}
an equation which captures the effects of gravitational radiation at both $\mathscr{I}^{\pm}$ of an AldS$_4$ spacetime.

A particularly interesting application of the flux formula \eqref{eq: mod_H_fluxes_two_ends} is the case of a closed system where the net flux must be zero. An example of such an AldS$_4$ spacetime would be one where both $\mathscr{I}^{\pm}$ are Cauchy slices. Due to the zero net flux condition we have 
\begin{equation}
    \int_{B^{+}_{12}} \mathbf{F}^+_{\xi_{(0)}^+} = \int_{B^{-}_{12}} \mathbf{F}^-_{\xi_{(0)}^-} \iff Q^+_{\xi}[C_2] -  Q^+_{\xi}[C_1] = Q^-_{\xi}[C_2] -  Q^-_{\xi}[C_1]\,,
\end{equation}
which is an equation which tells us that even though the quantities defined at cross-sections of $\mathscr{I}^{\pm}$ will change due to the presence of the radiation, the difference between neighbouring charges remains constant. In the absence of any radiation this result reduces to \eqref{eq: Delta_Q_vanishing}, a stronger condition which also requires the charges on each Cauchy slice to be radially invariant.
\subsubsection{Interior contribution}

\begin{figure}[H]
\begin{center}  
\includegraphics[width=0.5\linewidth]{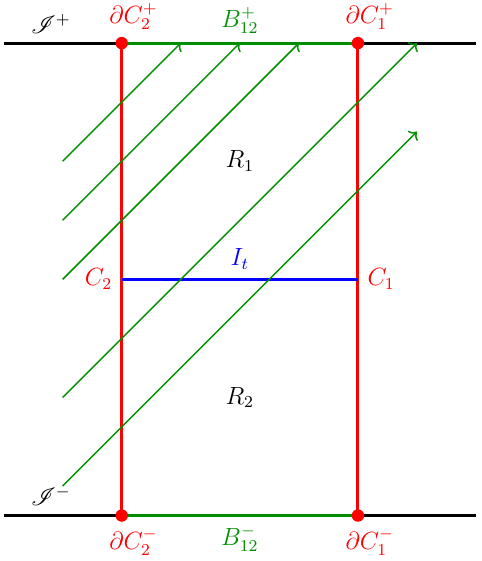}
\caption{Penrose diagram of an AldS spacetime with flux due to gravitational radiation passing through the conformal boundary component $\mathscr{I}^{+}$ and an interior hypersurface $I_t$.} \label{fig: two_ends_interior_flux}
\end{center} 
\end{figure}

The second important two-ended case occurs when one of the hypersurfaces is in the interior of the spacetime. We will consider the setup depicted in Figure \ref{fig: two_ends_interior_flux} 
\noindent which together with the analysis from Section \ref{sec: fluxes_scriminus} will be sufficient to consider all cases ({\it e.g.} $\mathscr{I}^-$ to an interior hypersurface or two interiors). In order to derive the full expression, we first focus on the case of the interior surface alone as this is the novel part of this section. We recall that on the interior surface Hamilton's equations read 
\begin{equation}
    \delta H^t_{\xi} = \int_{C \cap I_t} \delta \mathbf{Q} - i_{\xi} \mathbf{\Theta}\,,
\end{equation}
where due to the flux through $I_t$ we can no longer use equation \eqref{eq: theta_pullback_bulk}. We can modify the Hamiltonian on the interior slice by following the prescription \eqref{eq: mod_H}, finding
\begin{equation}
    \delta \mathcal{H}^t_{\xi} = \int_{C \cap I_t} \delta \mathbf{Q} - i_{\xi} ( \mathbf{\Theta} - \bm{\theta})\,,
\end{equation}
where we need to select $\bm{\theta}$ using the criterion in \eqref{eq: modification_criterion}. Using \eqref{eq: theta_pullback_spacelike}, the pullback of the symplectic current to $I_t$ is given by 
\begin{equation}
    \bar{\bm{\omega}}(\psi, \delta_1 \psi, \delta_2 \psi) = \left( \delta_1 \pi^{ij} \wedge \delta_2 \gamma_{ij} \right)  d\mu\,,
\end{equation}
which allows us to identify that the modification term must be
\begin{equation} \label{eq: Theta_w=0}
\bm{\theta} (\psi, \delta \psi) =  \left( \pi^{ij} \delta \gamma_{ij} + \delta W(\psi) \right) d\mu\,,
\end{equation}
where $W$ is an arbitrary local functional of the dynamical field $\gamma_{ij}$ on $I_t$ and represents the ambiguity in this prescription.

Unlike the case of $\mathscr{I}^{\pm}$, where it was argued in \eqref{eq: flux_ambiguity} that one must set the ambiguity to zero, for an interior surface there is no way to fix the ambiguity. As an example of a choice of $W$, we take 
\begin{equation}
    W = 0\,, 
\end{equation}
which makes the interior contribution to the overall Hamiltonian
\begin{equation}
    \mathcal{H}^t_{\xi} = \int_{C \cap I_t} \left(  \mathbf{Q} - \frac{1}{8\pi G} K \sqrt{\gamma} \, i_{\xi} d\mu \right) = - \int_{I_t \cap C } \pi^{k}_{l} \xi^l \varepsilon_{k i j} \, dx^i \wedge dx^j\,,
\end{equation}
a result in agreement with the charge \eqref{eq: Q_t_dirichlet} corresponding to the choice of Dirichlet boundary conditions,  precisely the spacelike analogue of the Brown-York quasilocal charge \cite{Brown:1992br}, a natural choice for the contribution to the charge arising at the interior corner. However, due to the flux we no longer have invariance of this quantity, with the flux formula now reading 
\begin{equation} \label{eq: t_flux}
    \left. \mathcal{H}^t_{\xi} \right|_{C_2 \cap I_t} - \left. \mathcal{H}^t_{\xi} \right|_{C_1 \cap I_t} = -\int_{I_t} \mathbf{F}^t_{\xi} = - \int_{I_t} \pi^{ij} \delta_{\xi} \gamma_{ij} d\mu = - \int_{I_t} \pi^{ij} L_{\xi} \gamma_{ij} \, d\mu\,, 
\end{equation}
where we used the fact that $\xi^t = 0$ in the third equality. This gives a possible modification to the interior contribution in order to define a charge. 

For the example above, the full modified Hamiltonian corresponding to the situation depicted in Figure \ref{fig: two_ends_interior_flux} is 
\begin{equation}
   \left. \mathcal{H}_{\xi} \right|_{C^U} = Q_{\xi}^+[C] - Q_{\xi}^t[C]\,, 
\end{equation}
where $Q^+_{\xi}[C]$ is given in \eqref{eq: Q_pm} and $Q^t_{\xi}[C]$ in \eqref{eq: Q_t_dirichlet}. Note that interior contribution to the modified Hamiltonian now comes with the opposite sign to that of the case of no radiation considered in \eqref{eq: H_t_difference} due to the fact that $I_t$ lies to the past of $B_{12}^{+}$. The corresponding flux formula for this setup reads
\begin{equation}
      \left. \mathcal{H}_{\xi} \right|_{C^U_2} -  \left. \mathcal{H}_{\xi} \right|_{C^U_1} = - \int_{B^{+}_{12}} \mathbf{F}^+_{\xi_{(0)}^+} + \int_{I_t} \mathbf{F}^t_{\xi}\,,
\end{equation}
where the fluxes on the right hand side are defined in equations \eqref{eq: fp_fluxes} and \eqref{eq: t_flux} respectively. 

\subsubsection{Remarks and future directions}

We note that one of the unsatisfying aspects of the interior flux formula is the lack of a unique choice for the modification ambiguity $W$. We previously considered the case of $W=0$ but if we leave this arbitrary term in the formulae then we obtain 
\begin{equation}
     \mathcal{H}^t_{\xi} \rightarrow  \mathcal{H}^t_{\xi} + \frac{1}{2} \int_{C \cap I_t} W \xi^i \varepsilon_{ijk} dx^j \wedge dx^k\,,
\end{equation}
for the modified Hamiltonian, and 
\begin{equation}
     \left. \mathcal{H}^t_{\xi} \right|_{C_2 \cap I_t} - \left. \mathcal{H}^t_{\xi} \right|_{C_1 \cap I_t} \rightarrow   \left. \mathcal{H}^t_{\xi} \right|_{C_2 \cap I_t} - \left. \mathcal{H}^t_{\xi} \right|_{C_1 \cap I_t} + \int_{I_t} L_{\xi} W d\mu\,, 
\end{equation}
for the flux. This result somewhat parallels the discussion of Section \ref{sec: interior}, where the lack of uniquely defined boundary conditions at the interior hypersurface meant that one could not uniquely identify the contribution to the Hamiltonian arising from $C \cap I_t$. This ambiguity arose from the general existence criterion \eqref{eq: theta_pullback_bulk} and can also be described by the ambuguity $W$, even though in that case there is no flux!

In order to demonstrate this, we show that a choice of $W \neq 0$ can transform between the Dirichlet and Neumann boundary conditions as discussed in Section \ref{sec: interior}. Starting with the generic expression for the pullback of the symplectic potential as given in \eqref{eq: theta_pullback_spacelike}, we have 
\begin{equation}
    \bm{\Theta} - \bm{\theta} = \delta  \left( \frac{K \sqrt{\gamma}}{8\pi G}   - W \right) d\mu\,,
\end{equation}
which is a total variation and thus will always lead to a well defined modified Hamiltonian. We note choosing $W=0$ recovers the Dirichlet charge \eqref{eq: Q_t_dirichlet} and $W = (K \sqrt{\gamma})/(8\pi G)$ the Neumann charge \eqref{eq: Q_t_neumann}, both of which are radially invariant when the respective boundary conditions are imposed at $I_t$. We see that the ambiguity is still undetermined even in the absence of flux and it is thus not possible for us to fix the ambiguity when flux is present. This issue is part of the wider question of attempting to define a quasi-local mass in general relativity \cite{Wang:2008jy}, as there is no well-defined notion of local energy density (due to the equivalence principle). 

We conclude this section with a comment on a possible way in which one could begin to identify $W$. This approach would be to impose some additional conditions on the spacetime which restrict the passage of gravitational radiation. A particularly interesting condition is that of the `no-incoming radiation' condition of \cite{Ashtekar:2019khv}, which mimics physical systems such as compact binaries by imposing that no radiation passes through the null surface $\mathscr{I}_{\text{rel}}$, the future event horizon of past timelike infinity $i^-$\footnote{$i^{\pm}$ are the endpoints of the worldline of the radiation source}. If such a condition is imposed in our formalism, then the past event horizon of future timelike infinity $i^+$ will have the same flux as $\mathscr{I}^+$ by causality, and thus one can partially fix $W$ by demanding that the flux through that null hypersurface is equivalent to the well-defined flux through $\mathscr{I}^+$. Realising this construction via our formalism will require an extension to describe null hypersurfaces in the bulk, as the current technology merely describes spacelike slices $I_t$.

\section{Examples of spacetimes with radially invariant gravitational charges} \label{sec: Examples}

\subsection{dS\texorpdfstring{$_4$}{4}}

A warm-up in studying the charges of the more complicated solutions is to study pure $dS_4$. As we shall see, this solutions does not possess non-trivial conserved charges, but will serve as an illustration of the techniques that one needs to employ in order to compute the charges of other AldS$_4$ spacetimes. 

We start by presenting the dS$_4$ solution in global coordiantes 
\begin{equation} \label{eq: dS_global}
    ds^2 = \frac{\ell^2}{\cos^2 \eta} (-d\eta^2 + d\Omega_3^2)\,,
\end{equation}
where $\eta \in (-\pi/2, \pi/2)$ is the conformal time coordinate and $d\Omega_3^2$ is the metric of the unit round $S^3$. In these coordinates, $\mathscr{I}^{\pm}$ are located at $\eta = \pm \pi/2$. 

In order to extract the conserved charges of this solution, we perform the transformation from global coordinates to Starobinsky gauge at both $\mathscr{I}^{\pm}$. In order to do this one merely needs to solve 
\begin{equation}
    \frac{d\eta^2}{\cos^2\eta} = \frac{d\rho_{\pm}^2}{\rho_{\pm}^2} \implies  \frac{d\eta}{\cos \eta} = \mp \frac{d\rho_{\pm}}{\rho_{\pm}}\,, 
\end{equation}
where the signs in the equation above are chosen to account for the ranges of the coordinates, recall $\rho_+ < 0$ and $\rho_- >0$. Solving this equation gives
\begin{equation} \label{eq: rho_pm_ds}
    \rho_{\pm} = \mp 2 \frac{\cos\left( \frac{\eta}{2} \right) \mp \sin \left( \frac{\eta}{2} \right)}{\cos\left( \frac{\eta}{2} \right) \pm \sin \left( \frac{\eta}{2} \right)}\,,   
\end{equation}
where the factor of 2 is a choice of integration constant. Using this map we find that the metric \eqref{eq: dS_global} takes the form
\begin{equation}
    ds^2 = \frac{\ell^2}{\rho_{\pm}^2} \left( - d\rho_{\pm}^2 + \frac{1}{\cos^2 (\eta(\rho_{\pm}))} d\Omega^2_3 \right)\,,
\end{equation}
which upon expanding $\cos^2\eta$ near the components of the conformal boundary gives near $\mathscr{I}^{\pm}$
\begin{equation}
    \left. ds^2 \right|_{\mathscr{I}^{\pm}} = \frac{\ell^2}{\rho_{\pm}^2} \left[ -d\rho_{\pm}^2 + \left(1 + \rho_{\pm}^2 + \mathcal{O}(\rho_{\pm}^4) \right) d\Omega_3^2 \right]\,,
\end{equation}
where we note that in particular $g^{(3)}_{\pm} =0$ and thus all charges vanish at both $\mathscr{I}^{\pm}$. An important note about the Starobinsky coordinates used to cover the neighbourhood of each boundary is that they are directly related via an inverse relation. Using \eqref{eq: rho_pm_ds} we find immediately 
\begin{equation} \label{eq: rho_pm_relationship}
    \rho_+ = -\frac{4}{\rho_-}\,,
\end{equation}
and thus we see that $\mathscr{I}^+$ is located at $\rho_- = \infty$ and similarly $\mathscr{I}^-$ is at $\rho_+ = -\infty$. From the perspective of analytic continuation, either of these can be obtained from the continuation of a single copy of $AdS_4$ using the rules in equation \eqref{eq: Triple_Wick_explicit} although we note that due to the relationship \eqref{eq: rho_pm_relationship} above, this analytic continuation is valid for only one of the boundaries, one cannot analytically continue $AdS_4$ to obtain both boundaries simultaneously (as expected, since the global structure of $dS_4$ is different from that of $AdS_4$).

\subsection{Schwarzchild-dS\texorpdfstring{$_4$}{4}}

In this subsection we will compute all of the charges of the Schwarzchild-dS$_4$ (SdS) solution, whose metric is given by \begin{equation} \label{eq: SdS_radial}
ds^2 =- h(r) dt^2 + \frac{1}{h(r)} dr^2 + r^2 d\Omega_2^2\,, \qquad h(r) = 1 - \frac{\Lambda}{3} r^2 - \frac{2m}{r}\,.
\end{equation}
In order for $h(r)$ to possess two positive roots, we require
\begin{equation} \label{eq: Nariai_bound}
    m < \frac{\ell}{3\sqrt{3}}\,,
\end{equation}
{\it i.e.} the mass parameter is below the Nariai bound. We denote the two roots as $h(r_h) = h(r_c) = 0$ where $r_h < r_c$. The smaller root corresponds to the black hole horizon and the larger the cosmological horizon. As we are interested in the conformal boundary of the solution, we first restrict consideration to $r \in (r_c, \infty)$, as $\mathscr{I}^+$ is located at $r = \infty$.

In order to compute the charges of this solution, we first map this metric into the Starobinsky gauge \cite{Starobinsky:1982mr} up to the required order. To do this, we follow a similar route to the computation of \cite{Poole:2018koa} which maps SAdS into Fefferman-Graham coordinates \cite{Fefferman:1985zza}, namely we first perform the map inverting the radial coordinate
\begin{equation}
r = - \frac{\ell^2}{z}\,,
\end{equation} 
then expand $z$ in the following power series in $\rho_+$
\begin{equation}
z = \rho_+ - \frac{\rho_+^3}{4 \ell^2} - \frac{m \rho_+^4}{3\ell^2} + \mathcal{O}(\rho^5)\,,
\end{equation}
which is sufficient to read off all of the useful information in the Starobinsky-SdS expansion. As a final step, we work with dimensionless coordinates 
\begin{equation}
\tilde{\rho}_+ = \frac{\rho_+}{\ell}\,, \qquad \tilde{t} = \frac{t}{\ell}\,,
\end{equation}
but we will drop the tildes from here on. The steps above produce a Starobinsky expansion of the following form 
\begin{equation}
ds^2 = \frac{\ell^2}{\rho_+^2} \left( - d\rho_+^2 + g_{(0)}^+ + \rho_+^2 g_{(2)}^+ + \rho_+^3 g_{(3)}^+ + \ldots \right)\,,
\end{equation}
where
\begin{align}
g_{(0)}^+ & = dt^2 + d\Omega_2^2 \label{eq: g_0_SdS}\,, \\
g_{(2)}^+ & = - \frac{1}{2} (dt^2 -d\Omega_2^2)\,, \\
g_{(3)}^+ & = -\frac{2m}{3 \ell} (2dt^2 - d\Omega_2^2)\,.
\end{align}
This expansion allows us to construct the holographic energy-momentum tensor 
\begin{equation}
T^+ = - \frac{3 \ell}{16 \pi G} g^{(3)} = \frac{2m}{16\pi G} (2dt^2 - d\Omega_2^2)\,,
\end{equation}
and the charges associated to this solution are thus 
\begin{equation} \label{eq: SdS_charges}
Q^+_{\xi} = \oint_{\partial C} T^+_{ij} n^i \xi^j \sqrt{g_{\Omega_2}} \, d^2x = \frac{1}{16 \pi G} \int_{0}^{\pi} \int_{0}^{2\pi} 4m \xi^t(t, \theta, \phi) \sin \theta \, d\phi d\theta\,,
\end{equation}
where we used $n^i = (\frac{\rho_+^2}{\ell^2} + \mathcal{O}(\rho_+^4)) (\partial_t)^i$, $\sqrt{g_{\Omega_2}} = \frac{\ell^2}{\rho_+^2} \sin \theta + \mathcal{O}(1)$ and all higher order terms vanish in the limit as $\rho_+ \rightarrow 0$ (one can also work directly on the $\rho_+ = 0$ hypersurface in the unphysical spacetime). The key point to note in the above formula is that all of the charges for the SdS solution only depend upon the $t$-component of the vector field $\xi$, and thus the computation boils down to computing $\xi^t$ for the conformal Killing vectors of (\ref{eq: g_0_SdS}). 

In order to compute these vectors, we use the fact that the SdS metric is asymptotically dS, {\it i.e.} it has a conformally flat $g_{(0)}$. To see this, we start with the metric in (\ref{eq: g_0_SdS}) and perform the following coordinate transformation 
\begin{equation} \label{eq: coord_trans_flat}
x = \frac{\sinh t}{\cos \theta + \cosh t}\,, \qquad y = \frac{\cos \phi \sin \theta}{\cos \theta + \cosh t}\,, \qquad z = \frac{\sin \phi \sin \theta}{\cos \theta + \cosh t}\,,
\end{equation}
which brings the line element into the form 
\begin{equation}
g^{(0)} = \Omega^2 (dx^2 + dy^2 + dz^2)\,, \qquad \Omega = \cosh t + \cos \theta\,,
\end{equation}
demonstrating that this metric is conformally equivalent to the standard flat metric on $\mathbb{R}^3$. Recall that conformally equivalent metrics share the same conformal Killing vectors\footnote{To see this, note that if $\xi$ is a CKV of a metric $g$, 
${\cal L}_\xi g = 2 \phi g$, then it is also a CKV of the metric $g'= \Omega g$, ${\cal L}_\xi g' = 2 \psi g'$, with $\psi=\phi+ {\cal L}_\xi \log \Omega$.}. Thus the CKV of $g^{(0)}$ is that of $\mathbb{R}^3$.
The set of conformal Killing vectors on $\mathbb{R}^3$ are well-known, see {\it e.g.} \cite{Ginsparg:1988ui, DiFrancesco:1997nk} and separate into the usual translations 
\begin{equation} 
\xi^i_T = a^i\,,
\end{equation}
rotations 
\begin{equation}
\xi^i_R = r^i_{\phantom{i} j} x^j\,, \qquad (r_{i j} = -r_{ji})\,,
\end{equation}
dilatation
\begin{equation} 
\xi^i_D = \lambda x^{i}\,,
\end{equation}
and special conformal transformations (SCTs)
\begin{equation}
\xi^{i}_{SCT} = b^i x_j x^j - 2 x^i b_j x^j\,,
\end{equation}
where $a^i, r_{ij}, \lambda, b^i$ are all spacetime constants. Counting degrees of freedom we have $3+3+1+3 =10$ CKVs for which we would like to compute the corresponding charges. The first step is to compute the $t$-components of all of the CKVs under the coordinate transformation (\ref{eq: coord_trans_flat}), for which the following derivatives are useful 
\begin{equation}
\frac{\partial t}{\partial x} = 1 + \cos \theta \cosh t\,, \qquad \frac{\partial t}{\partial y} = - \cos \phi \sin \theta \sinh t\,, \qquad \frac{\partial t}{\partial z} = -\sin \phi \sin \theta \sinh t\,,
\end{equation}
allowing us to compute 
\begin{align}
&(\partial_x)^t  = 1+ \cos \theta \cosh t\,, \label{eq: non_zero_Q1} \\
&(\partial_y)^t  = -\cos \phi \sin \theta \sinh t\, \\
&(\partial_z)^t  = -\sin \phi \sin \theta \sinh t\,, \\
&(x \partial_y - y \partial_x)^t  = -\cos \phi \sin \theta \cosh t\,, \\
&(x \partial_z - z \partial_x)^t  = -\sin \phi \sin \theta \cosh t\,, \\
&(y \partial_z - z \partial_y)^t  = 0\,, \\
&(x\partial_x + y\partial_y + z\partial_z)^t = (\cos \theta \sinh t)\,, \\
&([-x^2 +y^2+z^2] \partial_x -2xy \partial_y -2xz \partial_z)^t  = 1- \cos \theta \cosh t\,,  \label{eq: non_zero_Q2} \\
&([-y^2 +x^2+z^2] \partial_x -2xy \partial_x -2yz \partial_z)^t  = - \cos \phi \sin \theta \sinh t\,, \\
&([-x^2 +x^2+y^2] \partial_x -2xz \partial_x -2yz \partial_y)^t  = - \sin \phi \sin \theta \sinh t\,.
\end{align}
As (\ref{eq: SdS_charges}) involves an integral over $S^2$, the only CKV that can lead to a non-zero charge are the ones which contain parts that are independent of the angles and these are the vectors given in (\ref{eq: non_zero_Q1}) and (\ref{eq: non_zero_Q2}). The charges corresponding to these two vectors are 
\begin{align}
&Q_{\partial_x}  =  \frac{1}{16 \pi G} \int_{0}^{\pi} \int_{0}^{2\pi} 4m(1+\cos \theta \cosh t) \sin \theta \, d\phi d\theta = \frac{m}{G}\,, \\
&Q_{(-x^2 +y^2+z^2) \partial_x -2xy \partial_y -2xz \partial_z} =  \frac{m}{G}\,,
\end{align}
{\it i.e.} the $x$-translation and the $x$-SCT give non-vanishing and equivalent charges. This is reminiscent of that encountered in \cite{Poole:2018koa} where the ``mass aspect" for an AdS black brane solution was found to be a linear combination of the time translation and SCT aspects. Indeed, the same result holds in the dS case, which we can observe simply by noting
\begin{equation}
\frac{1}{2} \left[ (\partial_x)^t + ([-x^2 +y^2+z^2] \partial_x -2xy \partial_y -2xz \partial_z)^t  \right]= 1\,,
\end{equation}
which reproduces precisely the Schwarzchild mass aspect in the integrand of (\ref{eq: SdS_charges}).

\subsubsection{SdS charges at \texorpdfstring{$\mathscr{I}^-$}{A}} \label{sec: SdS_scri_minus}
The previous section analyses the charges of the SdS solution near $\mathscr{I}^+$. We note that one can use the Starobinsky expansion from the original coordinates \eqref{eq: SdS_radial} in order to cover $\mathscr{I}^-$ in much the same way one does $\mathscr{I}^+$, although now the direction of the ``time" along the boundary $\mathscr{I}^-$ is globally opposite to that at $\mathscr{I}^+$. In order to demonstrate this carefully, we will now discuss the ``Kruskalisation" \cite{Bazanski, Bicak:1995vc} of the SdS solution in order to obtain coordinates regular across both the past and future cosmological horizons $\mathscr{H}^-_{\mathscr{C}}$ and $\mathscr{H}^+_{\mathscr{C}}$ respectively.  

First we transform to coordinates which are smooth across the past cosmological horizon $\mathscr{H}^-_{\mathscr{C}}$. In order to find such coordinates, one first defines the tortoise coordinate in the usual manner 
\begin{equation} \label{eq: SdS_tortoise}
    r^* = \int \frac{dr}{h(r)}\,,  
\end{equation}
and then introduces the \textit{advanced} time coordinate via
\begin{equation}
    v = t + r_*\,,
\end{equation}
bringing the metric into the form 
\begin{equation} \label{eq: SdS_ingoing_EF}
    ds^2 = -h(r) dv^2 + 2 dv dr + r^2 d \Omega_2^2\,,  
\end{equation}
where $\mathscr{I}^{-}$ is now given by $r = \infty$ and the coordinate $v$ is finite along $\mathscr{I}^{-}$. 

In order to understand more intricately how the two-ended charge computation will work for SdS, we will transform the metric \eqref{eq: SdS_ingoing_EF} into a form which is regular across both the past and future cosmological horizons $\mathscr{H}^{\pm}_\mathscr{C}$, following the original paper \cite{Bazanski} which performs this calculation in detail. We begin by noting \cite{Bazanski, Bicak:1995vc} that 
\begin{equation}
    h(r) = \frac{(r-r_h)(r_c -r)(r+r_h+r_c)}{r(r_h^2 + r_h r_c +r_c^2)}\,,
\end{equation}
where $h(r_h)=h(r_c)=h(-r_h-r_c) = 0$, and $r_h < r_c$, with $r_h$ the black hole horizon and $r_c$ the cosmological horizon. We also note that the explicit form of the tortoise coordinate \eqref{eq: SdS_tortoise} is given by \cite{Bicak:1995vc}
\begin{equation} \label{eq: SdS_tortoise_explicit}
    r^* = \delta_h \log \frac{|r-r_h|}{r+r_h + r_c} - \delta_c \log \frac{|r_c-r|}{r+r_h + r_c} + \delta_h \left[ \log \left( \frac{r_c}{r_h}  \right) - \frac{1}{2} \right]\,,
\end{equation}
where 
\begin{equation}
    \delta_h = \frac{r_h}{1-\Lambda r_h^2} > 0\,, \qquad \delta_c = - \frac{r_c}{1-\Lambda r_c^2} > 0\,.
\end{equation}
The positivity  of $\delta_h$ and $\delta_c$ is a direct result of $\left.(r h)' \right|_{r_h} > 0$ and $\left.(r h)' \right|_{r_c} < 0$, respectively.
The constant term in \eqref{eq: SdS_tortoise_explicit} matches that of \cite{Bicak:1995vc}, where it was chosen in order to recover a sensible flat limit ($\Lambda \rightarrow 0$) from the SdS solution. We recall the definition of the \textit{retarded} time coordinate 
\begin{equation}
    u = t - r^*\,,
\end{equation}
which can be used to analyse $\mathscr{I}^+$ as $u$ is finite along $\mathscr{I}^+$. In order to cover both boundaries simultaneously we introduce the coordinates 
\begin{equation}
    U = \left( \frac{r_c}{r_h} \right)^{\frac{\delta_h}{2\delta_c}} \exp\left( \frac{u-\delta_h}{2\delta_c} \right)\,, \quad  V = - \left( \frac{r_c}{r_h} \right)^{\frac{\delta_h}{2\delta_c}} \exp\left(- \frac{v+\delta_h}{2\delta_c} \right)\,,
\end{equation}
in which the metric \eqref{eq: SdS_ingoing_EF} becomes 
\begin{equation}
    ds^2 = - \frac{4\Lambda \delta_c^2}{3r} (r-r_h)^{1+ \frac{\delta_h}{\delta_c}} (r+ r_h + r_c)^{2- \frac{\delta_h}{\delta_c}} dU dV + r^2 d\Omega_2^2\,,
\end{equation}
and the location of $\mathscr{I}^{\pm}$ is now 
\begin{equation}
    U V = 1\,,
\end{equation}
with $\mathscr{I}^+$ being $U > 0, V > 0$ and $\mathscr{I}^-$ being $U < 0, V < 0$. In order to compare these coordinates with those used in equation (37) of \cite{Bazanski}, we note that 
\begin{align}
\tilde{u} + \tilde{v} = V\,, \qquad \tilde{u} - \tilde{v} = -U\,.
\end{align}
These coordinates cover both $\mathscr{I}^{\pm}$ so they should be suitable in analysing a 2-ended hypersurface stretching between the surfaces. 

\begin{figure}[H]
\begin{center}  
\includegraphics[width=0.75\linewidth]{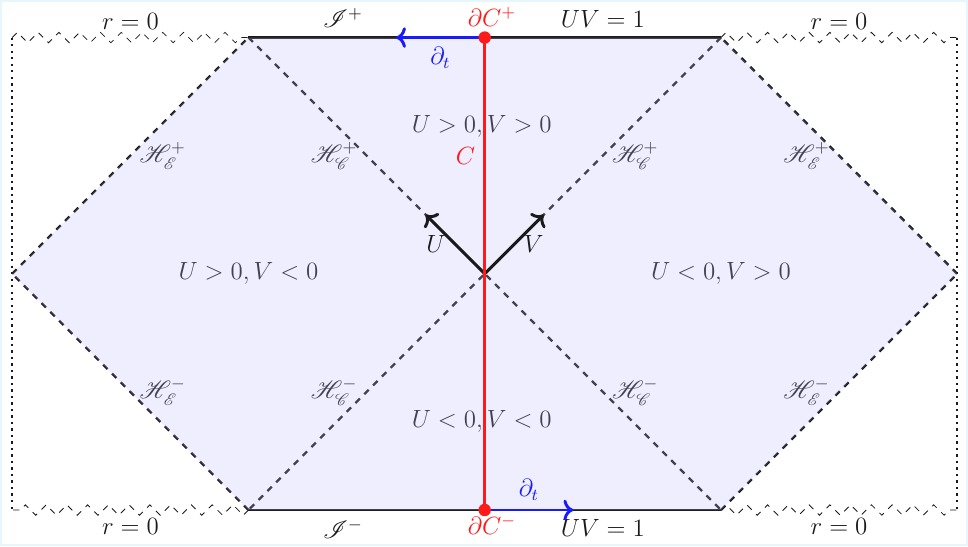}
\caption{Penrose diagram for the SdS solution with the shaded region being that covered by the Kruskal-type coordinates $(U,V)$. The red timelike hypersurface C is used to compute our two-ended Hamiltonian. We indicate the direction of increasing $t$ at $\mathscr{I}^{\pm}$ with blue arrows and the signs of the coordinates $(U,V)$ which change when one crosses the cosmological horizon. We have also inlcuded the location of the black hole singularity at $r=0$, which lies outside our region of interest. The maximally extended SdS Penrose diagram contains an infinite sequence of singularities and spacelike infinities \cite{Gibbons:1977mu}, of which we show one copy above. An alternative completion of the spacetime is given by identifying the vertical dotted lines on the left and right to wrap once around the Einstein static universe \cite{Bazanski}.} \label{fig: SdS_cosmological}
\end{center} 
\end{figure}

We will now consider the correct vector field to use in order to compute the charges. We know that the correct vector field at $\mathscr{I}^+$ which we should use to compute the charges is $\partial_t$, so all we need to do is identify $\partial_t$ in coordinates $(U,V)$ at $\mathscr{I}^+$ and then we will have a ``global"\footnote{Due to the presence of event and cosmological horizons, the SdS solution does not admit global coordinates, see \cite{Bazanski}. Here by ``global'' we mean coordinates which extend over the shaded region of Figure \ref{fig: SdS_cosmological}.} vector field with which to compute the charges. We use the map (39) in \cite{Bazanski} and obtain 
\begin{equation} \label{eq: t_to_UV}
    t = - 2 \delta_c \text{arctanh} \left( \frac{\tilde{v}}{\tilde{u}} \right) = - 2 \delta_c  \text{arctanh} \left( \frac{V+U}{V-U} \right)\,,
\end{equation}
and so 
\begin{equation} \label{eq: Killing_global}
    \partial_t = \frac{\partial V}{\partial t} \partial_V + \frac{\partial U}{\partial t} \partial_U = \frac{1}{2\delta_c} \left( U \partial_U  - V \partial_V \right)\,,
\end{equation}
where we note that $\partial_t$ is also the correct vector field to compute the charges at $\mathscr{I}^-$. As explained in \cite{Bazanski}, the necessary map between $\mathscr{I}^+$ to $\mathscr{I}^-$ is $(\tilde{u}, \tilde{v}) \rightarrow -(\tilde{u}, \tilde{v})$
under which \eqref{eq: t_to_UV} remains invariant. 

 We can thus compute the contribution to the two-ended ``mass" charge coming from the past end of the timelike slice. All we need is to perform the Starobinsky expansion in a neighbourhood of $\mathscr{I}^-$ and evaluate the charge associated with $\xi = \partial_t$. Starting with \eqref{eq: SdS_radial} and setting $\ell =1$, we begin with the map 
\begin{equation}
r = \frac{1}{z_-}\,,
\end{equation}
then perform the expansion in small $\rho_-$ of the form
\begin{equation}
    z_- = \rho_- - \frac{1}{4} \rho_-^3 + \frac{m}{3} \rho_-^4 + \mathcal{O}(\rho_-^5)\,,
\end{equation}
which gives a Starobinsky expansion of the form
\begin{equation}
    ds^2 = - \frac{d \rho_-^2}{\rho_-^2} + \frac{1}{\rho_-^2} \left( g^-_{(0)} + \rho_-^2 g^-_{(2)} + \rho_-^3 g_{(3)}^- + \ldots \right)\,,
\end{equation}
with 
\begin{align}
g_-^{(0)} & = dt^2 + d\Omega_2^2 \label{eq: g_0^-_SdS}\,, \\
g_-^{(2)} & = - \frac{1}{2} (dt^2 -d\Omega_2^2)\,, \\
g_-^{(3)} & = \frac{2m}{3} (2dt^2 - d\Omega_2^2)\,.
\end{align}
Using this expansion, we can obtain the energy-momentum tensor at $\mathscr{I}^-$ via \eqref{eq: EM_past} \begin{equation}
    \langle T \rangle^{\mathscr{I}^-} = - \frac{3 }{16 \pi G} g^-_{(3)} = -\frac{m}{8\pi G} (2dt^2 - d\Omega_2^2)\,,
\end{equation}
and thus via \eqref{eq: Q_pm} we have 
\begin{equation}
     Q^{-}_{\xi}[C] = \frac{1}{2} \int_{\partial C^{-}}T^{- k}_{\phantom{+}l} \xi_{-(0)}^l \epsilon^{(0)-}_{kij} dx^i \wedge dx^j = \frac{1}{2} \int_{\partial C^-} T^{- t}_{\phantom{-} t} \epsilon^{(0)-}_{t ij} dx^i \wedge dx^j = -\frac{m}{G}\,,
\end{equation}
which contributes to the total two-ended Hamiltonian via \eqref{eq: H_two_ended} {\it i.e.}
\begin{equation} \label{eq: SdS_mass_two_ended}
     H_{\xi} |_C  = Q_{\xi}^+ - Q_{\xi}^- = 2 \frac{m}{G}\,,
\end{equation}
{\it i.e.} including the contribution from the past end, one finds a doubling of the charge associated with the Killing vector given in equation \eqref{eq: Killing_global}. 

In order to understand the physical picture of this result, it is helpful to note the direction of increasing $t$ at $\mathscr{I}^{\pm}$ (marked with the blue arrows in Figure \ref{fig: SdS_cosmological}) where we note that the directions are ``opposite'' from the perspective of the Penrose diagram. One can see these directions by direct inspection of \eqref{eq: Killing_global}, noting that $\partial_t$ is tangential to $\mathscr{I}^{\pm}$. In order to see this latter result, note that a normal vector to the $UV=1$ surface is given by 
\begin{equation}
    n_{\mu} = \partial_{\mu} (U V) = V \delta^{U}_{\mu} + U \delta^V_{\mu}\,,
\end{equation}
and thus 
\begin{equation}
    (\partial_t)^{\mu} n_{\mu} = \frac{1}{2\delta_c} (UV-VU) = 0\,,
\end{equation}
justifying the blue arrows shown in Figure \ref{fig: SdS_cosmological}. This helps with the interpretation of \eqref{eq: SdS_mass_two_ended} in being a ``doubling'' of the one-ended Hamiltonian, where now the evolution is taken in opposite directions at each end.

\subsection{Kerr-dS\texorpdfstring{$_4$}{4}} 

Another asymptotically dS$_4$ solution of natural interest is the Kerr-dS$_4$ (KdS) solution. The metric for the KdS solution can be written in Boyer-Lindquist-type coordinates as \cite{Carter:1968ks, Plebanski:1976gy, Gibbons:1977mu, griffiths2012exact}
\begin{equation}
\begin{split} \label{eq: Kerr_dS_BL_coords}
    ds^2 =  - \frac{\Delta_r}{\Xi^2 \varrho^2} & \left( dt - a \sin^2 \theta \, d\phi^2 \right)^2 + \frac{\varrho^2}{\Delta_r} dr^2 + \frac{\varrho^2}{\Delta_\theta}d\theta^2\\
    & \phantom{\qquad \quad}+ \frac{\Delta_{\theta} \sin^2 \theta}{\Xi^2 \varrho^2} \left(a \, dt -(r^2 + a^2) \, d\phi \right)^2\,, 
    \end{split}
\end{equation}
where 
\begin{align}
    \varrho & = r^2 + a^2 \cos^2 \theta\,, \\
    \Delta_r & = (r^2 + a^2)\left(1 - \frac{\Lambda}{3} r^2 \right) - 2 m r = (r^2 + a^2)\left(1 - \frac{r^2}{\ell^2} \right) - 2 m r\,, \\
    \Delta_{\theta} & = 1 + \frac{\Lambda}{3}a^2 \cos^2 \theta =  1 + \frac{a^2}{\ell^2} \cos^2 \theta\,, \\
    \Xi & = 1 + \frac{\Lambda}{3} a^2= 1 + \frac{a^2}{\ell^2}\,.
\end{align}
The spacetime has horizons located at the roots 
\begin{equation}
    \Delta_r = 0\,, 
\end{equation}
a quartic polynomial which has (up to) four real roots. We will be interested in the case when four real roots are admitted, with the heirarchy of roots being 
\begin{equation}
    r_{c_1} < 0 < r_{h_1} < r_{h_{2}} < r_{c_2}\,,
\end{equation}
where $r_{c_{1,2}}$ denote the outer and inner cosmological horizons and $r_{h_{1,2}}$ the outer and inner event horizons. For more information on the conformal diagrams of such a spacetime see \cite{Gibbons:1977mu, griffiths2012exact}. 

\begin{figure}[H]
\begin{center}  
\includegraphics[width=0.75\linewidth]{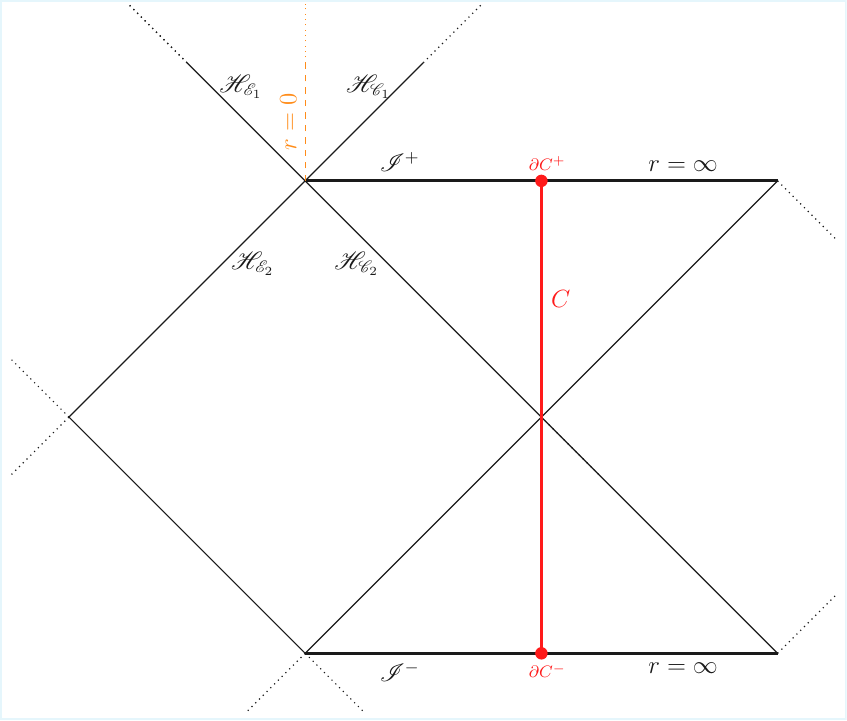}
\caption{Penrose diagram depicting the region of interest in the KdS solution in order to compute the charges at $\mathscr{I}^\pm$. The dotted lines indicate regions of the spacetime which will not be important for our calculation. See {\it e.g.} \cite{Gibbons:1977mu, griffiths2012exact, Akcay:2010vt} for the full analytic extension of the KdS spacetime.} \label{fig: KdS}
\end{center} 
\end{figure}

We want to compute the charges of such spacetimes using the formalism developed in Section \ref{sec: Charges}. In particular, we will be interested in the scenario depicted in Figure \ref{fig: KdS}
\noindent which will be sufficient to compute the charges of the KdS spacetime at $\mathscr{I}^\pm$. We will first focus on the one-ended case at $\mathscr{I}^+$, before computing the contribution at $\mathscr{I}^-$ using the methods applied to the SdS solution in Section \ref{sec: SdS_scri_minus}. We will see that the contribution from the past end results in a doubling of the one-ended Hamiltonian as in \eqref{eq: SdS_mass_two_ended}.

\paragraph{Charges of Kerr-dS:} In order to extract the charges of KdS, we first want to transform coordinates from Boyer-Lindquist coordinates \eqref{eq: Kerr_dS_BL_coords} into Starobinsky gauge \eqref{eq: Starobinsky_gauge} and read off the values of both the representative of the conformal class at $\mathscr{I}^+$, $g^{(0)}_{ij}$, as well as the energy-momentum tensor, $g^{(3)}_{ij}$. The specific steps of the derivation follow closely to the Kerr-AdS (KAdS) case considered in Section 6.1 of \cite{Papadimitriou:2005ii}, and as such the first step is to perform a simple rescaling of the time coordinate 
\begin{equation} \label{eq: t_rescaling}
    \tilde{t} = \frac{t}{\Xi}\,,
\end{equation}
and then writing \eqref{eq: Kerr_dS_BL_coords} and dropping the tilde, we obtain
\begin{equation} \label{eq: KNdS_rescaled}
ds^2 = \varrho^2 \left[ \frac{dr^2}{\Delta_r}+ \frac{d \theta^2}{\Delta_{\theta}} \right] + \frac{\Delta_{\theta} \sin^2 \theta}{\varrho^2} \left[ a dt - (r^2 +a^2) \frac{d \phi}{\Xi}  \right]^2- \frac{\Delta_r}{\varrho^2} \left[ dt - a \sin^2 \theta \frac{d \phi}{\Xi}  \right]^2\,.
\end{equation} 
In order to make sure that the metric we consider is just the analytically continued version of the KAdS metric of \cite{Papadimitriou:2005ii} under $\ell^2 \rightarrow -\ell^2$. We first expand the metric \eqref{eq: KNdS_rescaled} for large $r$ and find 
\begin{align}
\begin{split}
        ds^2 & = \frac{r^2}{\ell^2}\left[ 1 - \left( 1 - \frac{a^2}{\ell^2} \sin^2 \theta \right) \frac{\ell^2}{r^2} + \frac{2 m \ell^2}{r^3} +\mathcal{O}\left(\frac{1}{r^4} \right) \right] \left( dt - \frac{a \sin^2 \theta}{\Xi} d\phi \right)^2 \\
         & \quad - \frac{\ell^2}{r^2} \left[ 1 + \left( 1- \frac{a^2}{\ell^2} \sin^2 \theta \right) \frac{\ell^2}{r^2} - \frac{2 m \ell^2}{r^3} + \mathcal{O} \left( \frac{1}{r^4} \right) \right] dr^2 \\
         & \quad + \frac{r^2}{\Delta_{\theta}} \left( 1 + \frac{a^2}{r^2} \cos^2 \theta \right) d\theta^2 \\
         & \quad + \frac{r^2 \Delta_{\theta} \sin^2 \theta}{\Xi^2} \left[ d\phi^2 + \frac{a^2}{r^2}\left( (1+ \sin^2 \theta) d\phi^2  - \frac{2 \Xi}{a} d\phi dt \right) + \mathcal{O}\left( \frac{1}{r^4} \right) \right]\,.
\end{split}
\end{align}
This expression is clearly not in the Starobinsky form and will require some extra work to be put in that gauge due to the $\theta$-dependence of the component $g_{rr}$. We introduce new coordinates  
\begin{align}
\begin{split}
r & = \bar{r} - \frac{1}{\bar{r}} f(\bar{\theta}) + \mathcal{O}(\bar{r}^{-3})\,, \\
\theta & = \bar{\theta} - \frac{1}{\bar{r}^4} h(\bar{\theta}) +  \mathcal{O}(\bar{r}^{-6})\,,
\end{split}
\end{align}
and forcing the $d\bar{r} d\bar{\theta}$ cross-terms to vanish as well as the $\theta$-independence of $g_{rr}$ fixes 
\begin{align}
\begin{split}
f(\bar{\theta}) & = -\frac{a^2}{4} \cos^2 \bar{\theta} + f_0\,, \\
g(\bar{\theta}) & = -\frac{1}{8}\ell^2 a^2 \Delta_{\bar{\theta}} \sin \bar{\theta} \cos \bar{\theta}\,,
\end{split} 
\end{align}
where we follow \cite{Papadimitriou:2005ii} in setting the constant of integration $f_0 =0$.
These transformations result in a large $\bar{r}$ expansion of the form 
\begin{align}
\begin{split}
ds^2 = & -  \frac{\ell^2}{\bar{r}^2}  \left[1+ \left(1-\frac{a^2}{\ell^2} \right) \frac{\ell^2}{\bar{r}^2} - \frac{2m \ell^2}{\bar{r}^3} + \mathcal{O}\left( \frac{1}{\bar{r}^4} \right) \right] d\bar{r}^2 \\
& + \frac{\bar{r}^2}{\Delta_{\bar{\theta}}} \left[ 1+ \frac{3a^2}{2\bar{r}^2} \cos^2 \bar{\theta} + \mathcal{O}\left( \frac{1}{\bar{r}^4} \right) \right] d\bar{\theta}^2 \\
& + \frac{\bar{r}^2}{\ell^2} \left[1- \left( 1 - \frac{a^2}{\ell^2} + \frac{a^2}{2\ell^2} \cos^2 \bar{\theta} \right) \frac{\ell^2}{\bar{r}^2} + \frac{2 m \ell^2}{\bar{r}^3} +  \mathcal{O}\left( \frac{1}{\bar{r}^4} \right) \right] \left(dt - \frac{a \sin^2 \bar{\theta}}{\Xi} d\phi  \right)^2 \\
& + \frac{\bar{r}^2 \Delta_{\bar{\theta}} \sin^2 \bar{\theta}}{\Xi^2} \left[ d\phi^2 + \frac{a^2}{\bar{r}^2} \left( \left(2-\frac{1}{2} \cos^2 \bar{\theta} \right) d\phi^2 - \frac{2 \Xi}{a} d\phi dt \right) + \mathcal{O}\left( \frac{1}{\bar{r}^4} \right) \right]\,,
\end{split}
\end{align}
which is almost in the desired form of \eqref{eq: Starobinsky_gauge}, but needs a little work in order to fix the ``time-time" (here $d\bar{r}^2$ as $\mathscr{I}^+$ lies outside the cosmological horizon) component of the metric. The components of the induced metric are 
\begin{align}
\begin{split} \label{eq: induced_metric_kds}
\gamma_{\bar{\theta} \bar{\theta}} & = \frac{\bar{r}^2}{\Delta_{\bar{\theta}}}  \left[ 1+ \frac{3 a^2}{2 \bar{r}^2} \cos^2 \bar{\theta} +  \mathcal{O}\left( \frac{1}{\bar{r}^4} \right)  \right]\,, \\
\gamma_{t t} & = \frac{\bar{r}^2}{\ell^2} \left[ 1 - \left(1-\frac{a^2}{\ell^2} + \frac{a^2}{2\ell^2} \cos^2 \bar{\theta} \right) \frac{\ell^2}{\bar{r}^2} + \frac{2m\ell^2}{\bar{r}^3} +  \mathcal{O}\left( \frac{1}{\bar{r}^4} \right)  \right]\,, \\
\gamma_{t \phi} & = - \frac{\bar{r}^2 a \sin^2 \bar{\theta}}{\ell^2 \Xi} \left[1+ \left( 1+ \frac{1}{2} \cos^2 \bar{\theta} \right) \frac{a^2}{\bar{r}^2} + \frac{2m \ell^2}{\bar{r}^3} +  \mathcal{O}\left( \frac{1}{\bar{r}^4} \right) \right] \,, \\
\gamma_{\phi \phi}& = \frac{\bar{r}^2 \sin^2 \bar{\theta}}{\Xi} \left[ 1 + \left( 1+ \frac{1}{2} \cos^2 \bar{\theta} \right) \frac{a^2}{\bar{r}^2} + \frac{2ma^2 \sin^2 \bar{\theta}}{\bar{r}^3 \Xi} + \mathcal{O}\left( \frac{1}{\bar{r}^4} \right) \right]\,,
\end{split}
\end{align}
which we will use to read off the Starobinsky data after transforming the $d\bar{r}^2$ component. In order to do this, we first define the canonical time coordinate $\tau$ via 
\begin{equation} \label{eq: tau_to_r_kds}
d\tau = \ell \left[ 1+\frac{1}{2} \left(1- \frac{a^2}{\ell^2} \right) \frac{\ell^2}{\bar{r}^2} - \frac{m \ell^2}{\bar{r}^3} +  \mathcal{O}\left( \frac{1}{\bar{r}^4} \right) \right] \frac{d\bar{r}}{\bar{r}}\,,
\end{equation}
which brings the metric into the form of \eqref{eq: synchonous_gauge}
and finally the (dimensionful) Starobinsky time coordinate\footnote{This coordinate is defined such that $\rho_+$ increases as one moves towards $\mathscr{I}^+ = \{\rho_+ = 0\}$. This means that $\rho_+<0$.} as 
\begin{equation}
\rho_+ = - \ell \exp \left( - \frac{\tau}{\ell} \right)\,, 
\end{equation}
for which we can combine the above two transformations in order to write 
\begin{equation} \label{eq: r_bar_rho}
    \bar{r} = -\frac{\ell^2}{\rho_+} - \frac{\rho_+}{4\ell^2} (\ell^2 -a^2)-\frac{m \rho_+ ^2}{3 \ell^2}+\mathcal{O}(\rho_+ ^3)\,.
\end{equation}
These transformations bring the metric into the desired form of 
\begin{equation}
ds^2 = \frac{\ell^2}{\rho_+^2} \left[ d\rho_+^2 + \left(g^{(0)}_{ij} + \rho_+^2 g^{(2)}_{ij} + \rho_+^3 g^{(3)}_{ij} + \mathcal{O}(\rho_+^4) \right) dx^i dx^j \right]\,,
\end{equation}
where all components $g_{ij}^{(1,2,3)}$ can be read off directly by applying the mapping \eqref{eq: r_bar_rho} to the induced metric components given in \eqref{eq: induced_metric_kds}. In particular, we find $g^{(0)}_{ij}, g^{(3)}_{ij}$ (written in line-element form) as
\begin{equation} \label{eq: g_0_KNdS} 
ds_{(0)}^2 = dt^2 - \frac{2 a \sin^2 \bar{\theta}}{\Xi} dt d\phi + \frac{\ell^2}{\Delta_{\bar{\theta}}} d\bar{\theta}^2 +  \frac{\ell^2 \sin^2 \bar{\theta}}{\Xi} d\phi^2\,,
\end{equation} 
for the representative of the conformal class at $\mathscr{I}^+$, and 
\begin{equation}
ds_{(3)}^2 = -\frac{4m}{3\ell^4} dt^2 + \frac{8ma \sin^2 \bar{\theta}}{3\ell^4 \Xi} dt d\phi + \frac{2m}{3\ell^2 \Delta_{\bar{\theta}}}d\bar{\theta}^2 + \frac{ 2 m \sin^2 \bar{\theta} }{3\ell^2 \Xi^2} \left( \Xi - 3 \frac{a^2 \sin^2 \bar{\theta}}{\ell^2} \right) d\phi^2\,,
\end{equation}
for the energy-momentum tensor. 
We note that one can directly compare our analysis with the KAdS case of \cite{Papadimitriou:2005ii} via the simple analytic continuation $\ell^2 \rightarrow -\ell^2$ as discussed in Section \ref{sec: Triple_Wick}. Our convention for the energy-momentum tensor is 
\begin{equation} \label{eq: Tij_dimensionful}
 T_{ij}  = - \frac{3 \ell^2}{2\kappa^2} g_{ij}^{(3)}\,,
\end{equation}
which has picked up an additional factor of $\ell$ relative to \eqref{eq: EM_future} due to the dimensionful coordinate $\rho_+$. The non-zero components of this tensor are
\begin{align}
\begin{split}
T_t^t = \frac{2 m}{\ell^2 \kappa^2}\,, \qquad T_{\bar{\theta}}^{\bar{\theta}} = T_{\phi}^{\phi} = - \frac{m}{\ell^2 \kappa^2}\,, \qquad T_{\phi}^t = -\frac{3am \sin^2 \bar{\theta} }{\ell^2 \Xi \kappa^2}\,,
\end{split}
\end{align}
which is a convenient form in order to compute the charges. Recall that charges corresponding to an asymptotic conformal Killing vector, $\xi$, are given in equation \eqref{eq: Final_Hamiltonian} and thus for the KdS solution we have
\begin{equation}
H_{\xi} =  \int_{0}^{2\pi} \int_{0}^{\pi} \sqrt{g_{(0)} }\, T_i^t \xi_{(0)}^i \,  d\bar{\theta} \, d \phi\,,
\end{equation}
which we want to evaluate for charges corresponding to the mass and angular momentum. 

\paragraph{Angular momentum:} Using the Killing vector $\xi = - \partial_{\phi}$ we obtain 
\begin{align}
J_+ = H_{-\partial_{\phi}} & =  2\pi \int_{0}^{\pi} \frac{3a m \sin^3 \bar{\theta}}{\Xi^2 \kappa^2} \, d\bar{\theta} =  \frac{8 \pi a m}{\Xi^2 \kappa^2}\,.
\end{align} 

\paragraph{Mass:} This charge requires a slight redefinition of coordinates in order to identify the correct Killing vector, as one can immediately notice from equation (\ref{eq: g_0_KNdS}) that the representative of the conformal class is not the standard metric on $\mathbb{R} \times S^2$. In order to show that the standard metric is in fact in the conformal class, we consider the following coordinate transformation from $(t, \phi, \bar{\theta})$ to $(t', \phi', \bar{\theta}')$: 
\begin{equation}
t' = t\,, \qquad \phi' = \phi - \frac{a}{\ell^2} t \,, \qquad \Xi \tan^2 \bar{\theta}' = \tan^2 \bar{\theta} \,,
\end{equation} 
which brings the metric (\ref{eq: g_0_KNdS}) into the form
\begin{equation}
ds_{(0)}^2 = \frac{\cos^2 \bar{\theta}}{\cos^2 \bar{\theta}'} \left( dt'^2 + \ell^2 d\Omega_2'^2\right)\,,
\end{equation}
which is clearly conformal to the standard metric on $\mathbb{R} \times S^2$. We will thus use the Killing vector $\partial_{t'}$ in order to define the mass, which can be computed in terms of the old coordinates using the chain rule:
\begin{equation} \label{eq: t'_kds}
\partial_{t'} = \frac{\partial t}{\partial t'} \partial_t +  \frac{\partial \phi}{\partial t'} \partial_{\phi}  = \partial_t + \frac{a}{\ell^2} \partial_{\phi} \,.
\end{equation}
Now applying this to the charge formula we find the following expression for the mass 
\begin{align}
\begin{split}
M_+ = H_{\partial_t + \frac{a}{\ell^2} \partial_{\phi}}  = -2 \pi \int_{0}^{\pi} \left( - \frac{2m \sin \bar{\theta} }{\kappa^2 \Xi } + \frac{3a^2 m \sin^3 \bar{\theta}}{\ell^2 \Xi^2 \kappa^2} \right) d\bar{\theta}  = \frac{8 m \pi}{\kappa^2 \Xi^2}\,.
\end{split}
\end{align}

Finally we note that this pair is sufficient to compute the charges for the SdS ($a=0$) and dS ($a=m=0$) solutions, namely 
\begin{equation}
(M_{\text{SdS}}, J_{\text{SdS}}) = \left( \frac{8 m \pi}{\kappa^2} , 0 \right)\,, \qquad (M_{\text{dS}}, J_{\text{dS}}) = (0,0) \,,
\end{equation}
in agreement with \eqref{eq: SdS_mass_two_ended} as $\kappa^2 = 8\pi G$. 

\subsubsection{KdS charges at \texorpdfstring{$\mathscr{I}^-$}{A}} \label{sec: KdS_scri_minus} We have entirely focused on charges at $\mathscr{I}^+$ in the preceding analysis. As in the SdS case disucssed in Section \ref{sec: SdS_scri_minus}, the contributions to the total Hamiltonians arising from $\mathscr{I}^-$ are straightforward to calculate. The computation of these charges can be performed using the ``Kruskalisation" of the KdS solution in order to develop a coordinate chart which covers both $\mathscr{I}^{\pm}$ (as we did for SdS), see {\it e.g.} \cite{Hawking:1973uf, Akcay:2010vt, griffiths2012exact, Chrusciel:2012gz, Borthwick:2018qsb} for more information on such a procedure. Alternatively, one can directly evaluate the Starobinsky expansion at $\mathscr{I}^-$, which is what we shall do. 

The steps in performing the Starobinksy expansion at $\mathscr{I}^-$ are similar to those of $\mathscr{I}^+$ as discussed in the previous section, however we now replace the large $\bar{r}$ expansion \eqref{eq: r_bar_rho} with
\begin{equation}
    \bar{r} = \frac{\ell^2}{\rho_-} + \frac{\rho_-}{4\ell^2}(\ell^2 -a^2) - \frac{m \rho_-^2}{3\ell^2}+ \mathcal{O}(\rho_-^3)\,,
\end{equation}
and thus 
\begin{equation}
    g_{(3)ij}^- = -g_{(3)ij}^+\,,
\end{equation}
which upon using \eqref{eq: EM_past} gives us 
\begin{equation}
    \langle T_{ij} \rangle^{\mathscr{I}^-} = - \frac{3 \ell^2}{16 \pi G} g^-_{(3)ij} = - \langle T_{ij} \rangle^{\mathscr{I}^+} \,,
\end{equation}
and so for the angular momentum charge at $\mathscr{I}^-$ we find 
\begin{align}
\begin{split}
    J_- & =  \frac{1}{2} \int_{\partial C^-} T^{-k}_{\phantom{--}l} (\partial_{\phi})^l \epsilon^{(0)-}_{kij} dx^i \wedge dx^j \\
    & = - \frac{1}{2} \int_{\partial C^-} T^{k}_{\phantom{-}l} (\partial_{\phi})^l \epsilon^{(0)}_{kij} dx^i \wedge dx^j =  \frac{8 \pi a m}{\Xi^2 \kappa^2}\,,
    \end{split}
\end{align}
and thus the total angular momentum for the hypersurface stretching between $\mathscr{I}^+$ and $\mathscr{I}^-$ is 
\begin{equation}
    J = J_+ + J_- = 16 \pi \frac{a m}{(\Xi \kappa)^2} = 2\frac{am}{\Xi^2 G}\,.
\end{equation}
By an identical argument for the mass charge we have
\begin{equation}
    M = M_- + M_+ = 2 M_+ = 16 \pi \frac{ m }{(\kappa \Xi)^2} = 2 \frac{m}{\Xi^2 G}\,, 
\end{equation}
where we note that by an analogous argument to the Kruskalisation of SdS, we need to use the same vector field \eqref{eq: t'_kds} in order to compute the contribution to the mass at $\mathscr{I}^-$, noting that the directions of the vector field at $\mathscr{I}^{\pm}$ will be analogous to those displayed in Figure \ref{fig: SdS_cosmological}. 

\section{Robinson-Trautman-dS\texorpdfstring{$_4$}{RT4}} \label{sec: RTdS}

\subsection{Overview}

The final exact solution we will examine is that of the Robinson-Trautman de Sitter (RTdS) solution. The original solution \cite{Robinson:1960zzb} satisfies Einstein's equations with no cosmological constant. The solution represents gravitational radiation from isolated compact objects and has many remarkable properties, see  \cite{stephani_kramer_maccallum_hoenselaers_herlt_2003, carmeli1972group, griffiths2012exact} for textbook discussions.  

This solution is also of interest in the presence of a cosmological constant of either sign, see \cite{Chrusciel:1992tj} for $\Lambda = 0$, \cite{Bicak:1995vc} for $\Lambda > 0$ and \cite{Bicak:1997ne} for $\Lambda < 0$. In recent years, the case of $\Lambda < 0$ has become the topic of much interest due to its role in the AdS/CFT correspondence and non-equilibrium dynamics \cite{Bakas:2014kfa, Bakas:2015hdc, Ciambelli:2017wou, Skenderis:2017dnh, Ciambelli:2024kre, Arenas-Henriquez:2025rpt}, as well as describing non-trivial gravitational radiation in AdS,\footnote{The Robinson-Trautman equation acts as a ``Bondi mass loss" equation and is insensitive to the sign of $\Lambda$. For more information, see Section 3.3 of \cite{Bakas:2014kfa}.} an important first step in understanding fully the issue of gravitational waves in $\Lambda \neq 0$ spacetimes. 

As we expect the RT solution to provide an interesting example of non-trivial gravitational radiation in AldS spacetimes, it is natural to apply the techniques of Section \ref{sec: Charges} to see if flux formulae associated with the passage of gravitational waves arise. Before doing this, we begin with a brief overview of the spacetime, starting with the line element for the RTdS metric 
\begin{equation} \label{eq: RT_metric} 
ds^2 = - F(r,u,z,\bar{z}) du^2 -2 du dr +2 r^2 e^{\Phi(z, \bar{z};u)} dz d\bar{z}\,.
\end{equation}
The coordinate $u$ is a null retarded time coordinate $u \in (u_0, \infty)$ where $u_0> -\infty$ is an initial null hypersurface, $r$ is an affine radial distance and $(z, \bar{z})$ are K\"ahler coordinates for the distorted 2-dimensional spheres which arise as surfaces of constant $u$ and $r$. The front factor $F$ is given by
\begin{equation}
F = r \partial_u \Phi - \Delta \Phi - \frac{2m}{r} - \frac{\Lambda}{3} r^2\,,
\end{equation}
and 
\begin{equation}
    \Delta = e^{-\Phi} \partial_{z} \partial_{\bar{z}} = \frac{1}{2} \nabla^2\,,
\end{equation}
denotes the Laplacian of the distorted 2-spheres. In order for the metric to satisfy the field equations \eqref{eq: Einstein}, $\Phi$ is required to satisfy the \textit{Robinson-Trautman equation}
\begin{equation} \label{eq: RT_equation}
3 m \partial_u \Phi + \Delta \Delta \Phi = 0\,,
\end{equation}
an equation which has a close association to the Calabi flow equation on $S^2$ \cite{Calabi+1982+259+290, Calabi1985}, as was noted in \cite{Tod_1989}. In order to see this, one may consider the 2-metric
\begin{equation} \label{eq: 2-metric}
    ds_2^2 = 2 e^{\Phi} dz d\bar{z} \,, 
\end{equation}
which we note corresponds to the $r=1, u = \text{constant}$ surface in \eqref{eq: RT_metric}. The Calabi flow equation on $S^2$ takes the form 
\begin{equation} \label{eq: Calabi_flow}
    \partial_{u} g_{z \bar{z}} = \partial_{z} \partial_{\bar{z}} R_2\,,
\end{equation}
where $R_2$ is the Ricci curvature scalar of the metric \eqref{eq: 2-metric}. One may evaluate this equation explicitly as 
\begin{equation}
    \partial_u \Phi + 2 \Delta \Delta \Phi = 0\,.
\end{equation}
This is the Robinson-Trautman equation with $m = 1/6$, which may be achieved by a choice of units. The Calabi flow is a well-studied subject in mathematics, and one can use several results concerning the flow to derive properties of Robinson-Trautman metrics. We will soon see how such properties will allow us to construct interesting examples of non-trivial gravitational charges for this spacetime.

We note that the RTdS solution reduces to the SdS solution \eqref{eq: SdS_radial} when one chooses 
\begin{equation}
\Phi= \Phi_0 \equiv -2 \log \left(1+ \frac{z \bar{z}}{2}\right)\,,
\end{equation}
and makes use of the map 
\begin{equation} \label{eq: complex_to_spherical}
    z = \sqrt{2} \cot\left( \frac{\theta}{2} \right) e^{i \phi} \, , \quad \bar{z} = \sqrt{2} \cot\left( \frac{\theta}{2} \right) e^{-i \phi} \, ,
\end{equation}
between the K\"ahler and spherical coordinates. We also note that in considering this limit we want the SdS solution to possess two distinct horizons $(r_h, r_{c})$. This means that we will again impose the inequality \eqref{eq: Nariai_bound} for the RTdS solutions. 

The relationship between the RTdS and SdS is deeper than just a particular choice of $\Phi$ in RTdS reducing to SdS. It has been shown in \cite{Chrusciel:1992cj} (see also \cite{Bicak:1995vc}) that the RTdS solution approaches the SdS metric exponentially fast as $u \rightarrow \infty$ (equivalently the Calabi flow reaches a fixed point of the unit round $S^2$). This extension will generically not be smooth, but can be made ``arbitrarily smooth" by considering the extremal value of $\ell \rightarrow 3\sqrt{3} m$ in the Nariai bound \eqref{eq: Nariai_bound} \cite{Bicak:1995vc}. The smoothness of this extension will not be important for our purposes as we shall always work in the  RTdS spacetime and will not consider the explicit extension beyond $u = \infty$. The fact that the solution settles down to SdS at late times will however be of importance.

\begin{figure}[H]
\begin{center}
\includegraphics[height=8cm]{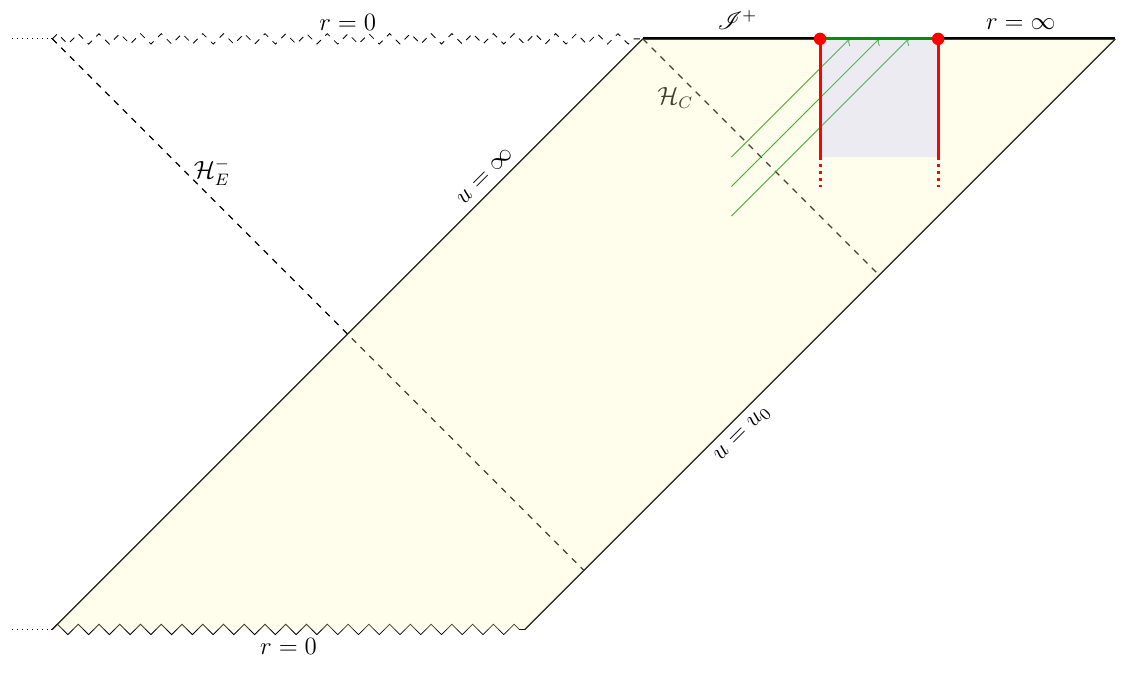}   
\end{center}
\caption{The Penrose diagram for the Robinson-Trautman de Sitter spacetime is highlighted in yellow. The spacetime starts at some initial retarded time $u = u_0$ and evolves to the SdS solution at $u = \infty$. We have included the completion of the black hole interior of SdS solution past the $u=\infty$ hypersurface, although this is not strictly part of the RTdS spacetime (one can extend through this surface, but not smoothly for generic $\Lambda$ \cite{Bicak:1995vc}). We will consider the flux through a region of $\mathscr{I}^+$, via calculation of the gravitational charges associated with the red timelike hypersurfaces.}
\label{fig: RT_dS_Penrose_diagram} 
\end{figure}

These solutions are also an interesting class as they are explicitly AldS by virtue of the conformal class induced at the conformal boundary not being conformally flat \cite{Bakas:2014kfa}. In what follows we will examine these spacetimes at $\mathscr{I}^+$. Our formulae at $\mathscr{I}^-$ will not be applicable for this spacetime as it does not possess a $\mathscr{I}^-$. The fact that the metric does not have a $\mathscr{I}^-$ (and in fact has a naked singularity at $r=0$) is due to the fact that the metric is only supposed to accurately describe the gravitational field of a radiating body \textit{outside} the source and does not include the (possibly very complicated) interior description, which should be included past $u=u_0$. This information is depicted via the Penrose diagram for the RTdS solution in Figure \ref{fig: RT_dS_Penrose_diagram}. As we shall shortly see more explicitly, these spacetimes generically do not possess conformal Killing vectors at $\mathscr{I}^+$ \cite{deFreitas:2014lia} and thus we will use the modification procedure discussed in Section \ref{sec: Modified_H} in order to construct modified Hamiltonians. 

\subsection{Gravitational charges at future infinity}

The aim is now to use this metric in order to compute the Hamiltonians \eqref{eq: Final_Hamiltonian} which occur at $\mathscr{I}^+$. In particular, we wish to evaluate
\begin{equation} \label{eq: RT_general_charge}
H_{\xi}  = \frac{1}{2} \int_{\partial C} T^{k}_{l} \xi_{(0)}^l \epsilon^{(0)}_{kij} dx^i \wedge dx^j\,,
\end{equation}
a formula which will require us to obtain the Starobinsky expansion for RTdS spacetimes. Such a map has already been performed in partial detail in \cite{Bicak:1995vc} (recovering $g_{(0)}$ but not $g_{(3)}$ as required for the Hamiltonians) and in full detail for RTAdS in \cite{Bakas:2014kfa} (see also \cite{Arenas-Henriquez:2025rpt} for a recent discussion). Here we will provide the full calculation for RTdS in order to obtain all necessary information required for computation of the Hamiltonians. 

The map from Robinson-Trautman into Starobinsky coordinates takes the direct form given in \cite{Bicak:1995vc}
or can be done in stages following \cite{Bakas:2014kfa}. We will follow the second approach and start by transforming the line element \eqref{eq: RT_metric} via the initial map 
\begin{equation} \label{eq: RT_FG_initial}
    u = t - r_*\,, 
\end{equation}
where the tortoise coordinate $r_*$ is that of the SdS solution \eqref{eq: SdS_tortoise}. The full map then takes the form of
\begin{align}
    \begin{split} \label{eq: RT_FG_main}
        r_* & \rightarrow - \rho -\frac{1}{2}( \partial_t \hat{\Phi} )\rho^2 - \left[ \frac{\Lambda}{9} + \frac{3}{16} (\partial_t \hat{\Phi})^2 + \frac{1}{4} \partial_t^2 \hat{\Phi} + \frac{\Lambda}{12} \hat{\Delta} \hat{\Phi}\right] \rho^3 - \frac{1}{432} \Big[ 8m\Lambda^2 \\
        & \qquad +27 (\partial_t \hat{\Phi})^3 +72 \partial_t\hat{\Phi} (\Lambda + \partial_t^2 \hat{\Phi} ) +36 \partial_t^3 \hat{\Phi} +12 \Lambda (2\partial_t (\hat{\Delta}\hat{\Phi}) + 3  \hat{\Delta} \hat{\Phi} \partial_t \hat{\Phi})  \Big] \rho^4 \\
        & \qquad + \mathcal{O}(\rho^5) \, , \\
        t & \rightarrow t  -\frac{1}{2} (\partial_t \hat{\Phi}) \rho^2 - \frac{1}{6} \left[ \frac{2\Lambda}{3} + (\partial_t \hat{\Phi})^2  + \partial^2_t \hat{\Phi} + \frac{2\Lambda}{3} \hat{\Delta} \hat{\Phi} \right] \rho^3 - \frac{1}{32} \bigg[ 2 (\partial_t \hat{\Phi})^3 \\
        &\qquad + \partial_t \hat{\Phi} \left( \frac{16 \Lambda}{3} + 5 \partial_t^2 \hat{\Phi} \right) + 2 \partial_t^3 \hat{\Phi} + \frac{2\Lambda}{3} (4 \hat{\Delta} \hat{\Phi} \partial_t \hat{\Phi} + 3 \partial_t (\hat{\Delta} \hat{\Phi})) \bigg]\rho^4 + \mathcal{O}(\rho^5) \, ,  \\
        z & \rightarrow z + e^{-\hat{\Phi}} \partial_{\bar{z}} \left[ \frac{\Lambda}{18} (\partial_t \hat{\Phi}) \rho^3 + \frac{\Lambda}{48} \left( \partial_t^2 \hat{\Phi} - \frac{1}{4} (\partial_t \hat{\Phi})^2 + \frac{\Lambda}{3} \hat{\Delta} \hat{\Phi} \right) \rho^4 \right] + \mathcal{O}(\rho^5)\, , \\
        \bar{z} & \rightarrow \bar{z} +  e^{-\hat{\Phi}} \partial_{z} \left[ \frac{\Lambda}{18} (\partial_t \hat{\Phi}) \rho^3 + \frac{\Lambda}{48} \left( \partial_t^2 \hat{\Phi} - \frac{1}{4} (\partial_t \hat{\Phi})^2 + \frac{\Lambda}{3} \hat{\Delta} \hat{\Phi} \right) \rho^4 \right] + \mathcal{O}(\rho^5) \, ,
    \end{split}
\end{align}
where we note that our equation above is the dS analogue of equations (4.5)-(4.8) of \cite{Bakas:2014kfa}, including the reinstating of all factors of $\Lambda$ in the $\mathcal{O}(\rho^4)$ term in the $r_*$ map. We also use the notation of \cite{Bakas:2014kfa} in denoting $\hat{\Phi}$ as the boundary value of $\Phi$, {\it i.e.}
\begin{equation}
    \hat{\Phi}(t, z, \bar{z}) = \lim_{r_* \rightarrow 0} \Phi(u,z,\bar{z} )\,,
\end{equation}
where $\hat{\Phi}$ satisfies the boundary RT equation 
\begin{equation} \label{eq: boundary_RT}
    3 m \partial_t \hat{\Phi} + \hat{\Delta}\hat{\Delta} \hat{\Phi} = 0\,,
\end{equation}
and $\hat{\Delta}$ is the Laplacian of the constant-$t$ cross-sections of $\mathscr{I}^+$
\begin{equation}
    \hat{\Delta} = e^{-\hat{\Phi}} \partial_{z}\partial_{\bar{z}}\,. 
\end{equation}

After performing this change of variables, we have the following terms in the Starobinsky expansion \eqref{eq: Starobinsky_gauge} for the Robinson-Trautman spacetime 
\begin{subequations}
\begin{align}
    ds_{(0)}^2 & = dt^2 + \frac{6}{\Lambda} e^{\hat{\Phi}} dz d\bar{z} \,, \label{eq: RT_g_0} \\
    \begin{split} \label{eq: RT_dS_g_2}
    ds_{(2)}^2 & = \frac{1}{2} \left[ \frac{\Lambda}{3} \hat{\Delta} \hat{\Phi} - \frac{1}{4} (\partial_t \hat{\Phi})^2 - \partial_t^2 \hat{\Phi}  \right] dt^2 - \left[ \frac{3}{4\Lambda}  (\partial_t \hat{\Phi})^2 + \hat{\Delta} \hat{\Phi} \right]e^{\hat{\Phi}} dz d\bar{z} \\
    & \qquad - [(\partial_t \partial_{z} \hat{\Phi}) dt dz]_{\text{c.c.}}\,,
    \end{split}\\
    \begin{split}
     ds_{(3)}^2 & = - \frac{4 m \Lambda^2}{27}  dt^2 + \frac{4m \Lambda}{9} e^{\hat{\Phi}} dz d\bar{z} \\
     & \qquad - \frac{1}{3} \left[ \frac{2 \Lambda}{3} \partial_{z} (\hat{\Delta} \hat{\Phi} ) dt dz + \partial_t \left( \frac{1}{2} (\partial_{z} \hat{\Phi})^2 - \partial_{z}^2 \hat{\Phi} \right) dz^2 \right]_{\text{c.c.}}\,, \label{eq: rt_ds_g_3}
     \end{split}
\end{align}
\end{subequations}
where the subscript c.c. indicates that one also has to add the complex conjugate of the given term. This is all of the relevant information for the RTdS spacetime required for the computation of the charges and fluxes at $\mathscr{I}^+$. 

Inspection of the $g_{(0)}$ term \eqref{eq: RT_g_0} shows that the RTdS spacetime is merely asymptotically locally dS and not asymptotically dS \cite{Bakas:2014kfa}. This can be shown via computation of the Cotton-York tensor 
\begin{equation}
    C^{ij} = \epsilon_{(0)}^{ikl} \nabla^{(0)}_{k} \left( R^{\phantom{(0)} j}_{(0)l} - \frac{1}{4} \delta^j_l R_{(0)} \right)\,,
\end{equation}
where $\epsilon^{(0)}_{ijk}$ is the volume element of $g_{(0)}$, oriented with positive direction in spherical coordinates $(\theta, \phi)$ {\it i.e.}
\begin{equation} \label{eq: RT_volume_form}
   \epsilon^{(0)}_{t \theta \phi} = \sqrt{g_{(0)}} = \frac{3}{\Lambda} e^{\hat{\Phi}} \sin \theta \csc^4\left( \frac{\theta}{2} \right) \quad \iff \quad \epsilon^{(0)}_{tz \bar{z}} = - i \frac{3}{\Lambda} e^{\hat{\Phi}} \, .
\end{equation}
We can compute the components of the Cotton-York tensor to be 
\begin{align}
    \begin{split}
        C_{z z} = \bar{C}_{\bar{z} \bar{z}} = - \frac{i}{4} \partial_t \left( (\partial_{z} \hat{\Phi})^2 - 2 \partial_{z}^2 \hat{\Phi} \right)\,, \quad C_{t z} = \bar{C}_{t\bar{z}} = - \frac{i}{6} \Lambda \partial_{z} (\hat{\Delta} \hat{\Phi})\,, 
    \end{split}
\end{align}
with all others vanishing. The fact that the Cotton-York tensor is non-zero confirms that this spacetime is not asymptotically dS and merely asymptotically locally dS as the conformal class at $\mathscr{I}^+$ is not conformally flat. We will also observe that \eqref{eq: RT_g_0} generically admits no conformal Killing vectors and thus if we want to define notions of charges at $\mathscr{I}^+$ then we will be forced to use the modified Hamiltonians \eqref{eq: mod_H}.

We can perform the sanity check on our formula for the second non-vanishing term in the Starobinsky expansion \eqref{eq: RT_dS_g_2} which is given explicitly in terms of the leading order term $g_{(0)}$. We recall from \eqref{eq: g_2_dS} that 
\begin{equation}
    g_{(2)ij} = R^{}_{(0) ij} - \frac{1}{4} R_{(0)} g^{}_{(0) ij}\,,
\end{equation}
and computing components we find 
\begin{align}
    R^{}_{(0) tt} & = - \partial_t^2 \hat{\Phi} - \frac{1}{2} (\partial_t \hat{\Phi} )^2\,, \quad  \quad R^{}_{(0)z \bar{z}} = - \frac{1}{2} e^{\hat{\Phi}} \left( \frac{3}{\Lambda} \left( \partial_t^2 \hat{\Phi} + (\partial_t \hat{\Phi})^2 \right) + 2 \hat{\Delta} \hat{\Phi} \right)\,, \\ 
    R_{(0) tz}^{} & = \bar{R}_{t\bar{z}}^{(0)} =- \frac{1}{2} \partial_t \partial_{z} \hat{\Phi}\,, \qquad R_{(0)} = -2 \partial_t^2 \hat{\Phi} - \frac{3}{2} (\partial_t \hat{\Phi})^2 - \frac{2 \Lambda}{3} \hat{\Delta} \hat{\Phi}\,, 
\end{align}
which reproduce \eqref{eq: RT_dS_g_2} via \eqref{eq: g_2_dS}.

We can also derive the energy-momentum tensor at $\mathscr{I}^+$ using \eqref{eq: EM_future}, which as in the KdS case \eqref{eq: Tij_dimensionful} becomes 
\begin{equation}
 T_{ij}  = - \frac{3 \ell^2}{2\kappa^2} g_{ij}^{(3)}\,,
\end{equation}
due to the use of dimensionful time coordinate $\rho$. Using the values for the components of $g_{(3)}$ as given in \eqref{eq: rt_ds_g_3} we find 
\begin{align} 
    \kappa^2 T_{tt} & = \frac{2 m \Lambda}{3} \, , \qquad \kappa^2 T_{t z} = \frac{1}{2} \partial_{z} ( \hat{\Delta} \hat{\Phi} ) \,, \label{eq: TEM_RT} \\
    \kappa^2 T_{z \bar{z}} & = - m e^{\hat{\Phi}} \, , \qquad \kappa^2 T_{z z} = \frac{3}{4\Lambda} \partial_t \left( (\partial_{z} \hat{\Phi})^2 - 2\partial_z^2 \hat{\Phi} \right) \label{eq: TEM_RT_angles} \, ,
\end{align}
with 
\begin{equation} \label{eq: T_conjugates}
    T_{t \bar{z}} = \bar{T}_{t z} \,, \qquad T_{\bar{z} \bar{z}} = \bar{T}_{z z}\, .
\end{equation}
We can also verify that the energy-momentum tensor satisfies tracelessness \eqref{eq: em_traceless} $T_i^i=0$ and conservation \eqref{eq: em_conserved} $\nabla^{(0)}_i T^i_j =0$, the latter of which is due to the boundary RT equation \eqref{eq: boundary_RT} in the following manner
\begin{equation}
    \nabla_{i}^{(0)} T^{i}_j =  \frac{1}{\ell^2 \kappa^2} (\hat{\Delta} \hat{\Delta} \hat{\Phi} + 3m \hat{\Phi}_{,t} )\delta^t_j\,.
\end{equation}

Finally, we note the relation between the components of the energy-momentum tensor and the Cotton tensor 
\begin{align}
\begin{split} \label{eq: cotton_em_relations}
    C_{z z} & = - i \frac{\Lambda}{3} \kappa^2 T_{z z} \, , \qquad C_{\bar{z} \bar{z}} = i \frac{\Lambda}{3} \kappa^2 T_{\bar{z} \bar{z}}\, , \\
    C_{tz} & = - i \frac{\Lambda}{3} \kappa^2 T_{t z} \, , \qquad C_{t \bar{z}} = i \frac{\Lambda}{3}\kappa^2 T_{t \bar{z}}\, ,
    \end{split}
\end{align}
and as in the AlAdS case \cite{Bakas:2014kfa}, all components are real when written in spherical coordinates $(\theta, \phi)$. 

\subsubsection{Mass-like gravitational charges} 
We can now compute the charge expression \eqref{eq: Final_Hamiltonian} for a vector field $\xi_{(0)}^{\mu} = \xi^t_{(0)} \delta^{\mu}_t$ which is orthogonal to the constant-$t$ cross-sections of $\mathscr{I}^+$. We will see that the RTdS solution does not in general possess ACKVs and thus the existence criterion \eqref{eq: H_existence} will fail, and thus we may use the modification procedure discussed in Section \ref{sec: Modified_H} in order to define modified Hamiltonians, or (equivalently) use the formulae for charges derived in 
Sections \ref{sec: Noeth1}-\ref{sec: Noet2}, which satisfy flux-balance laws.

We start by recalling the metric induced at the conformal boundary \eqref{eq: RT_g_0} 
\begin{equation} \label{eq: RT_ds_boundary}
ds_{(0)}^2 = dt^2 + \frac{6}{\Lambda} e^{\hat{\Phi}} dz d\bar{z} \, , \qquad \epsilon^{(0)}_{tz \bar{z}}  = - i  \frac{3}{\Lambda} e^{\hat{\Phi}} = - i  e^{\hat{\Phi}_0} \frac{3}{\Lambda} e^{\hat{\Phi}-\hat{\Phi}_0} \, ,
\end{equation}
where we performed the somewhat unusual rewriting of the volume form in order to separate the volume of the unit round $S^2$,\footnote{The factor of $i$ that appears in (\ref{eq: RT_ds_boundary}) is related to the fact that $z, \bar{z}$ are complex. The relationship with the usual spherical polar coordinates is \eqref{eq: complex_to_spherical}.} where $\hat{\Phi}_0$ is given by  
\begin{equation}
e^{\hat{\Phi}_0} = \frac{1}{\left(1+ \frac{z \bar{z}}{2} \right)^2} \, .
\end{equation} 
 Using \eqref{eq: TEM_RT} together with \eqref{eq: RT_ds_boundary} we have
\begin{align}
\begin{split} \label{eq: H_Rt}
\mathcal{H}_{\xi} & = \frac{1}{2} \int_{\partial C} T^k_l \xi^l_{(0)} \epsilon^{(0)}_{kij} \, dx^i \wedge dx^j \\
& = \int_{\partial C} T^t_l \xi^l_{(0)} \epsilon^{(0)}_{t z \bar{z}} \,  
dz d\bar{z} \\
& = \frac{2 m}{\kappa^2} \int_{\partial C}  \xi_{(0)}^t e^{\hat{\Phi}-\hat{\Phi}_0} \, d\Omega_0\, ,
 \end{split}
\end{align}
where we write the unit round $S^2$ volume element as $d\Omega_0 = - i e^{\hat{\Phi}_0} dz d\bar{z}$. We also write the volume element on the generic distorted $S^2$ as $d\Omega = - i e^{\hat{\Phi}} dz d\bar{z}$. 

In order to evaluate \eqref{eq: H_Rt}, one needs to choose the vector field $\xi_{(0)}^t$. In order to gain intuition as to suitable choices of $\xi_{(0)}$, we study the form of the conformal Killing equations of the metric \eqref{eq: RT_g_0}. This will allow us to identify for which cases CKVs (and thus conserved charges) exist, as well as identifying suitable candidate vector fields to describe the passage of outgoing gravitational waves when no solutions exist. In order to simplify the presentation, {\color{black}we write our vector field as
\begin{equation} \label{eq: xi_ansatz}
    \xi_{(0)}^i \partial_i = f(t, z, \bar{z}) \partial_t\,,
\end{equation}
and attempt to solve the conformal Killing equations. We compute the components of the Lie derivative of the metric to be
\begin{align} 
\mathcal{L}_{\xi_{(0)}} g^{(0)}_{tt} & = 2 f_{,t} = 2 f_{,t} g^{(0)}_{tt} \, , \label{eq: Lie_tt} \\
\mathcal{L}_{\xi_{(0)}} g^{(0)}_{t z} & = \mathcal{L}_{\xi_{(0)}} g^{(0)}_{z t} = f_{,z} \, , \\
\mathcal{L}_{\xi_{(0)}} g^{(0)}_{t \bar{z}} & = \mathcal{L}_{\xi_{(0)}} g^{(0)}_{\bar{z} t} = f_{,\bar{z}}\, , \\
\mathcal{L}_{\xi_{(0)}} g^{(0)}_{z \bar{z}} & =  \mathcal{L}_{\xi_{(0)}} g^{(0)}_{\bar{z} z} = \frac{3}{\Lambda} \partial_t \left( e^{\hat{\Phi}} \right) f = f \hat{\Phi}_{,t} g^{(0)}_{z \bar{z}} \label{eq: Lie_zbz} \, , \\
\mathcal{L}_{\xi_{(0)}} g^{(0)}_{z z} & = \mathcal{L}_{\xi_{(0)}} g^{(0)}_{\bar{z} \bar{z}} = 0\, ,
\end{align}
which allows us to rewrite the conformal Killing equations as equations for $f$. We note that from the second and third lines above together with the fact that $g_{t z} = g_{t \bar{z}} = 0$ we have 
\begin{equation} \label{eq: f_angle_indep}
    f_{, z} = f_{,\bar{z}} = 0 \implies f = f(t)\,.
\end{equation}
The remaining equation now comes from demanding that the factors multiplying the components of the metric in \eqref{eq: Lie_tt} and \eqref{eq: Lie_zbz} are equivalent, giving us the equation
\begin{equation}
    2 f_{,t} = f \hat{\Phi}_{,t} \, ,
\end{equation}
which can be integrated 
\begin{equation} \label{f_t_int}
    f(t) = c_0(z, \bar{z}) \exp\left(\frac{1}{2} \hat{\Phi}\right) \,,
\end{equation}
where $c_0$ is an integration ``constant" that depends on angular variables, $z, \bar{z}$.  Since $f$ only depends on $t$ this solution is only possible when 
\begin{equation} \label{phi split}
\hat{\Phi}(t, z, \bar{z})=\varphi_{0}(t) +\varphi_1(z, \bar{z}) \,,  
\end{equation}
for some $\varphi_{0}(t)$ and $\varphi_1(z, \bar{z})$. Then the $z$ and $\bar{z}$ independence of $f$ imply that 
\begin{equation} \label{eq: c_0}
   c_0(z, \bar{z}) = c_1 e^{-\frac{1}{2} \varphi_1(z, \bar{z})}\,,
\end{equation}
where $c_1$ is a constant. Thus, RTdS spacetimes that admit asymptotic conformal Killing vectors \eqref{eq: xi_ansatz} must have $\hat{\Phi}$ of the form \eqref{phi split}.
For such $\hat{\Phi}$ the boundary Robinson-Trautman equation \eqref{eq: boundary_RT} implies,
\begin{equation}
    3 m \dot{\varphi}_0(t) + e^{-2 \varphi_0(t)} 
    e^{-\varphi_1(z, \bar{z})} \partial_z \partial_{\bar{z}} \left(
    e^{-\varphi_1(z, \bar{z})} \partial_z \partial_{\bar{z}} \varphi_1(z, \bar{z})\right)=0\, .
\end{equation}
This PDE separates,
\begin{align}
    3 m \dot{\varphi}_0(t) &= \lambda e^{-2 \varphi_0(t)}\, ; \label{sep_time} \\
    \partial_z \partial_{\bar{z}} \left(
    e^{-\varphi_1(z, \bar{z})} \partial_z \partial_{\bar{z}} \varphi_1(z, \bar{z})\right) &= - \lambda e^{\varphi_1(z, \bar{z})}\, , \label{sep_sphe}
\end{align}
where $\lambda$ is the separation constant. Integrating \eqref{sep_sphe} over the sphere yields $\lambda=0$ and thus from \eqref{sep_time} and \eqref{phi split} we obtain
\begin{equation} \label{eq: t_indep_Phi}
    \hat{\Phi}_{,t} = 0\,. 
\end{equation}
Then, either by solving \eqref{sep_sphe} (with $\lambda=0$) or by using the fact that at late times the Robinson-Trautman-dS spacetime approaches the SdS solution \cite{Bicak:1995vc} and thus 
(due to the time-independence of $\hat{\Phi}$)  the $\hat{\Phi}$  must be that of SdS, we find
\begin{equation}
    e^{\hat{\Phi}} = e^{\hat{\Phi}_0} =  \frac{1}{\left(1+ \frac{z \bar{z}}{2} \right)^2}\,.
\end{equation}

We have shown that in order for an RTdS spacetime to admit asymptotic  conformal Killing vectors of the form \eqref{eq: xi_ansatz} it must be the SdS solution. We also note that in this case the conformal Killing vector, $\mathring{\xi}_{(0)}$, is strictly a Killing vector of the boundary and is given by $\mathring{\xi}^i_{(0)} \partial_i = \partial_t$. In the setting of a generic RTdS spacetime there are no conformal Killing vectors and} we will be using the results of Section 
\ref{sec: grav_charges} to define gravitational charges that satisfy flux-balance laws. Clearly there are (infinitely) many candidate vector fields, $\xi_{(0)}^t$ which one may use to define a ``mass''-like charge for the RTdS spacetime. In order to attach a physical meaning to the flux, it must be associated with a quantity which is conserved quantity. For example, if there is a solution of the conformal Killing equation in a neighborhood of the boundary, then one can defined a charge which is conserved there but its value changes outside this neighborhood, because flux carrying this charge enters or leaves that portion of the boundary. Taking this into account, we note that a sensible candidate must limit to the SdS time translation in the late time limit {\it i.e.}
\begin{equation} \label{eq: vector_sds_limit}
    \lim_{t \rightarrow \infty} \xi_{(0)}^t = 1\,,
\end{equation}
which is the only a priori assumption that we will place on our vector fields.

\subsubsection{Angular-momentum-like gravitational charges} One can also construct ``angular momentum'' charges corresponding to vector fields $\varphi_{(0)}$, which are tangential to cross-sections of $\mathscr{I}^+$. Since these vectors are tangential to cross-sections, the existence property \eqref{eq: H_existence} is satisfied and thus we do not require use of the modification procedure. However, we note that these charges will not necessarily be radially invariant along the $t$-direction since they generically have $\left. \boldsymbol{\omega} (\psi, \delta \psi, \mathcal{L}_{\xi} \psi) \right|_{B_{12}} \neq 0$. We write these Hamiltonians as 
    \begin{align}
\begin{split} \label{eq: J_Rt}
H_{\varphi} & = \frac{1}{2} \int_{\partial C} T^k_l \varphi^l_{(0)} \epsilon^{(0)}_{kij} \, dx^i \wedge dx^j \\
& = \int_{\partial C} T^t_l \varphi^l_{(0)} \epsilon^{(0)}_{t z \bar{z}} \,  
dz d\bar{z} \\
& = \frac{1}{2\kappa^2} \frac{3}{\Lambda} \int_{\partial C}  \varphi^A_{(0)} \partial_A \left( \hat{\Delta} \hat{\Phi} \right) e^{\hat{\Phi}-\hat{\Phi}_0} \, d\Omega_0\, ,
 \end{split}
\end{align}
where captial Roman indices above denote a summation over the angular ($z, \bar{z}$) directions. We expect there to be no asymptotic conformal Killing vectors for the generic RTdS spacetime in the same manner as we found for the vector field $\xi$ which was orthogonal to the cross-sections. We put an ansatz of the form 
\begin{equation} \label{eq: ang_mom_vector_ansatz}
    \varphi_{(0)}^A \partial_A = \varphi_{(0)}^{z}(t,z, \bar{z}) \partial_z + \varphi_{(0)}^{\bar{z}}(t,z, \bar{z}) \partial_{\bar{z}} 
    %= g(t, z, \bar{z}) \partial_{z}+ h(t, z, \bar{z}) \partial_{\bar{z}} 
    \,, 
\end{equation}
with the late time condition of settling down to the axial Killing vector of the SdS solution
\begin{equation} \label{eq: late_time_angular_momentum}
    \lim_{t \rightarrow \infty} \varphi_{(0)}^A \partial_A = \partial_{\phi} = i \left( z \partial_{z} - \bar{z} \partial_{\bar{z}}\right)\,,
\end{equation}
{\it i.e.} $\lim_{t \rightarrow \infty} \varphi_{(0)}^{z} = i z$, $\lim_{t\rightarrow \infty} \varphi_{(0)}^{\bar{z}} = - i \bar{z}$. We evaluate the components of the Lie derivative of the metric as 
\begin{align}
\mathcal{L}_{\varphi_{(0)}} g^{(0)}_{tt} & = 0 \, , \label{eq: Lie_tt_phi} \\
\mathcal{L}_{\varphi_{(0)}} g^{(0)}_{t z} & = \mathcal{L}_{\varphi_{(0)}} g^{(0)}_{z t} = \frac{3}{\Lambda} e^{\hat{\Phi}} \partial_t \left( \varphi^{\bar{z}}_{(0)} \right) \, , \\
\mathcal{L}_{\varphi_{(0)}} g^{(0)}_{t \bar{z}} & = \mathcal{L}_{\varphi_{(0)}} g^{(0)}_{\bar{z} t} = \frac{3}{\Lambda} e^{\hat{\Phi}} \partial_t \left( \varphi^{z}_{(0)} \right)\, , \\
\mathcal{L}_{\varphi_{(0)}} g^{(0)}_{z \bar{z}} & =  \mathcal{L}_{\varphi_{(0)}} g^{(0)}_{\bar{z} z} = \frac{3}{\Lambda} \partial_A \left( \varphi^A_{(0)} e^{\hat{\Phi}} \right) \label{eq: Lie_zbz_phi} \, , \\
\mathcal{L}_{\varphi_{(0)}} g^{(0)}_{z z} & = \frac{6}{\Lambda} e^{\hat{\Phi}} \partial_{z} \left( \varphi^{\bar{z}}_{(0)} \right)\,, \\ \mathcal{L}_{\varphi_{(0)}} g^{(0)}_{\bar{z} \bar{z}} & = \frac{6}{\Lambda} e^{\hat{\Phi}} \partial_{\bar{z}} \left( \varphi^{z}_{(0)} \right) \, ,
\end{align} 
and we again attempt to solve the conformal Killing equation. The first equation above tells us that we need to strictly solve the Killing equation rather than the conformal Killing equation. In doing this, the second, third, fifth and sixth lines above reduce the ansatz of \eqref{eq: ang_mom_vector_ansatz} to holomorphic and anti-holomorphic transformations on $S^2$,
\begin{equation} \label{eq: almost_killing_phi}
\varphi_{(0)}^A \partial_A = \varphi_{(0)}^{z}(z) \partial_z + \varphi_{(0)}^{\bar{z}}(\bar{z}) \partial_{\bar{z}}\,. 
\end{equation}
This only leaves the vanishing of \eqref{eq: Lie_zbz_phi} to be solved. We note that when \eqref{eq: Lie_zbz_phi} is satisfied, the Hamiltonian \eqref{eq: J_Rt}  appears to vanish. Indeed, upon integration by parts
\begin{equation} \label{eq: H_phi}
    H_{\varphi} = \frac{1}{2\kappa^2} \frac{3}{\Lambda} \int_{\partial C}  \varphi^A_{(0)} \partial_A \left( \hat{\Delta} \hat{\Phi} \right) e^{\hat{\Phi}-\hat{\Phi}_0} \, d\Omega_0 =   \frac{i}{2\kappa^2} \frac{3}{\Lambda} \int_{\partial C}  \partial_A  \left(  \varphi^A_{(0)} e^{  \hat{\Phi}}  \right) \hat{\Delta} \hat{\Phi}  \, dz d\bar{z} = 0\,.
\end{equation}
However, one has to be careful with the total derivatives. Writing $\varphi_{(0) n}^{z}(z) = z^{n+1}$, the holomorphic vector fields $l_n = \varphi_{(0) n}^{z}(z) \partial_z$ are globally defined on $S^2$ only when $n=-1, 0, 1$ (as is well-known, see for example \cite{Ginsparg:1988ui}) - all other cases are singular either $z=0$ or at $z \to \infty$. Unless $\hat{\Phi}$ is such that the factor that multiplies $\varphi_{(0) n}^{z}$ goes to zero sufficiently fast at the positions of the singularities, the integral in \eqref{eq: H_phi} is ill-defined. For generic $\hat{\Phi}$ only the transformations with $n=-1, 0, 1$ are well-defined. Similar comments apply to the anti-holomorphic transformations.

An example of RTdS spacetimes with a Killing vector of this form is that of axisymmetric solutions {\it i.e.} those for which $\partial_{\phi} \hat{\Phi} = 0$ and thus admit a Killing vector of the form $\mathring{\varphi}^A_{(0)} = \partial_{\phi}$. One can evaluate the Hamiltonian \eqref{eq: J_Rt} as 
\begin{equation}
    H_{\mathring{\varphi}} =   \frac{1}{2\kappa^2} \frac{3}{\Lambda} \int_{\partial C}   \partial_{\phi} \left( \hat{\Delta} \hat{\Phi} \right) e^{\hat{\Phi}-\hat{\Phi}_0} \, d\Omega_0 = 0\,, 
\end{equation}
demonstrating that axisymmetric Robinson-Trautman metrics have zero angular momentum.

For generic $\hat{\Phi}$ which satisfies \eqref{eq: boundary_RT} there will be no Killing vectors and one will again be forced to deal with quantities which admit some flux across $\mathscr{I}^+$. As in the case of the ``mass'' \eqref{eq: H_Rt}, the angular momentum charge in RTdS spacetime thus has an infinite family of candidates based on vector fields which satisfy the late time condition \eqref{eq: late_time_angular_momentum}.

\subsection{Fluxes}

As we have discussed, there are generically no asymptotic conformal Killing vectors associated with RTdS spacetimes. This means that the gravitational charges \eqref{eq: H_Rt}, \eqref{eq: J_Rt} will not remain constant as one varies the coordinate $t$ along $\mathscr{I}^+$, but rather they would satisfy the flux-balance law
\begin{equation} 
   \left. H_{\xi} \right |_{C_2} -  \left. H_{\xi} \right |_{C_1} = - \int_{B_{12}} \mathbf{F}_{\xi_{(0)}}\,,
\end{equation}
where $B_{12}$ is the strip of $\mathscr{I}^+$ between $C_1$, see Figure \ref{fig: ds_flux}, and $C_2$ and the flux is given by
\begin{equation} \label{eq: flux_formula_rt}
\bm{F}_{\xi} %= \bm{\theta}(\psi, \mathcal{L}_{\xi_{(0)}} \psi) 
= -\frac{1}{2}  T^{ij} \mathcal{L}_{\xi_{(0)}} g^{(0)}_{ij}\,  \bm{\epsilon}^{(0)}\,,
\end{equation}
and $\xi_{(0)}$ is the boundary value of the bulk vector $\xi$, as in \eqref{eq: xi_ASG}.
We will now analyse the flux formulae for the mass type vector field \eqref{eq: xi_ansatz} and the angular momentum type vector field \eqref{eq: ang_mom_vector_ansatz} individually.
The flux for the Hamiltonian for a general linear combination of these fields can then be deduced via the suitable linear combination of the fluxes.

There are two important cases that we discuss below:
\begin{itemize}
    \item {\bf The integral of the flux vanishes}. This implies that the gravitational charge is {\bf conserved}. We discussed earlier in Section \ref{sec: grav_charges} that a sufficient condition is that the boundary metric admits CKVs. In this case the flux \eqref{eq: flux_formula_rt} vanishes point-wise. 
    However, a sufficient condition for conservation is that the integral vanishes and this only requires that ${\bf F}_\xi$ is exact over the compact directions.
    \item {\bf The integral of the flux is sign-definite}. In this case the charges are not conserved, but they are {\bf monotonic}. An important example is the Bondi mass-loss formula, and we will discuss generalisations.     
\end{itemize}

\subsubsection{Mass fluxes}
We first compute the flux formula for the generic vector field  \eqref{eq: xi_ansatz}. Using equations \eqref{eq: Lie_tt}-\eqref{eq: Lie_zbz} together with the expressions for the energy-momentum tensor \eqref{eq: TEM_RT}-\eqref{eq: T_conjugates} and \eqref{eq: RT_volume_form} for the volume form we can write the flux form as 
\begin{equation} \label{eq: flux_RTdS_explicit}
\bm{F}_{\xi} =  \frac{i}{2\kappa^2} \left[  \partial_{z} \left( \hat{\Delta} \hat{\Phi} \right) f_{, \bar{z}} + \partial_{\bar{z}} \left( \hat{\Delta} \hat{\Phi}  \right) f_{,z}  +  4m e^{\hat{\Phi}} f_{,t} -2 m e^{\hat{\Phi}} f \hat{\Phi}_{,t} \right]dt \wedge dz \wedge d\bar{z}\, ,
\end{equation} 
and the flux-balance equation \eqref{eq: flux_formula_rt} becomes
\begin{align}
\begin{split}
\mathcal{H}_{\xi}|_{t_1} - \mathcal{H}_{\xi}|_{t_0} & = - \int_{B_{12}}
%{\Delta \subset \mathscr{I}^+} 
\bm{F}_{\xi} \\
& = \frac{i}{\kappa^2} \int_{B_{12}}
%{\Delta \subset \mathscr{I}^+} 
e^{\hat{\Phi}} \left[ ( \hat{\Delta} \hat{\Delta} \hat{\Phi} ) f -  2m  f_{,t} + m  f \hat{\Phi}_{,t}   \right] \, dt  dz  d\bar{z} \,,
\end{split}
\end{align}
where we integrated by parts in the angular coordinates and ignored total derivatives to simplify the first term on the second line. This term is now in the ideal form to apply the boundary RT equation \eqref{eq: boundary_RT}. Application of this, followed by some further manipulations allows us to bring the flux formula into the form 
\begin{equation} \label{eq: flux_general}
- \int_{B_{12}} \bm{F}_{\xi} = -  \frac{2mi}{\kappa^2} \int_{B_{12}} \partial_t \left( e^{\hat{\Phi}} f \right)  \, dt dz d\bar{z}  \, ,  
\end{equation}
which is in agreement with the obvious formula for flux between two cuts which one could obtain via Stokes' theorem. 
\subsubsection{Angular momentum fluxes}

We now compute the flux form \eqref{eq: flux_formula_rt} for the vector field given by \eqref{eq: ang_mom_vector_ansatz}. The general expression for the flux form is given by 
\begin{align}
    \begin{split} \label{eq: F_phi}
        \mathbf{F}_{\varphi} & = - \frac{i}{2\kappa^2 \Lambda} \left\{ 2 m \Lambda \left[ \partial_{z} \left( e^{\hat{\Phi}} \varphi_{(0)}^{z} \right) + \partial_{\bar{z}} \left(e^{\hat{\Phi} } \varphi_{(0)}^{\bar{z}}\right)  \right] - \right. \\
       & \qquad \qquad \qquad \left.3 \left[ (\varphi_{(0)}^{z})_{,\bar{z}} \left( \hat{\Phi}_{,z} \hat{\Phi}_{, tz} - \hat{\Phi}_{,tz z} \right) + (\varphi_{(0)}^{\bar{z}})_{,z} \left( \hat{\Phi}_{,\bar{z}} \hat{\Phi}_{,t\bar{z}} - \hat{\Phi}_{,t\bar{z}\bar{z}} \right) \right. \right. \\
      & \qquad \qquad \qquad \quad \left. \left.  +   (\varphi_{(0)}^{z})_{,t} e^{\hat{\Phi}} \partial_{z} \left( \hat{\Delta} \hat{\Phi} \right) + (\varphi_{(0)}^{\bar{z}})_{,t}  e^{\hat{\Phi}}  \partial_{\bar{z}} \left( \hat{\Delta} \hat{\Phi} \right)   \right]  \right\} \, dt \wedge dz \wedge d\bar{z} \,,
    \end{split}
\end{align}
which we can simplify further by integrating over a strip of the conformal boundary 
\begin{align}
\begin{split}
- \int_{B_{12}} \mathbf{F}_{\varphi} & = - \frac{i}{2\kappa^2} \frac{3}{\Lambda} \int_{B_{12}} \left[  (\varphi_{(0)}^{z})_{,t} e^{\hat{\Phi}} \partial_{z} \left( \hat{\Delta} \hat{\Phi} \right) + \varphi_{(0)}^{z} \partial_t \left(  e^{\hat{\Phi}} \partial_{z} \left( \hat{\Delta} \hat{\Phi} \right)\right) \right. \\
& \left. \qquad \qquad \qquad  +  (\varphi_{(0)}^{\bar{z}})_{,t} e^{\hat{\Phi}} \partial_{\bar{z}} \left( \hat{\Delta} \hat{\Phi} \right) + \varphi_{(0)}^{\bar{z}} \partial_t \left(  e^{\hat{\Phi}} \partial_{\bar{z}} \left( \hat{\Delta} \hat{\Phi} \right)\right) \right] \, dtdz d\bar{z} \\
& = - \frac{i}{2\kappa^2} \frac{3}{\Lambda} \int_{B_{12}} \partial_t \left( \varphi_{(0)}^{z} e^{\hat{\Phi}} \partial_{z} \left( \hat{\Delta} \hat{\Phi} \right) + \varphi_{(0)}^{\bar{z}} e^{\hat{\Phi}} \partial_{\bar{z}} \left( \hat{\Delta} \hat{\Phi} \right) \right) \, dt dz d\bar{z} \\ 
& = \frac{1}{2\kappa^2} \frac{3}{\Lambda} \int_{B_{12}} \partial_t \left( \varphi_{(0)}^A \partial_A  \left( \hat{\Delta} \hat{\Phi} \right) e^{\hat{\Phi} - \hat{\Phi}_0} \right) \, dt d\Omega_0\,,
\end{split}
\end{align}
giving the expected result when comparing with the Hamiltonians \eqref{eq: J_Rt}. In deriving the above equation, we used the fact that total derivatives in $z, \bar{z}$ integrate to zero. This fact meant that the first line of \eqref{eq: F_phi} vanishes when integrated over the angular directions. We then use integration by parts in $z$ and $\bar{z}$ on the second line of \eqref{eq: F_phi} and after some rearrangement arrive at the above. Note that we did not require the Robinson-Trautman equation \eqref{eq: boundary_RT} in order to arrive at the expected result. 

\subsection{Examples of conserved and monotonic gravitational charges} \label{sec: RTdS_examples}

We now give several interesting examples of conserved and monotonic gravitational charges for the RTdS solution. We shall present several vector fields associated with conserved charges, despite not admitting any non-trivial conformal Killing vectors. We shall see that these charges are invariant not by virtue of the flux form \eqref{eq: flux_RTdS_explicit} vanishing, but rather it being a total derivative over the cross-sections {\it i.e.}
\begin{equation} \label{eq: F_cross-section_exact}
    \mathbf{F} = d_2 \mathbf{a} \wedge dt\,,
\end{equation}
where $d_2$ is the exterior derivative associated with the metric \eqref{eq: 2-metric_calabi} and $\mathbf{a}$ a one-form. Our analysis suggests that in order to look for conserved charges of general spacetimes, one should search for solutions with flux of the form \eqref{eq: F_cross-section_exact} rather than the more restrictive condition of $\mathbf{F}=0$. We will also give examples of charges  that are monotonically decreasing in boundary coordinate $t$. We provide a further example of a non-monotonic charge in Appendix \ref{app: Quadratic_flux}, where we compare our formalism with the linearised results of \cite{Chrusciel:2020rlz, Kolanowski:2020wfg, Chrusciel:2021ttc, Bonga:2023eml}.

\subsubsection{Conserved charges}

\paragraph{Cross-section area}

The first choice we make is 
\begin{equation}
    \xi^t_{(0)} = f = 1\,,
\end{equation}
which satisfies \eqref{eq: vector_sds_limit} and gives the gravitational charge \eqref{eq: H_Rt}
\begin{equation} \label{eq: area_Hamiltonian}
    \mathcal{H}_{\xi} = \frac{2m}{\kappa^2} \int_{\partial C} e^{\hat{\Phi}-\hat{\Phi}_0} \, d\Omega_0 = \frac{2m}{\kappa^2} \int_{\partial C} \, d\Omega = \frac{2 m}{\kappa^2} A_2\,,
\end{equation}
where we note that $A_2$ is the area of the cross-section $\partial C$. Evaluation of the flux form \eqref{eq: flux_RTdS_explicit} gives 
\begin{equation}
    \int_{B_{12}} \mathbf{F}_{\xi} = - \frac{i}{\kappa^2} m \int_{B_{12}} e^{\hat{\Phi}} \hat{\Phi}_{,t} \, dt dz d\bar{z} =  \frac{i}{3\kappa^2} \int_{B_{12}} \partial_{z}\partial_{\bar{z}} \hat{\Delta} \hat{\Phi} \, dt dz d\bar{z} = 0 \,, 
\end{equation}
where we used the boundary RT equation \eqref{eq: boundary_RT} in going from the second to the third expression and then the fact that the resulting integrand is a total derivative over a closed surface in order to reach the conclusion of zero flux. We note that the conservation of the area follows directly from the Robinson-Trautman equation, and is in fact a general property of two-metrics evolving under Calabi flow \cite{Calabi+1982+259+290, Calabi1985}.

\paragraph{Euler characteristic}

The second choice we make is 
\begin{equation} \label{eq: xi_Euler}
    \xi^t_{(0)} = f = - \hat{\Delta} \hat{\Phi} \,,
\end{equation}
which satisfies \eqref{eq: vector_sds_limit} by virtue of $\hat{\Delta} \hat{\Phi}_0 = -1$. We note that this choice is motivated by the scalar curvature of the metric
\begin{equation} \label{eq: 2-metric_calabi}
    ds_2^2 = 2 e^{\hat{\Phi}(z, \bar{z};t)} dz d\bar{z} \,, 
\end{equation}
which is found to be 
\begin{equation}
    R_2 = - 2 \hat{\Delta} \hat{\Phi}\,. 
\end{equation}
The gravitational charge \eqref{eq: H_Rt} is 
\begin{equation} \label{eq: Euler_Hamiltonian}
    \mathcal{H}_{\xi} = \frac{ m}{\kappa^2} \int_{\partial C} R_2 \, d\Omega = \frac{m}{G} \chi = 2 \frac{m}{G}\,,
\end{equation}
which is proportional to the Euler characteristic $\chi=2$ by the Gauss-Bonnet theorem. Clearly this quantity is constant, a fact which can be confirmed using the integral of the flux form \eqref{eq: flux_RTdS_explicit} for a vector field given by \eqref{eq: xi_Euler}. Several applications of the RT equation yield 
\begin{equation}
    \int_{B_{12}} \mathbf{F}_{\xi} = - 2 m \frac{i}{\kappa^2} \int \partial_{z} \partial_{\bar{z}} \partial_t \hat{\Phi} \, dt dz d\bar{z} = 0\,,
\end{equation}
again via Stokes' theorem for integration over the closed $(z, \bar{z})$-surface. 

\paragraph{Angular momentum and Lorentz charges}

We now turn to study some interesting examples of the angular momentum like Hamiltonians \eqref{eq: J_Rt}. A natural first choice of vector field to analyse is 
\begin{equation}
    \varphi_{(0)}^A \partial_A =  \partial_{\phi} = i \left( z \partial_{z} - \bar{z} \partial_{\bar{z}}\right)\,,
\end{equation}
where we recall that this is the unique vector field \eqref{eq: almost_killing_phi} which solves all components of the Killing equation other than the $z \bar{z}$ component (given by \eqref{eq: Lie_zbz_phi}) and satisfies the asymptotic property \eqref{eq: late_time_angular_momentum}. The Hamiltonian is given by 
\begin{align}
\begin{split} \label{eq: ang_mom_exact}
    H_{\varphi} & = \frac{1}{2\kappa^2} \frac{3}{\Lambda} \int_{\partial C} \left[ z \partial_{z} \left( \hat{\Delta} \hat{\Phi}  \right) - \bar{z} \partial_{\bar{z}} \left(\hat{\Delta} \hat{\Phi} \right) \right] e^{\hat{\Phi}} \, dz d\bar{z}  \\
    & = \frac{1}{4\kappa^2} \frac{3}{\Lambda} \int_{\partial C} \left[ - \partial_{\bar{z} } \left( z (\hat{\Phi}_{,z} )^2 \right) +  \partial_{z} \left( \bar{z} (\hat{\Phi}_{,\bar{z}} )^2 \right) \right] \, dz d\bar{z} \\
    & = 0 \,,
    \end{split}
\end{align}
which is obviously radially invariant. As a consistency check we analyse the flux form 
\begin{equation}
    \int_{B_{12}} \mathbf{F}_{\varphi} = - \frac{ i m }{\kappa^2} \int_{B_{12}} \left[ \partial_{z} \left(i e^{\hat{\Phi}} z \right) - \partial_{\bar{z}} \left(i e^{\hat{\Phi}} \bar{z} \right) \right] \, dtdz d\bar{z}  =0 \,, 
\end{equation}
where we used Stokes' theorem in order to ignore the total derivatives. This example illustrates that even when $\partial_{\phi}$ is not a symmetry of the Robinson-Trautman spacetime, it gives a vanishing Hamiltonian charge. 

In fact, even if we were to relax the asymptotic property \eqref{eq: late_time_angular_momentum} then the corresponding Hamiltonian still vanishes. Explicitly, we take 
\begin{equation} \label{eq: meromorphic_vf}
    \varphi_{(0)}^A \partial_A = \varphi_{(0)}^{z}(z) \partial_z + \varphi_{(0)}^{\bar{z}}(\bar{z}) \partial_{\bar{z}} 
    %= g(z) \partial_{z} + h(\bar{z}) \partial_{\bar{z}}
    \,,  
\end{equation}
which is the sum of holomorphic, $\varphi_{(0)}^{z}(z) \partial_z$  and anti-holomorphic, $\varphi_{(0)}^{\bar{z}}(\bar{z}) \partial_{\bar{z}}$ vector fields that are globally well-defined: 
\begin{equation}
    \varphi_{(0)}^{z}(z) = a_0 + a_1z + a_2 z^2 \,, \quad \varphi_{(0)}^{\bar{z}}(\bar{z}) = b_0 + b_1 \bar{z} + b_2 \bar{z}^2\,, 
\end{equation}
where the coefficients $a_{0,1,2}, b_{0,1,2}$ are constants. This six dimensional space of vectors forms a basis for the Lorentz algebra and thus
we will call globally defined vectors of the form \eqref{eq: meromorphic_vf} \textit{Lorentz vector fields}. The Hamiltonian corresponding to any of these vectors reads 
\begin{align}
\begin{split} \label{eq: ang_mom_general}
    H_{\varphi} & = -\frac{i}{2\kappa^2} \frac{3}{\Lambda} \int_{\partial C} \left[ \varphi_{(0)}^{z} \partial_{z} \left( \hat{\Delta} \hat{\Phi}  \right) + \varphi_{(0)}^{\bar{z}} \partial_{\bar{z}} \left(\hat{\Delta} \hat{\Phi} \right) \right] e^{\hat{\Phi}} \, dz d\bar{z} \\
    & = \frac{i}{2\kappa^2} \frac{3}{\Lambda} \int_{\partial C} \left[ \partial_{\bar{z}} \left( (\varphi_{(0)}^{z})_{,z} \hat{\Phi}_{,z} \right)+ \partial_{z} \left( (\varphi_{(0)}^{\bar{z}}),_{\bar{z}} \hat{\Phi}_{,\bar{z}} \right) \right] \, dz d\bar{z} \\
    & = 0\,,
    \end{split}
\end{align}
and we thus conclude that the Lorentz charges of the RTdS metric vanish. We can again verify this via examination of the integrated flux form 
\begin{equation}
    \int_{B_{12}} \mathbf{F}_{\varphi} = - \frac{ i m }{\kappa^2} \int_{B_{12}} \partial_A \left(  \varphi_{(0)}^A e^{\hat{\Phi}} \right)  \, dtdz d\bar{z}  =0 \,. 
\end{equation}
We note that even though \eqref{eq: meromorphic_vf} is not a Killing vector of \eqref{eq: RT_g_0}, it does satisfy all of the Killing equations other than \eqref{eq: Lie_zbz_phi} which is a total derivative. The fact that $T^{z \bar{z}} {\epsilon}_{(0)}$ (present in equation \eqref{eq: flux_formula_rt}) is constant then immediately forces the flux form to be exact and thus the flux vanishes. One could construct additional charges with non-zero value and flux properties. We leave the investigation of such charges and their physical properties for future work.

Another interesting future direction is to relax the condition that the vector field \eqref{eq: meromorphic_vf} is globally defined and allow for local vector fields with meromorphic singularities. This would parallel the discussion of extended BMS vector fields for asymptotically flat spacetimes considered in \cite{Barnich:2009se, Barnich:2010eb, Barnich:2010ojg} (in turn inspired by \cite{Belavin:1984vu}). Such charges would certainly be relevant 
for the cases where the solution of the RT equation is such that the integrals over $S^2$ that define the charges are finite despite the singularity in the vector field, and they may also have a role to play more generally.

\subsubsection{Monotonic charges}

We now give several examples of monotonic charges w.r.t. evolution in $t$: 
\begin{equation} \label{eq: monotonicity}
    \frac{d}{dt} \mathcal{H}_{\xi} \leq 0\,.
\end{equation}

\paragraph{Bondi mass}

The first of our monotonically decreasing charge is one which we will refer to as the \textit{Bondi mass} \cite{Bondi:1962px}. The Bondi mass was originally  defined in \textit{asymptotically flat} spacetime, although it has been shown \cite{Bakas:2014kfa} that an analogous quantity can be defined in RTAdS spacetime, exhibiting physically sensible properties such as being a monotonically decreasing function of the boundary time. Here we will show that such a quantity can be constructed as gravitational charge for RTdS following the analysis in Section \ref{sec: grav_charges}, and that it satisfies the monotonicity property \eqref{eq: monotonicity}. 

The choice of vector field is 
\begin{equation} \label{eq: almost_CKV} 
\xi^t_{(0)} = f = \exp \left(\frac{1}{2} (\hat{\Phi} - \hat{\Phi}_0) \right) = \frac{1}{\hat{\sigma}(z, \bar{z} ; t)} \,,
\end{equation}
where \eqref{eq: vector_sds_limit} is clearly satisfied and where our use of the function $\hat{\sigma}$ matches \cite{Bakas:2014kfa}, where it was introduced to measure the discrepancy between the general RT metric and the SdS solution ($\hat{\sigma}=1$ is SdS). One can check that the variation induced by this vector field rescales the $tt$ and $z \bar{z}$ components of the metric by the same factor {\it i.e.} 
\begin{align}
\mathcal{L}_{\xi_{(0)}} g^{(0)}_{tt} & = \exp \left(\frac{1}{2} (\hat{\Phi} -  \hat{\Phi}_0) \right)( \partial_{t} \hat{\Phi} ) g_{tt}^{(0)}\,, \label{eq: rr_lie_deriv} \\
\mathcal{L}_{\xi_{(0)}} g^{(0)}_{z \bar{z}} & = \exp \left(\frac{1}{2} (\hat{\Phi} -  \hat{\Phi}_0) \right)( \partial_{t} \hat{\Phi} ) g^{(0)}_{z \bar{z}}\,,
\end{align}
but it fails to be a conformal Killing vector by virtue of 
\begin{equation} \label{eq: rzeta_lie_deriv} 
\mathcal{L}_{\xi_{(0)}} g^{(0)}_{t z} =  \partial_{z} \exp \left(\frac{1}{2} (\hat{\Phi} - \hat{\Phi}_0) \right) =   \partial_{z}\left( \frac{1}{\hat{\sigma}} \right)  \neq 0\,,
\end{equation}
and
\begin{equation} \label{eq: tzetabar_lie_deriv} 
\mathcal{L}_{\xi_{(0)}} g^{(0)}_{t \bar{z}} =  \partial_{\bar{z}} \exp \left(\frac{1}{2} (\hat{\Phi} - \hat{\Phi}_0) \right) =   \partial_{\bar{z}}\left( \frac{1}{\hat{\sigma}} \right)  \neq 0\, .
\end{equation}
Inserting the vector field \eqref{eq: almost_CKV} into \eqref{eq: H_Rt} we find the modified Hamiltonian
\begin{equation} \label{eq: Bondi_mass_charge}
\mathcal{H}_{\xi}= \frac{2m}{\kappa^2} \int_{C \cap \mathscr{I}^+} \exp \left(\frac{1}{2} (3 \hat{\Phi} - 3 \hat{\Phi}_0) \right) \, d\Omega_0 = \frac{m}{4\pi G} \int_{C \cap \mathscr{I}^+}\frac{1}{\hat{\sigma}^{3}}  \, d\Omega_0 = \mathcal{M}_{\text{Bondi}} \, ,
\end{equation} 
which is precisely the Bondi mass integral as given in equation (3.30) of \cite{Bakas:2014kfa}. 

The flux formula in this case gives 
\begin{equation} \label{eq: Bondi_mass_loss}
- \int_{B_{12}} \bm{F}_{\xi} =  %-\frac{2mi}{\kappa^2} \int_{\Delta \subset \mathscr{I}^+}  \partial_t \left(e^{(3\hat{\Phi}-\hat{\Phi}_0)/2} \right) \, dt dz d\bar{z}%
 \frac{2m}{\kappa^2} \int_{B_{12}}  \partial_t \left( \frac{1}{\hat{\sigma}^3} \right) \, dt d\Omega_0 \leq 0\,,
\end{equation}
where the final inequality comes from \cite{Singleton2020, Chrusciel:1992cj, Chrusciel:1992rv} (see Appendix A of \cite{Chrusciel:1992cj} for a proof of this result). The aformentioned work was for the asymptotically flat RT spacetime and the inequality was used in order to prove that the  Bondi mass is monotonically decreasing in the RT spacetime, but the monotonicity proof only uses the RT equation \eqref{eq: RT_equation} which is independent of $\Lambda$. The proof thus extends immediately to the $\Lambda \neq 0$ cases and thus the flux formula gives a monotonically decreasing quantity. 

\paragraph{Calabi functional}

A second interesting example of a monotonic charge arises from the choice of 
\begin{equation}
    \xi_{(0)}^t = f = (\hat{\Delta} \hat{\Phi} )^2\,,
\end{equation}
which again satisfies the asymptotic condition \eqref{eq: vector_sds_limit} and gives the modified Hamiltonian 
\begin{equation}
    \mathcal{H}_{\xi} = \frac{2m}{\kappa^2} \int_{\partial C} (\hat{\Delta} \hat{\Phi} )^2 \, d\Omega = \frac{m}{2\kappa^2} \int_{\partial C} (R_2 )^2 \, d\Omega =  \frac{m}{2\kappa^2} \mathcal{C} \,,
\end{equation}
where we read off the Calabi functional $\mathcal{C}$ on the right hand side. Examination of the flux form gives
\begin{equation} 
 - \int_{B_{12}} \bm{F}_{\xi} = \frac{m}{2\kappa^2} \frac{d}{dt} \mathcal{C} \leq 0\,,
\end{equation}
where we used the fact that the Calabi functional is monotonically decreasing along the flow. For a nice proof of this, see {\it e.g.} Appendix A of \cite{Chrusciel:1992cj}. We note again that this monotonicity only depends on the Calabi flow (RT equation) which is independent of $\Lambda$. 

\paragraph{Other monotonically decreasing quantities}

In the two preceding examples we have been able to recover well-known monotonically decreasing quantities associated with the RT spacetime as modified Hamiltonians at $\mathscr{I}^+$. One may be interested in constructing other quantities that obey monotonicity, with a further example being the choice 
\begin{equation}
    \xi_{(0)}^t = f = \hat{\Phi}\,,
\end{equation}
which we note does not satisfy \eqref{eq: vector_sds_limit} and as such it is not a good candidate to describe the loss of energy through $\mathscr{I}^+$. In any case, this vector field leads to a modified Hamiltonian of the form 
\begin{equation}
    \mathcal{H}_{\xi} = \frac{2m}{\kappa^2} \int_{\partial C} \hat{\Phi} \, d\Omega\,,
\end{equation}
which admits a monotonically decreasing flux via \eqref{eq: flux_RTdS_explicit}
\begin{equation}
    - \int \mathbf{F}_{\xi} = - \frac{2}{3\kappa^2} \int (\hat{\Delta} \hat{\Phi} )^2 \, dtd\Omega \leq 0\,.
\end{equation}

While this monotonic charge is not related to the loss of energy of the system (due to the aformentioned failing of \eqref{eq: vector_sds_limit}), this construction highlights how one may use the flux-balance law in order to construct potentially novel monotonically decreasing functionals associated with the spacetime. 

We also note that the construction of gravitational charges as discussed in this work may allow one to understand global aspects of the RTdS metrics using their properties. Interesting global issues for RTdS metrics include the presence and properties of a past apparent horizon (see \cite{Tod_1989} for $\Lambda = 0$ and \cite{Bakas:2014kfa} for $\Lambda <0$ cases) and whether the Bondi mass of the spacetime satisfies a Penrose inequality \cite{Penrose:1969pc} (see \cite{KPTod_1986} for $\Lambda =0$ and \cite{Bakas:2014kfa} for $\Lambda <0$) or Thorne's hoop conjecture \cite{1972mwm..book..231T} (see \cite{Gibbons:2009xm, Cvetic:2011vt} for $\Lambda =0$, \cite{Bakas:2014kfa} for $\Lambda <0$). The study of these topics for RTdS spacetimes is at a less-advanced stage relative to their $\Lambda \leq 0$ cousins, although this work shows that several of the equations describing these global issues may now be expressed in terms of gravitational charges (such as \eqref{eq: Bondi_mass_charge}) associated with RTdS spacetimes. It would be interesting to investigate whether the properties of the charges discussed in this work would help to make headway in understanding global aspects of RTdS metrics.

\section{Conclusions and outlook} \label{sec: conclusions}

We have presented a comprehensive analysis of gravitational charges in asymptotically locally de Sitter spacetimes. There are similarities and differences relative to asymptotically locally AdS spacetimes. The  local analysis near the conformal boundary is linked to that of AdS by analytic continuation. In the AldS case, however, the spatial sections are compact and conformal infinity is now at past and future infinity, and the gravitational charges are defined using two-ended timelike hypersurfaces. In general, the overall Hamiltonian receives 
non-trivial contributions from both $\mathscr{I}^+$ and $\mathscr{I}^-$, and while classically causality and the presence of a cosmological horizon implies that $\mathscr{I}^+$ and $\mathscr{I}^-$ may be treated separately, the global nature of the Hamiltonian may become important when analysing the symmetry constraints on the transition amplitude from $\mathscr{I}^-$ and $\mathscr{I}^+$ at the quantum level.

We provided a first principle derivation of the gravitational charges and the flux-balance law they satisfy, and we illustrated our analysis using several examples. A necessary and sufficient condition for conservation of the charge is that the gravitational flux vanishes. This includes the case where the conformal boundary admits conformal Killing vectors, but (surprisingly) we also find examples where the conservation is not linked with an asymptotic conformal Killing vector. In such cases, the flux density is not vanishing but it is exact over compact directions so that its integral vanishes.

Examples of such conserved charges appeared in our analysis of the RTdS spacetime at $\mathscr{I}^+$. The RTdS possesses a non-trivial conformal structure at infinity, which (in general) does not admit conformal Killing vectors. Nevertheless, by studying the vanishing of the flux we were able to establish the existence of two conserved quantities. To understand their meaning, recall that the RTdS solution is obtained by solving the RT equation, which is the same equation that  governs the Calabi flaw on $S^2$. It is a mathematical fact that the Calabi flow is area preserving and it does not change the topology of space, {\it i.e.} the genus of the surface is invariant under the flow. The two conserved charges are precisely the area of the deformed $S^2$ and its genus. It would be interesting to find other solutions that exhibit analogous conserved charges.

The flux-balance formulae allow us also to systematically investigate the existence of monotonic charges, i.e. charges that change monotonically: the existence of such charges is related to flux of definite sign. In the context of radiation, a particularly important example of such a charge is the Bondi mass. We have indeed identified a monotonically decreasing charge that can be identified with the Bondi mass (and it is equal to the standard Bondi mass for the $\Lambda=0$ RT solution). It is well-known that the Calabi flow possesses a functional that is decreasing along the flow. We have identified the corresponding gravitational charge, and also identified yet another monotonic charge, which appears to be new. It would be interesting to understand the implications of the existence of this charge for the Calabi flow. 

We found several interesting functionals as gravitational charges for the RTdS spacetime, but a full systematic classification of the asymptotic symmetry structure is still lacking. The asymptotic conditions of \eqref{eq: vector_sds_limit} and \eqref{eq: late_time_angular_momentum} that the RTdS spacetime reduces to SdS as $u\rightarrow \infty$ are useful but not restrictive enough to identify unique quantities which describe the arrival of gravitational radiation at $\mathscr{I}^+$. In choosing our vector fields in Section \ref{sec: RTdS_examples} we often made use of the fact that several, but not all, of the components of the conformal Killing equation were satisfied. A potentially important observation is that the vector field \eqref{eq: almost_CKV} which gives the Bondi mass Hamiltonian \eqref{eq: Bondi_mass_charge} is an example of an \textit{off-shell} conformal Killing vector of \eqref{eq: RT_g_0} in that it solves the conformal Killing equation as long as $\hat{\Phi}$ does not satisfy the boundary RT equation. The physical properties of the gravitational charges associated with such a vector field suggest that the strategy of constructing off-shell conformal Killing vectors and analysing the corresponding charges may be a fruitful avenue in understanding the properties of gravitational radiation in AldS spacetimes more generally.

In this paper we focused our analysis on pure gravity in four dimensions. Natural extensions include the addition of matter, including Maxwell, scalar fields, {\it etc.}. Extending to higher dimensions would also be interesting. In odd dimensions, we would need to deal with the analogue of the holographic conformal anomaly \cite{Henningson:1998gx, Papadimitriou:2005ii}. 
In our analysis, we presented three different derivations of the conserved charges: two based on the first and second Noether theorem and the third on covariant phase space. We have shown that the expressions for the charges were equivalent, but for this to be the case we needed to include a corner term in the definition of the charge localized in the intersection of $\mathscr{I}^\pm$ and the hypersurface $C$. This allowed us to relax a technical assumption made in \cite{Papadimitriou:2005ii}. It would be interesting to work out the corner contribution in other dimensions as well.  

We discussed quantities that satisfy flux-balance laws with respect to temporal translations. It would be important to further improve the discussion of spacetime fluxes when the past end lies in the spacetime interior as discussed in Section \ref{sec: interior}. In particular, it would be useful to understand better the boundary condition at interior surfaces that are induced dynamically by the time evolution. Related to this, it would be good to understand the physics behind the fixing of the ambiguity $W$ as presented in \eqref{eq: Theta_w=0}, for example, by making reasonable physical assumptions on the nature of sources and outgoing radiation, {\it etc.}. This would allow one to construct a description of bulk gravitational radiation in AldS spacetimes.

One way of excluding incoming radiation is suggested in \cite{Ashtekar:2019khv} (the surface should be a non-expanding horizon) and this condition may be expressible in terms of the ambiguity $W$ that one adds to the past end. In order to explore this fully, one would need to extend the discussion of spacelike separated past ends contained in this work to include lightlike separated past ends. This would allow us to consider null surfaces such as the cosmological horizon as the past end \cite{Ashtekar:2019khv}. The works \cite{Chrusciel:2020rlz, Chrusciel:2021ttc, Chrusciel:2023umn} consider Hamiltonians associated with null surfaces in AldS spacetimes and may be of use in this endeavor. 

There have been several recent works that generalize concepts from 
electromagnetic radiation, such as the Poynting vector, {\it etc.}, to the gravitational setting \cite{Fernandez-Alvarez:2019kdd, Fernandez-Alvarez:2020hsv, Wylleman:2020ubq, Fernandez-Alvarez:2021yog, Fernandez-Alvarez:2023wal, Fernandez-Alvarez:2024bkf, Ciambelli:2024kre, Arenas-Henriquez:2025rpt, Fernandez-Alvarez:2025ivs}. It would be interesting to understand the relation to our flux-balance formulae. Their approach appears to relate radiation to linear independence of energy-momentum and Cotton tensors {\it i.e.} for $\Lambda >0$ the condition given is that ``there is no gravitational radiation on the open portion $B_{12} \subset \mathscr{I}^+$ if and only if $C_{ij}$ and $T_{ij}$ commute''. For the RTdS spacetime \eqref{eq: cotton_em_relations} can be used to show that the tensors do not commute and thus the RTdS spacetime has gravitational radiation, consistent with our observations as well as the recent work \cite{Arenas-Henriquez:2025rpt}. It would be interesting to explore more deeply the relation between this condition and our analysis, in particular to study whether the condition can be utilised to understand the space of conformal Killing vectors of $\mathscr{I}^+$ for a spacetime. This may then allow for a direct connection between the super-Poynting vector and the language of charges and fluxes as developed in this work.  

The definition of gravitational charges makes use of the $T^t_i$ components of the energy-momentum tensor. In non-relativistic physics the space-space components of $T_{ij}$ encode the stresses that act on bodies: the force in the $i$-direction that acts on a surface whose normal is $j$ with diagonal elements representing normal stress and off-diagonals shear. It would be interesting to extract in our case the physics associated with these components of $T_{ij}$. In the example of the RTdS spacetime, and from the perspective of a boundary fluid, such components correspond to the viscous part of the energy-momentum tensor \cite{Bakas:2014kfa, Ciambelli:2017wou, Skenderis:2017dnh, Arenas-Henriquez:2025rpt} and describe higher order corrections to the ideal hydrodynamic behaviour. 

Another future direction is to investigate whether one can use the technology introduced in this paper to study more advanced aspects of radiation, such as the gravitational memory effect in de Sitter. In asymptotically flat spacetime it has been shown that the memory effect can be understood as a vacuum transition between spacetimes related by asymptotic symmetries \cite{Flanagan:2014kfa, Strominger:2014pwa, Pasterski:2015tva}. It would be interesting to see if one can characterise the memory effect in AldS spacetimes in a similar manner. 
The study of the memory effect in de Sitter settings is a topic of some recent interest \cite{Kehagias:2016zry, Tolish:2016ggo, Ferreira:2016hee, Hamada:2017gdg, Bonga:2020fhx, Enriquez-Rojo:2022ntu, Revof:2025mgz} and it would be interesting to examine how our formalism relates to these approaches.

Last but not least, it would be interesting to extract that implication of the charges at the quantum level. 

\acknowledgments

We would like to thank Abhay Ashtekar, Antony J. Speranza and Robert M. Wald for discussions. A.P. is supported by the Brain Pool Fellowship RS-2025-25457100 at Kyung Hee University, by the National Science and Technology Council, R.O.C. (NSTC 112-2112-M-002-024-MY3 and NSTC 113-2112-M-002-040-MY2), and by National Taiwan University. A.P. acknowledges the hospitality of The University of Oxford and The University of Birmingham, as well as the conferences ``Quantum Extreme Universe: Matter, Information, and Gravity'' at The Okinawa Institute of Science and Technology and ``NTU-NCTS Holography \& Quantum Information Workshop'' at National Taiwan University, where preliminary results from this work were presented. K.S. is supported in part by the STFC consolidated grant ST/X000583/1 ``New Frontiers in Particle Physics, Cosmology and Gravity." KS acknowledges the hospitality of the  Simons Center for Geometry and Physics at Stony Brook and the International Centre for Mathematical Sciences in Edinburgh. M.T. is supported in part by the EPSRC grant EP/Y028872/1 Mathematical Foundations of Intelligence. M.T. acknowledges the hospitality of the International Centre for Mathematical Sciences in Edinburgh and the Galileo Galilei Institute in Florence. 

\appendix

\section{Useful results on the covariant phase space} \label{sec: CPS_proofs}

We provide proofs for a number of results concerning the covariant phase space that were stated in the main text.

\subsection{Symplectic current is closed on-shell} \label{sec: symplectic_current_closed}

We can rewrite equation (\ref{eq: omega_closed}) as 
\begin{equation}
d \boldsymbol{\omega} \approx 0\,,   
\end{equation}
a result which we will prove following \cite{Lee:1990nz, Wald:1999wa} (a nice recap is also given in \cite{Papadimitriou:2005ii}). We start by noting that two variations of the Lagrangian gives
\begin{align}
\begin{split}
\delta_1 \delta_2 \mathbf{L} & = \delta_1 (\mathbf{E} \delta_2 \psi + d \bm{\Theta}(\psi, \delta_{2} \psi)) \\
& = \delta_1 \mathbf{E}  \delta_2 \psi + \mathbf{E}  \delta_1 \delta_2 \psi + \delta_1 d \bm{\Theta}(\psi, \delta_{2} \psi) \\
& \approx d  \delta_1 \bm{\Theta}(\psi, \delta_{2} \psi)\,,
\end{split}
\end{align}
where we used (\ref{eq: Lagrangian_variation}) as well as the commutativity of the variation $\delta_1$ and the spacetime exterior derivative $d$. By an almost identical line of argument we can write 
\begin{align}
\delta_2 \delta_1 \mathbf{L} \approx d \delta_2 \bm{\Theta}(\psi, \delta_{1} \psi)\,, 
\end{align}
and now using the fact that the variation operators commute ($[\delta_1 , \delta_2]=0$) we are able to write 
\begin{align} \label{eq: closed_omega}
\begin{split}
0 = \delta_1 \delta_2 \mathbf{L} - \delta_2 \delta_1 \mathbf{L} & \approx d \delta_1 \bm{\Theta}(\psi, \delta_{2} \psi) - d  \delta_2 \bm{\Theta}(\psi, \delta_{1} \psi) =d \boldsymbol{\omega}(\psi, \delta_1 \psi, \delta_2 \psi)\,,
\end{split}
\end{align}
which completes the proof.

\subsection{Variation of Noether current} \label{sec: Noether_current_variation}
The variation of the Noether current form is given by
\begin{align}
\begin{split} \label{eq: delta_J_off_shell}
\delta \mathbf{J} [\xi] & = \delta \bm{\Theta} ( \psi, \mathcal{L}_{\xi} \psi ) - \delta (i_{\xi} \mathbf{L} ) \\
& = \delta \bm{\Theta} ( \psi, \mathcal{L}_{\xi} \psi ) - i_{\xi} \delta \mathbf{L} \\
& =  \delta \bm{\Theta} ( \psi, \mathcal{L}_{\xi} \psi ) - i_{\xi} [d \bm{\Theta}(\psi, \delta \psi)+\mathbf{E}(\psi) \delta \psi ] \\
& = \delta \bm{\Theta} ( \psi, \mathcal{L}_{\xi} \psi ) - \mathcal{L}_{\xi} \bm{\Theta}(\psi, \delta \psi) + d i_{\xi} \bm{\Theta}(\psi, \delta \psi) - i_{\xi} \mathbf{E}(\psi) \delta \psi \\
& = \boldsymbol{\omega} (\psi, \delta \psi, \mathcal{L}_{\xi} \psi) + d i_{\xi} \bm{\Theta}(\psi, \delta \psi)- i_{\xi} \mathbf{E}(\psi) \delta \psi\,.
\end{split}
\end{align}
In going from the first to the second line we have used the property of $\xi$ being fixed and from the second to the third line we have used (\ref{eq: Lagrangian_variation}). In third to the fourth lines we have applied the identity $\mathcal{L}_{\xi}=i_{\xi} d + d i_{\xi}$ for the Lie derivative acting on differential forms and finally we have applied (\ref{eq: omega_def}) in order to reinstate $\boldsymbol{\omega}$. One can now easily see that by going on-shell, $\mathbf{E} \approx 0$, equation (\ref{eq: J_omega_rel}) is immediately obtained.

 More generally than the on-shell result (\ref{eq: Q_def}), it has been shown in \cite{Iyer:1995kg} that we can write
\begin{equation} \label{eq: off_shell_J}
    \mathbf{J}[\xi]=d \mathbf{Q}[\xi]+ i_{\xi} \mathbf{C}(\psi)\,,
\end{equation}
where $\mathbf{C}$ is a $d$-form which vanishes on-shell (physically this form vanishing is fulfillment of the constraint equations of the theory \cite{Lee:1990nz}). We can use (\ref{eq: off_shell_J}) to simplify (\ref{eq: delta_J_off_shell}) via
\begin{align} \label{eq: off_shell_omega}
\begin{split}
\boldsymbol{\omega} (\psi, \delta \psi, \mathcal{L}_{\xi} \psi) & =  \delta (d \mathbf{Q}[\xi]+ i_{\xi} \mathbf{C}(\psi) ) - d (i_{\xi} \bm{\Theta}(\psi, \delta \psi)) + i_{\xi} \mathbf{E}(\psi) \delta \psi \\
& = d [\delta \mathbf{Q}[\xi] - i_{\xi} \bm{\Theta}(\psi, \delta \psi)] + i_{\xi} \delta \mathbf{C}(\psi) + i_{\xi} \mathbf{E}(\psi) \delta \psi\,,
\end{split}
\end{align}
and now going on-shell at the level of both the equations of motion and the linearised equations of motion, the final two terms vanish and one immediately arrives at equation (\ref{eq: on_shell_omega}).

\section{Holographic counterterms from the Hamiltonian approach} \label{sec: counterterms}

We derive the explicit form of the counterterm action (\ref{eq: S_ct}). In doing this, we demonstrate that this method is completely equivalent to the Lagrangian approach to holographic renormalisation for AldS spacetimes as initially discussed in \cite{Skenderis:2002wp} and more recently considered in the context of asymptotic charges in \cite{Compere:2020lrt, Kolanowski:2021hwo}. 

In order to compute the counterterm action (\ref{eq: S_ct}), we merely need to compute the eigenfunctions of the dilatation operator $K_{(0)}, K_{(2)}$. The first one of these is easily computed using (\ref{eq: K_eigenfunctions}) and taking the trace of the first equation, resulting in 
\begin{equation}
    K_{(0)} = 3\,.
\end{equation}

The second term, $K_{(2)}$, requires a little more work as simply tracing over the second equation of (\ref{eq: K_eigenfunctions}) gives an expression which is not manifestly covariant in terms of the induced metric on the constant $\tau$ hypersurface $\Sigma_{\tau}$. In order to derive a covariant expression for $K_{(2)}$, we plug the expansions (\ref{eq: dilatation_expansions}) into the first equation of (\ref{eq: Einstein_ADM}) and solve order by order in dilatation weight. The equation at weight 2 reads\footnote{$R[\gamma] = R_{(2)}[\gamma]$ by construction.} 
\begin{equation}
    2 K_{(0) i}^{\phantom{(0)} j} K_{(2) j}^{\phantom{(2)} i} - 2 K_{(0)} K_{(2)} = R[\gamma]\,,
\end{equation}
and using the first equation in (\ref{eq: K_eigenfunctions}), we arrive at the result 
\begin{equation}
    K_{(2)} = - \frac{R[\gamma]}{4}\,,
\end{equation}
and thus the counterterm action (\ref{eq: S_ct}) is
\begin{equation}
    S_{\text{ct}} =  \frac{1}{8 \pi G}  \int_{\Sigma_{\tau_0}} d^3x  \sqrt{\gamma} \left( 2 - \frac{R[\gamma]}{2}\right)\,,
\end{equation}
precisely in agreement with those of \cite{Balasubramanian:1999re, Skenderis:2002wp, Compere:2020lrt, Kolanowski:2021hwo}.

\section{Finiteness of the Wald Hamiltonian at \texorpdfstring{$\mathscr{I}^+$}{SCRIM}} 
\label{app: Hamiltonian_proof}

We prove the formula (\ref{eq: Final_Hamiltonian}) for the Hamiltonian defined at a cross-section of $\mathscr{I}^+$, starting from the final line of (\ref{eq: Corner_Hamiltonian}). Clearly all we need to show is 
\begin{equation} \label{eq: vanishing_corner}
    I =  \int_{\partial C}  \left( v^k_{(2)} + \sum_{n=0}^2 ( K^{\phantom{(n)}k}_{(n)l} - \lambda_{(n)} \delta^k_l) \xi^l  \right) \epsilon_{kij} dx^i \wedge dx^j = 0\,,
\end{equation}
which we will now do. We note that this proof follows very closely to that of the AlAdS case considered in appendix C of \cite{Papadimitriou:2005ii}. Although now we will be able to extend the result to spacetimes without additional isometries due to the inclusion of the corner term induced by the variation of the holographic counterterms. We begin with the definition (\ref{eq: Noether_current}), which using (\ref{eq: Q_def}) we can write on-shell as
\begin{equation}
    d \mathbf{Q}[\xi] = \bm{\Theta} (\psi, \mathcal{L}_{\xi} \psi) - i_{\xi} \mathbf{L}_{\text{on-shell}}\,,
\end{equation}
or equivalently using (\ref{eq: sym_pot}) and (\ref{eq: Q_GR}) as
\begin{equation}
   - d * \mathbf{\Xi} = * \bm{v} - * \bm{\xi} L_{\text{on-shell}}\,,
\end{equation}
where we introduced $\mathbf{L}_{\text{on-shell}} = L_{\text{on-shell}} \bm{\epsilon}$.
We can take an additional Hodge star of this equation to obtain 
\begin{align}
- *  d * \mathbf{\Xi} = * * \left( \bm{v} - \bm{\xi} L_{\text{on-shell}} \right)\,,
\end{align}
or equivalently 
\begin{equation} \label{eq: divergence_Xi}
    \nabla_{\nu} \Xi^{\nu \mu} = v^{\mu} - \xi^{\mu} L_{\text{on-shell}}\,.
\end{equation}
We use the fact that the on-shell action can be written as
\begin{equation}
    L_{\text{on-shell}} = \frac{3}{8 \pi G} = \frac{1}{8 \pi G} (\dot{\lambda} + K \lambda )\,,
\end{equation}
and study the boundary component of (\ref{eq: divergence_Xi})
\begin{equation}
   \frac{1}{8 \pi G} \partial_{\tau} \left( \sqrt{\gamma} (K^i_j - \delta^i_j \lambda) \xi^j \right) = \partial_j \left(\sqrt{\gamma} \Xi^{ji} \right) - \sqrt{\gamma} v^i + \mathcal{O}(e^{-2\tau})\,, 
\end{equation}
where we made use of (\ref{eq: extrinsic_curvature}), (\ref{eq: xi_ASG_CKV}) and (\ref{eq: Xi_t_i}) in manipulating (\ref{eq: divergence_Xi}). The equation above will be the crucial result in proving (\ref{eq: vanishing_corner}). 
In order to do this, we will expand each side of the equation in eigenfunctions of $\delta_D$, the dilatation operator, and equate terms of the same order. We have already seen that the time derivative is very closely related to this operator (\ref{eq: time_der_dilatation}) and following \cite{Papadimitriou:2004ap, Papadimitriou:2005ii} we can formally expand this operator as 
\begin{equation}
\partial_{\tau} = \delta_D + (\text{terms which vanish on} \: \mathscr{I^+})\,.
\end{equation}
Using our expansions (\ref{eq: dilatation_expansions}) for $K^i_j$ and $\lambda$ as well as the dilatation weight expansions 
\begin{equation}
\Xi^{ij} = \Xi^{ij}_{(2)}+ \ldots\,, \qquad v^i = v^i_{(2)} + \ldots\,,
\end{equation}
we have
\begin{align}
   \sqrt{\gamma} (K_{(0)j}^{\phantom{(0)}i} - \delta^i_j \lambda_{(0)}) \xi^j & = 0\,, \\
   \sqrt{\gamma} (K_{(2)j}^{\phantom{(2)}i} - \delta^i_j \lambda_{(2)}) \xi^j  & = \partial_j \left(\sqrt{\gamma} \Xi_{(2)}^{ji} \right) - \sqrt{\gamma} v_{(2)}^i \label{eq: K-l-w2}\,,
\end{align}
the first of which one can check explicitly using equations (\ref{eq: lambda_n}) and (\ref{eq: K_eigenfunctions}), thus demonstrating that the zeroth order term in (\ref{eq: vanishing_corner}) vanishes identically. Plugging (\ref{eq: K-l-w2}) back into (\ref{eq: vanishing_corner}) we find 
\begin{equation}
    I =  \int_{\partial C} \partial_l \left(\sqrt{\gamma} \Xi_{(2)}^{lk} \right) \varepsilon_{k i j} dx^i \wedge dx^j = - \int_{\partial C} d *_3 \mathbf{\Xi} = 0\,,
\end{equation}
where we used the fact that $\partial (\partial C) = \emptyset$. This completes the proof that (\ref{eq: vanishing_corner}) vanishes and demonstrates that the Hamiltonians are finite. 

\section{Variations of induced quantities} \label{app: Variations}

In the main text we make use of the variations of several of the induced quantities defined on the hypersurface $\Sigma_{\tau}$, all which are induced by the variation of the bulk spacetime metric 
\begin{equation}
    \delta_{\xi} g_{\mu \nu} = \mathcal{L}_{\xi} g_{\mu \nu}\,.
\end{equation}
Here we list several useful formulae for variations of induced quantities, making use of (\ref{eq: Lie_metric}) for the components of the Lie derivative in the gauge (\ref{eq: metric_ADM})
\begin{align}
\delta_{\xi} \gamma_{ij} & = \delta_{\xi} g_{ij} = \mathcal{L}_{\xi} g_{ij} = L_{\xi} \gamma_{ij} + 2 K_{ij} \xi^{\tau}\,, \\ 
    \delta_{\xi} \sqrt{\gamma} & = \frac{1}{2} \sqrt{\gamma} \gamma^{ij} \delta_{\xi} \gamma_{ij} = \frac{1}{2} g^{ij} \mathcal{L}_{\xi} g_{ij} = \sqrt{\gamma} (D_i \xi^i + K \xi^{\tau} )\,, \\
    \delta_{\xi} \gamma^{ij} & = \delta_{\xi} g^{ij} = - g^{ki} g^{lj} \delta_{\xi} g_{kl} = - 2 \left(D^{(i}\xi^{j)} + K^{ij} \xi^{\tau}\right)\,, \\
    2 \delta_{\xi} K_{ij} & = \delta_{\xi} \partial_{\tau} \gamma_{ij} =  \left( \partial_{\tau} \delta_{\xi} \gamma_{ij} - \delta_{\dot{\xi}} \gamma_{ij} \right) =   \partial_{\tau} (L_{\xi} \gamma_{ij} + 2K_{ij} \xi^{\tau}) - (L_{\dot{\xi}} \gamma_{ij} + 2 K_{ij} \dot{\xi}^{\tau} )\,, \\
    \delta_{\xi} K & =  \delta_{\xi}\left( K_{ij} \gamma^{ij} \right) = \partial_{\tau} \left( D_i \xi^i + K \xi^{\tau} \right) - \left(D_{i} \dot{\xi}^i + \dot{\xi}^{\tau} K \right)\,, \\
    \delta_{\xi} \left(\sqrt{\gamma} K \right) & = \xi^{\tau} \partial_{\tau} \left( \sqrt{\gamma} K\right)+ \partial_i \partial_{\tau} \left(\sqrt{\gamma} \xi^i \right) - \sqrt{\gamma} D_{i} \dot{\xi}^i = \xi^{\tau} \partial_{\tau} \left( \sqrt{\gamma} K\right)+ \partial_i \left( \xi^i \partial_{\tau} \sqrt{\gamma} \right)\,. \label{eq: Variation_det_K}
 \end{align}   
In particular, we note that the final variation is precisely the integrand of the GHY term that we consider in (\ref{eq: variation_GHY}). The total derivative term does not contribute in (\ref{eq: variation_GHY}) due to the compactness of $\Sigma_{\tau}$ and thus one gets immediately the result presented in the main text.

 \section{Charge expression at \texorpdfstring{$\mathscr{I}^-$}{SCRIM}}   \label{sec: scri_-}

We give a brief derivation of the Hamiltonian at $\mathscr{I}^-$, indicating where the signs change from the charge at $\mathscr{I}^+$. We start with the Starobinsky expansion near the boundary of $\mathscr{I}^-$ (setting $\ell=1$)
\begin{equation} \label{eq: FG_dS_scrim}
    g = \frac{1}{\rho_-^2} \left( -d\rho_-^2 + g^-_{(0)} +\rho_-^2 g^-_{(2)} + \rho_-^3 g^-_{(3)} +\ldots \right)\,, 
\end{equation}
where $\rho_- =0$ is $\mathscr{I}^-$ and $\rho_- >0$. The volume element in these coordinates is oriented as 
\begin{equation}
    \bm{\epsilon} = \frac{1}{3!} \sqrt{-g} \varepsilon_{ijk} d \rho_- \wedge dx^i \wedge dx^j \wedge dx^k\,,  
\end{equation}
and the integration is performed as 
\begin{equation}
    \int_{M^-} d^4 x \, ( \ldots ) = \int_0 d\rho_-' \int_{\Sigma_{\rho_{-}'}} d^3x \, ( \ldots )\,.  
\end{equation}
Using the (orientation-reversing) diffeomorphism 
\begin{equation} \label{eq: orientation_reversing_diffeo}
    \rho_- = e^{- \tau_-}\,,
\end{equation}
the metric (\ref{eq: FG_dS_scrim}) becomes 
\begin{equation}
    g = - d\tau_-^2 + e^{2\tau_-} g^-_{(0)} + g^-_{(2)} + e^{-\tau_-} g^-_{(3)}  +\ldots\,,
\end{equation}
where $\mathscr{I}^-$ is at $\tau_- = \infty$ and $\rho_- = \infty$ is $\tau_- = -\infty$. Integration in these coordinates becomes 
\begin{equation}
    \int_{M^-} d^4x \, (\ldots) = - \int_{\infty}^{-\infty} d\tau_-' \int_{\Sigma_{\tau_-'}} d^3x \, (\ldots) = \int_{-\infty}^{\infty} d\tau_-' \int_{\Sigma_{\tau_-'}} d^3x \, (\ldots)\,,
\end{equation}
where we note that the vector $\partial_{\tau_-}$ is now \textit{past-pointing} and so is outward-pointing from the perspective of $\mathscr{I}^-$ embedded in the bulk spacetime. This makes the $\mathscr{I}^-$ analysis computationally equivalent to the case of $\mathscr{I}^+$ as studied in Sections \ref{sec: Theory and holographic renormalisation} through \ref{sec: charges} with the replacements 
\begin{equation} \label{eq: FG_scri+_to-}
    g^+_{(0)} \rightarrow  g^-_{(0)}\,, \qquad g^+_{(2)} \rightarrow  g^-_{(2)}\,, \qquad g^+_{(3)} \rightarrow - g^-_{(3)}\,,
\end{equation}
where we note in particular that the third term changes sign. These substitutions can be applied to all expressions at $\mathscr{I}^+$ in order to recover the analogous expressions at $\mathscr{I}^-$, in particular, if we define the $\mathscr{I}^-$ energy-momentum tensor by analogy with equation \eqref{eq: EM_future} to be
\begin{equation}
     \langle T_{ij} \rangle^{\mathscr{I}^-} =  {T}^{-}_{ij}  = - \frac{3 \ell}{16 \pi G} g^-_{(3)ij}\,,
\end{equation}
then the Hamiltonian associated with a one-ended hypersurface $C$ which ends at $\mathscr{I}^-$ is
\begin{equation} 
    H_{\xi}  = - \frac{1}{2} \int_{\partial C^-} T^{-k}_{\phantom{-}l} \xi_{(0)-}^l \epsilon^{(0)-}_{kij} dx^i \wedge dx^j\,,
\end{equation} 
which arises immediately from \eqref{eq: Final_Hamiltonian} and picks up an additional minus sign due to the sign flip in \eqref{eq: FG_scri+_to-}. We note that using these results we can immediately express the Hamiltonian associated with a \textit{two-ended} hypersurface $C$ as 
\begin{equation} \label{eq: two-ended-Hamiltonian}
    H_{\xi} = \frac{1}{2} \left(\int_{\partial C^+}T^{+k}_{\phantom{+}l} \xi_{+(0)}^l \epsilon^{(0)+}_{kij} dx^i \wedge dx^j -  \int_{\partial C^-} T^{-k}_{\phantom{-}l} \xi_{-(0)}^l \epsilon^{(0)-}_{kij} dx^i \wedge dx^j \right)\,,
\end{equation}
where $\partial C^{\pm} = C \cap \mathscr{I}^{\pm}$. 
\section{Pullback of the symplectic current to a spacelike hypersurface}
\label{app: pullback_omega}

Start from the generic expression for the pullback of the symplectic current 
\begin{equation} \label{eq: symplectic_current_pullback}
\bm{\omega}(\psi, \delta \psi, \mathcal{L}_{\zeta}\psi) |_{I_t} = ( \delta \pi^{ij} \mathcal{L}_{\zeta} \gamma_{ij} - \mathcal{L}_{\zeta} \pi^{ij} \delta \gamma_{ij}  ) \bm{\varepsilon}_3\,,
\end{equation}
for a generic vector $\zeta$ defined on the full spacetime $g_{\mu \nu}$. We begin by noting 
\begin{align}
\begin{split}
\mathcal{L}_{\zeta} \gamma_{ij} & = L_{\zeta} \gamma_{ij} + \zeta^t \dot{\gamma}_{ij} = D_i \zeta_j +D_j \zeta_i + \zeta^t \dot{\gamma}_{ij}\,, \\
\mathcal{L}_{\zeta} \pi^{ij} & = L_{\zeta} \pi^{ij} + \zeta^t \dot{\pi}^{ij} = \zeta^k D_k \pi^{ij} - \pi^{ik} D_k \zeta^j - \pi^{kj} D_k \zeta^i + (D_k \zeta^k) \pi^{ij} +  \zeta^t \dot{\pi}^{ij}\,,
\end{split}
\end{align}
where $D_i$ is the covariant derivative compatible with $\gamma_{ij}$, and we recall that $\pi^{ij}$ is a tensor density of weight $+1$. Applying these expressions to the symplectic current \eqref{eq: symplectic_current_pullback} gives us 
\begin{align}
\begin{split} \label{eq: omega_zeta}
\bm{\omega}(\psi, \delta \psi, \mathcal{L}_{\zeta}\psi) |_{I_t} & =  \big( \delta \pi^{ij} [ D_i \zeta_j +D_j \zeta_i + \zeta^t \dot{\gamma}_{ij}] \\
& \phantom{aaaaaa}  - \delta \gamma_{ij} [ D_k(  \zeta^k \pi^{ij} )- \pi^{ik} D_k \zeta^j - \pi^{kj} D_k \zeta^i + \zeta^t \dot{\pi}^{ij}]  \big) \bm{\varepsilon}_3\,,
\end{split}
\end{align}
at which point we want to use the ADM equations of motion \eqref{eq: Einstein_ADM} in order to substitute equations for $\dot{\gamma}_{ij}$ and $\dot{\pi}^{ij}$ as well as the linearised constraint equations in order to rearrange the terms involving $\delta \pi^{ij}$ and $ \delta \gamma_{ij}$. For the constraint equations, we follow the presentation of Appendix E.2 of \cite{Wald:1984rg} (the only difference here is that we have $\Lambda>0$) and they read
\begin{align}
-R[\gamma] + 4 \kappa^4 \gamma^{-1} & \pi^{ij} \pi_{ij} -  2 \kappa^4 \gamma^{-1} \pi^2 + 2 \Lambda = 0\,, \label{eq: constraint_1} \\
& D_i (\gamma^{-1/2} \pi^{ij} ) = 0\,,
\end{align}
where $\pi= \pi^i_i$. This means that the linearised constraint equations are 
\begin{align}
R^{ij}[\gamma] \delta \gamma_{ij}  + 4\kappa^4 \delta & ( \gamma^{-1}  \pi^{ij} \pi_{ij}) -  2 \kappa^4 \delta( \gamma^{-1} \pi^2) = 0\,, \label{eq: lin_constraint_1} \\
& D_i (\delta \pi^{ij} ) + \delta \Gamma^j_{ik} \pi^{ik} = 0\,. \label{eq: lin_constraint_2}
\end{align}
Similarly, the dynamical equations are given by
\begin{align}
\dot{\gamma}_{ij} & = 4 \kappa^2 \gamma^{-1/2} \left(\pi_{ij} - \frac{1}{2} \gamma_{ij} \pi \right)\,, \label{eq: gamma_evolution} \\
\begin{split}
2\kappa^2 \dot{\pi}^{ij} & = -\gamma^{1/2} \left( {R}^{ij}[\gamma] + 2\kappa^4 \gamma^{ij} \gamma^{-1} \pi^2 - 4\kappa^4 \gamma^{ij} \gamma^{-1} \pi^{kl} \pi_{kl} \right) \\
&\qquad - 8 \kappa^4 \gamma^{-1/2} \left( \pi^{ik} \pi_{k}^{j} - \frac{1}{2} \pi \pi^{ij} \right)\,, \label{eq: pi_evolution}
\end{split}
\end{align}
where we have applied (\ref{eq: constraint_1}) in writing the evolution equation for $\pi^{ij}$, accounting for the differences between this and that found in \cite{Wald:1984rg}. The aim is to use these equations in order to prove $D$-exactness of the symplectic current, and be able to extract the $\delta$-integrable and $\delta$-non-integrable parts once the spacetime integration has been performed. Rearranging the expression for (\ref{eq: omega_zeta}), and using the second of the linearised constraint equations (\ref{eq: lin_constraint_2}), together with the identity 
\begin{equation}
D_k (\delta \gamma_{ij} ) = \gamma_{lj} \delta \Gamma^{l}_{ki} - \gamma_{il} \delta \Gamma^l_{kj}\,, 
\end{equation}
we are able to re-express (\ref{eq: omega_zeta}) as 
\begin{align}
\begin{split} 
\bm{\omega}(\psi, \delta \psi, \mathcal{L}_{\zeta}\psi) |_{I_t} = \left( D_i (2\delta(\pi^{ij} \zeta_j ) - \zeta^i \pi^{jk} \delta \gamma_{jk} )+ \zeta^t[ \delta \pi^{ij}  \dot{\gamma}_{ij} - \delta \gamma_{ij} \dot{\pi}^{ij}] \right) \bm{\varepsilon}_3\,,
\end{split}
\end{align}
upon which we use the evolution equations of motion (\ref{eq: gamma_evolution}), (\ref{eq: pi_evolution}) together with the first linearised initial value constraint (\ref{eq: lin_constraint_1}) in order to show that the final term proportional to $\zeta^t$ is in fact zero, giving us a neat expression for the pullback of the symplectic current:
\begin{equation}
\bm{\omega}(\psi, \delta \psi, \mathcal{L}_{\zeta}\psi) |_{I_t} =  D_i \left(2\delta(\pi^{ij} \zeta_j ) - \zeta^i \pi^{jk} \delta \gamma_{jk} \right) \bm{\varepsilon}_3\,,
\end{equation}
which matches equation \eqref{eq: pullback_omega_2} in the main text. 

\section{Flux formula for Bondi time translation in linearised RTdS} \label{app: Quadratic_flux}

One of the examples of gravitational charges in Section \ref{sec: RTdS_examples} demonstrates that the Bondi mass is a monotonically decreasing charge at $\mathscr{I}^+$. In order to arrive at this result, we made the choice of vector field \eqref{eq: almost_CKV}, which is certainly not a unique choice of vector field satisfying the requirement \eqref{eq: vector_sds_limit}. 
In this appendix we consider a different choice of vector field, namely that of the \textit{Bondi time translation}
\begin{equation} \label{eq: Bondi_time_translation}
    \xi = \partial_{u_\text{B}} \,, 
\end{equation}
where $u_{\text{B}}$ is the \textit{retarded Bondi time} coordinate \cite{Bondi:1962px}. This vector field was already used in \cite{Kolanowski:2020wfg, Chrusciel:2020rlz, Chrusciel:2021ttc, Bonga:2023eml} to define a notion of Bondi mass for AldS spacetimes, and thus by considering this vector field within our framework we will be able to compare our result with the existing literature.  

We will follow \cite{Bonga:2023eml} most closely, a work which considers \textit{axisymmetric} RT solutions \textit{linearised} around the SdS solution (see also \cite{Robinson:1988ue} for a brief review of axisymmetric RT metrics with $\Lambda =0$). We begin by recalling their quadratic flux formula 
\begin{equation} \label{eq: RT_flux_BBP}
    \frac{dE}{du_{\text{B}}} = - \frac{\epsilon^2}{16 \pi G} \oint d^2 S \, \left[ \left( \dot{f}'' - \cot \theta \dot{f}' \right)^2 + \frac{3m}{\ell^2} \partial_{u_\text{B}} \left( f'^2 \right) \right]\,,  
\end{equation}
where $\epsilon$ is the perturbative parameter away from the Schwarzchild solution. We note that the expression above contains both a monotonically decreasing term (the first one) and a term which is not monotonically decreasing and depends on the cosmological constant $\Lambda = 3/\ell^2$. In order to understand this formula within our framework, we first provide an explanation of the notation above. The function $f$ is defined through the linearised map from RT into Bondi coordinates 
\begin{align}
\begin{split} \label{eq: RT_Bondi_linear}
        u_{\text{RT}} & = u_{\text{B}} - \epsilon f(u_{\text{B}}, \theta_{\text{B}}) + \mathcal{O}(r_\text{B}^{-3})\,, \\
        \theta_{\text{RT}} & = \theta_{\text{B}} + \mathcal{O}(r_{\text{B}}^{-1}) \,,
        \end{split}
\end{align}
and we also note that $\dot{f} = \partial_{u_\text{B}} f$, $f' = \partial_{\theta_\text{B}} f$ which coincide to leading order with the same derivatives in the RT coordinates, hence the quadratic flux formula \eqref{eq: RT_flux_BBP} can be taken to be read with either the Bondi or RT coordinates. We also note that $\oint d^2 S = \int d\Omega_0 = \int \sin \theta \, d\theta d\phi$, {\it i.e.} the integral over the round unit $S^2$.

It is of interest to see whether the Bondi time translation \eqref{eq: Bondi_time_translation} recovers the flux formula \eqref{eq: RT_flux_BBP} when input into our charge formula \eqref{eq: RT_general_charge}. We first express this vector field in terms of RT coordinates, which can be done schematically via 
\begin{equation}
    \partial_{u_\text{B}} = \frac{\partial x^{\mu}_{\text{RT}}}{\partial u_\text{B}} \partial_{\mu} = \frac{\partial u_{\text{RT}}}{\partial u_\text{B}} \partial_{u_{\text{RT}}} + \frac{\partial \theta_{\text{RT}}}{\partial u_\text{B}} \partial_{\theta_{\text{RT}}} + \frac{\partial r_{\text{RT}}}{\partial u_\text{B}} \partial_{r_{\text{RT}}}\,,
\end{equation}
for the axisymmetric case. We then express the vector in FG coordinates via the maps \eqref{eq: RT_FG_initial}, \eqref{eq: RT_FG_main} obtaining
\begin{align}
\begin{split}
     \partial_{u_\text{B}} & = \frac{\partial u_{\text{RT}}}{\partial u_\text{B}} \partial_{t} + \frac{\partial \theta_{\text{RT}}}{\partial u_\text{B}} \partial_{\theta_{\text{RT}}} + \frac{\partial r_{\text{RT}}}{\partial u_\text{B}} \partial_{r_{\text{RT}}} \\
     & = \frac{\partial u_{\text{RT}}}{\partial u_\text{B}} \partial_{t} + \frac{\partial \theta_{\text{RT}}}{\partial u_\text{B}} \partial_{\theta} - \frac{\partial r_{\text{RT}}}{\partial u_\text{B}} \frac{\partial r_*}{\partial r_{\text{RT}}} \partial_{\rho} + \mathcal{O}(\rho) \,,
     \end{split}
\end{align}
where the coordinates with no subscript correspond to boundary coordinates. We want to pull this vector field back to the boundary and since this is given by $\rho = \rho_0 (\rightarrow 0)$ we can ignore the $\partial_{\rho}$ component as well as the $\mathcal{O}(\rho)$ terms as we proceed. We are thus interested in computing the charge \eqref{eq: RT_general_charge} defined by the vector field
\begin{equation}
  \left.  \partial_{u_B} \right|_{\mathscr{I}^+} = \frac{\partial u_{\text{RT}}}{\partial u_B} \partial_{t} + \frac{\partial \theta_{\text{RT}}}{\partial u_B} \partial_{\theta} \, ,
\end{equation}
up to quadratic order in perturbation $\epsilon$ around the SdS solution. 

In order to do this, we use the non-linear transformation from RT to Bondi coordinates as given in Appendix B of \cite{Bonga:2023eml}, which we write as
\begin{align}
\begin{split} \label{eq: RT_to_Bondi}
    u_{\text{RT}} & = F(u_B, \theta_{B}) + \mathcal{O}(1/r_B) \,, \\
    \theta_{\text{RT}} & = H(u_B, \theta_B) + \mathcal{O}(1/r_B)\,,
    \end{split}
\end{align}
where we can ignore the subleading terms in the Bondi radial coordinate as these will vanish as $\rho \rightarrow 0$. We note from \cite{Bonga:2023eml} the constraints that must be satisfied by these functions 
\begin{equation} \label{eq: rt_bondi_constraint_1}
    H'^2 + \frac{P^2 F'^2}{\ell^2} = \csc^2 \theta \sin^2 H\,,
\end{equation}
which comes from demanding that the leading order part of the angular metric in Bondi gauge is the unit round $S^2$ metric, and
\begin{equation} \label{eq: rt_bondi_constraint_2}
    P ( \dot{F} H' - F' \dot{H}) = \csc^2 \theta \sin^2 H\,,
\end{equation}
which comes from demanding the correct asymptotic behaviour of the $g_{uu}$ component at leading order in Bondi gauge.\footnote{The asymptotics of the Bondi gauge in \cite{Bonga:2023eml} is different from the ones of {\it e.g.} \cite{Poole:2018koa, Compere:2019bua}. More precisely, \cite{Bonga:2023eml} has $g_{u A} = \mathcal{O}(r_{\text{B}}^2)$ but has $g_{AB} = r_{\text{B}}^2 \mathring{\gamma}_{AB} +\mathcal{O}(r_{\text{B}})$, where $\mathring{\gamma}_{AB}$ is the metric of the unit round $S^2$, whereas \cite{Compere:2019bua} uses the $\Lambda$-BMS gauge choice of $g_{AB} = r_{\text{B}}^2 \gamma_{AB} +\mathcal{O}(r_{\text{B}})$, where $\sqrt{\gamma} = \sqrt{\mathring{\gamma}}$, but has $g_{u A} = \mathcal{O}(1)$. These correpsond to different choices of boundary gauge fixing. It would also be interesting to examine the charge corresponding to the $\Lambda$-BMS choice.} Note that $P = P(F(u_B, \theta_B), H(u_B,\theta_B))$ and in matching our conventions we have $P = \hat{\sigma} = e^{(\hat{\Phi}_0-\hat{\Phi})/2}$. We note that the linear order of the map \eqref{eq: RT_to_Bondi} can be read off from \eqref{eq: RT_Bondi_linear}, with the result being 
\begin{align}
    \begin{split}
         F(u_B, \theta_B) & = u_B- \epsilon f(u_B, \theta_{B})  \,, \\
   H(u_B, \theta_{B}) & = \theta_B\,.
    \end{split}
\end{align}
We want to compute the mass aspect up to quadratic order in order to analyse the flux, so we want to extend the transformation above to $\mathcal{O}(\epsilon^2)$. We start with the general form of the transformation
\begin{align}
    \begin{split}
         F(u_B, \theta_B) & = 1- \epsilon f(u_B, \theta_{B}) + \epsilon^2 f_2(u_B, \theta_B) \,, \\
   H(u_B, \theta_{B}) & = \theta_B + \epsilon^2 h_2 (u_B, \theta_B)\,,
    \end{split}
\end{align}
and then use equations \eqref{eq: rt_bondi_constraint_1}, \eqref{eq: rt_bondi_constraint_2} to solve for $f_2, h_2$. We start with \eqref{eq: rt_bondi_constraint_1} and find at quadratic order in $\epsilon$ the equation
\begin{equation}
    \frac{f'^2}{\ell^2} + 2 h_2' - 2 h_2 \cot \theta_B = 0\,,
\end{equation}
which in particular tells us that 
\begin{equation}
    \oint d^2 S \, \frac{f'^2}{\ell^2} = -4 \oint d^2 S \, h_2'\,,
\end{equation}
where we made use of integration by parts on the sphere. With this in hand, we now rewrite \eqref{eq: rt_bondi_constraint_2} as 
\begin{equation}
    \dot{F} = \frac{1}{P H'}\left(   H'^2 + \frac{P^2 F'^2}{\ell^2} + P F' \dot{H} \right)\,,
\end{equation}
which is a useful result in evaluating the vector field since 
\begin{equation}
    \left.  \partial_{u_B} \right|_{\mathscr{I}^+} = \dot{F} \partial_{t} + \dot{H} \partial_{\theta} \, ,
\end{equation}
we note that since $\dot{H} = \mathcal{O}(\epsilon^2)$ and $\partial_\theta(\hat{\Delta} \hat{\Phi}) = \mathcal{O}(\epsilon)$ the angular contribution to the modified Hamiltonian \eqref{eq: J_Rt} is $\mathcal{O}(\epsilon^3)$ and it is thus suitable to consider merely 
\begin{equation}
      \left.  \partial_{u_B} \right|_{\mathscr{I}^+} = \dot{F} \partial_{t}\,,
\end{equation}
when computing up to $\mathcal{O}(\epsilon^2)$. Using this together with \eqref{eq: H_Rt}, we write the modified Hamiltonian as
\begin{equation}
    \mathcal{H}_{\xi_B} = \frac{2 m}{\kappa^2} \int_{\partial C}  \dot{F}  e^{\hat{\Phi}-\hat{\Phi}_0} \, d\Omega_0   = \frac{2 m}{\kappa^2} \oint  \frac{1}{P^3} \left(   H' + \frac{P^2 F'^2}{\ell^2 H'} + \frac{P F' \dot{H}}{H'} \right) \, d^2 S\, ,
\end{equation}
and now expanding the round bracket above up to $\mathcal{O}(\epsilon^2)$ we find 
\begin{equation} \label{eq: mod_Ham_Bondi}
    \mathcal{H}_{\xi_B} = \frac{2 m}{\kappa^2} \oint  \frac{1}{P^3} \left(   1+ \epsilon^2 \left[ \frac{f'^2}{\ell^2} + h_2' \right]\right) \, d^2 S = \frac{2 m}{\kappa^2} \oint  \frac{1}{P^3} \left(   1+ \epsilon^2\frac{3}{4} \frac{f'^2}{\ell^2} \right) \, d^2 S \, ,
\end{equation}
which is a promising-looking $\mathcal{O}(\epsilon^2)$ correction to the monotonically decreasing quantity given in \eqref{eq: Bondi_mass_charge}. 

In order to compare the above expression directly with \eqref{eq: RT_flux_BBP}, we first analyse the monotonically decreasing part. We begin with the expression for the ``Bondi mass'' part of the modified Hamiltonian
\begin{equation}
    \mathcal{M}_B = \frac{2 m}{\kappa^2} \oint  \frac{1}{P^3} \, d^2 S \, ,
\end{equation}
and we wish to look at the boundary time derivative of this quantity 
\begin{equation}
   \partial_t \mathcal{M}_B = -\frac{6 m}{\kappa^2} \oint  \frac{\dot{P}}{P^4} \, d^2 S \, ,
\end{equation}
clearly the important quantity above is $\dot{P}$, which we can compute using the quadratic approximation as 
\begin{equation}
   P = 1+ \epsilon p + \epsilon^2 p_2\,, \implies  \dot{P} = \epsilon \dot{p} + \epsilon^2 \dot{p}_2\,.
\end{equation}
In order to solve for $\dot{p}, \dot{p}_2$, we again use the RT equation \eqref{eq: RT_equation}, written in the notation of \cite{Bonga:2023eml} as
\begin{equation}
    3m \frac{\dot{P}}{P} + \frac{1}{8} \Delta_h R_h = 0\,,
\end{equation}
where $R_h$ is the Ricci scalar of $h = P^{-2} d\Omega_2^2$ and $\Delta_h$ is the Laplacian of the same metric. Solving this equation at $\mathcal{O}(\epsilon)$ gives 
\begin{align} \label{eq: p_dot}
   \dot{p} = - \frac{1}{12m} \left( p^{(4)}+\cot\theta  \left(2 p'''-\cot \theta  p''+\left(\csc ^2\theta +2\right) p'\right) \right)\,, 
\end{align}
and the expression for $\dot{p}_2$ is a little lengthy but arises in the same way by solving the Robinson-Trautman equation at $\mathcal{O}(\epsilon^2)$. We note that up to quadratic order the time derivative of the Bondi mass gives
\begin{equation}
     \partial_t \mathcal{M}_B = \frac{2m}{\kappa^2} \oint \left[ -3 \epsilon  \dot{p} -3 \epsilon ^2 \left(\dot{p}_2-4 p \dot{p}\right) \right] \, d^2 S\,,
\end{equation}
now using \eqref{eq: p_dot} and several integrations by parts, we see that the $\mathcal{O}(\epsilon)$ term above vanishes. The $\mathcal{O}(\epsilon^2)$ term gives 
\begin{equation}
     \partial_t \mathcal{M}_B = -\frac{\epsilon^2}{2\kappa^2} \oint  \left(p''-\cot (\theta ) p'\right)^2 \, d^2 S = -\frac{\epsilon^2}{16 \pi G} \oint  \left(\dot{f}''-\cot (\theta ) \dot{f}'\right)^2 \, d^2 S \,,
\end{equation}
which is precisely the monotonically decreasing part of \eqref{eq: RT_flux_BBP}. This means that our flux formula for the modified Hamiltonian \eqref{eq: mod_Ham_Bondi} is
\begin{equation}
    \partial_t{\mathcal{H}_{{\xi}_B}} = -\frac{\epsilon^2}{16 \pi G} \oint \left( \left(\dot{f}''-\cot (\theta ) \dot{f}'\right)^2 - 3 m \partial_t \left( \frac{f'^2}{\ell^2} \right) \right) \, d^2 S \,,
\end{equation}
which differs from \eqref{eq: RT_flux_BBP} but agrees with the flux formulae presented in \cite{Chrusciel:2020rlz, Kolanowski:2020wfg, Chrusciel:2021ttc}. 

Clearly there is some discrepancy in the literature concerning the precise flux formula for RT waves in de Sitter spacetime. However, we note that all formulae discussed so far are consistent at the level of the total energy radiated through the conformal boundary $\mathscr{I}^+$, a point which was already noted in \cite{Bonga:2023eml} when providing comparison with \cite{Ashtekar:2015lla, Kolanowski:2020wfg, Chrusciel:2021ttc}. We will now recount this argument for our Robinson-Trautman flux formula, arguing that all quadratic formulae result in overall non-positive flux of energy, assuming some basic properties on the falloff of the function $f'$. We begin by considering the total energy radiated through the boundary 
\begin{equation}
    \Delta \mathcal{H}_{\xi_{B}} = -\frac{\epsilon^2}{16 \pi G} \int_{t_0}^{\infty} dt \, \oint  \left( \left(\dot{f}''-\cot (\theta ) \dot{f}'\right)^2 - 3m \partial_t \left( \frac{f'^2}{\ell^2} \right) \right)  \, d^2 S \,,
\end{equation}
then we perform the $t$-integral over the total derivative term and use the fact that the spacetime must settle down to SdS as $t \rightarrow \infty$ which tells us $\lim_{t\rightarrow \infty} f' =0$, together with the assumption $f'(t_0)=0$. Putting these facts together we find  
\begin{equation}
    \Delta \mathcal{H}_{\xi_{B}} = -\frac{\epsilon^2}{16 \pi G} \int_{t_0}^{\infty} dt \, \oint   \left(\dot{f}''-\cot (\theta ) \dot{f}'\right)^2   \, d^2 S \,,
\end{equation}
which is manifestly non-positive. This argument shows that the linearised charge corresponding to the Bondi time translation vector field in RTdS spacetime admits a loss of energy through $\mathscr{I}^+$, a sensible result on physical grounds. 

\bibliography{refs}{}
\bibliographystyle{JHEP}
\end{document}